\documentstyle [12pt,epsf,epsfig]{article}
\begin{document}
\begin{titlepage}

\pagenumbering{arabic}

\begin{figure}[t]
\unitlength1cm
 \begin{minipage}[t]{2.cm}
    \vspace{-2.cm}
     \mbox{\hspace{-5.67cm}\epsfig{file=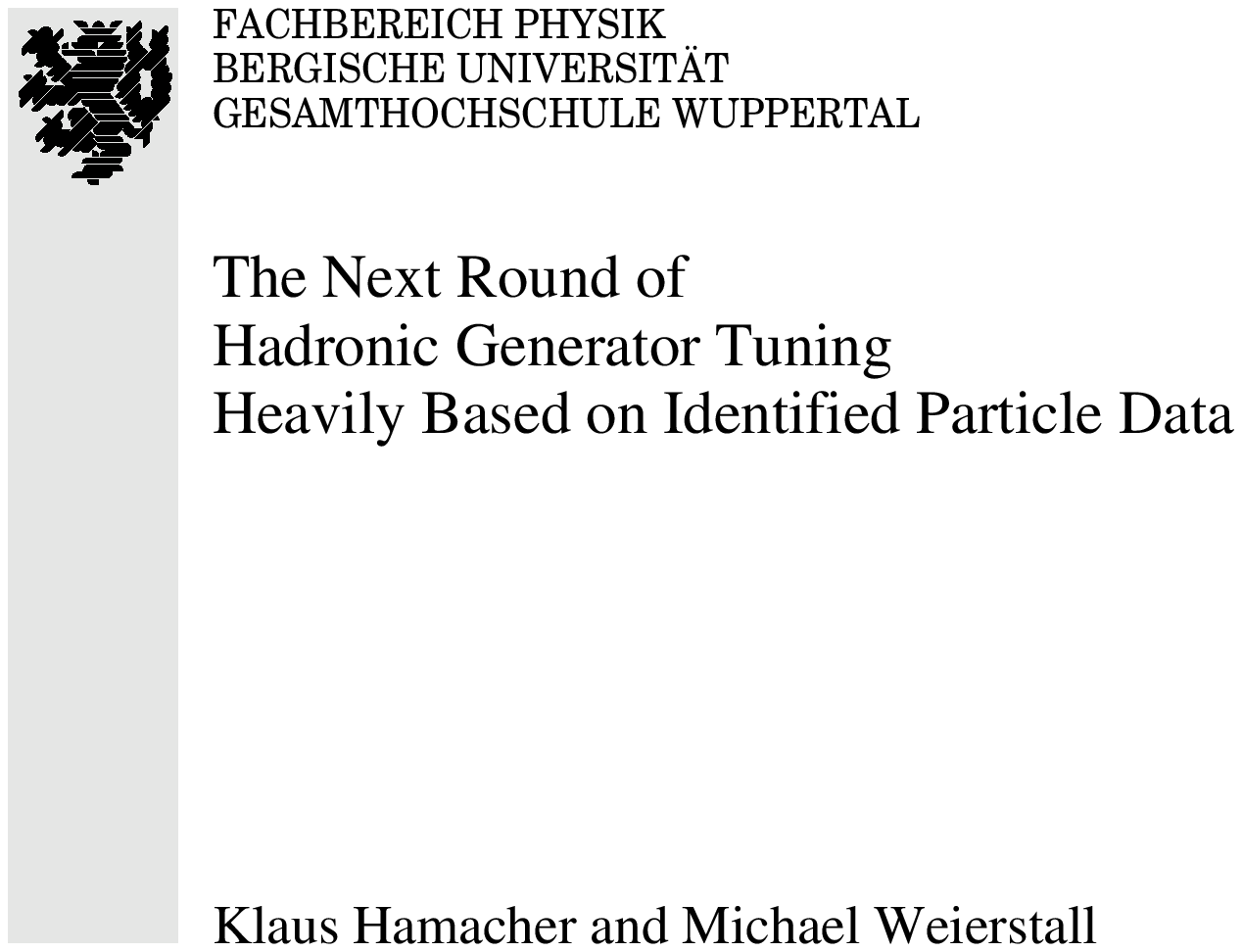,width=21.0cm}}
 \end{minipage}

\begin{minipage}[h]{13.cm}
{}~\\
\vspace{-10cm}
{}~\\
{}~\\
{}~\\
Wuppertal, Juni 1995\\
WU B 95-07\\
DELPHI 95-80 PHYS 515\\
hep-ex/951127\\
\end{minipage}
\end{figure}

\end{titlepage}

%=================================================================

\clearpage

\begin{titlepage}
\mbox{}
\end{titlepage}

\newpage

%==================================================================

%%% put your own definitions here:

\begin{titlepage}

\pagenumbering{arabic}
\begin{figure}[ht]

\unitlength1cm
 \begin{minipage}[t]{2.cm}
    \vspace{-3.cm}
     \mbox{\hspace{-5.67cm}\epsfig{file=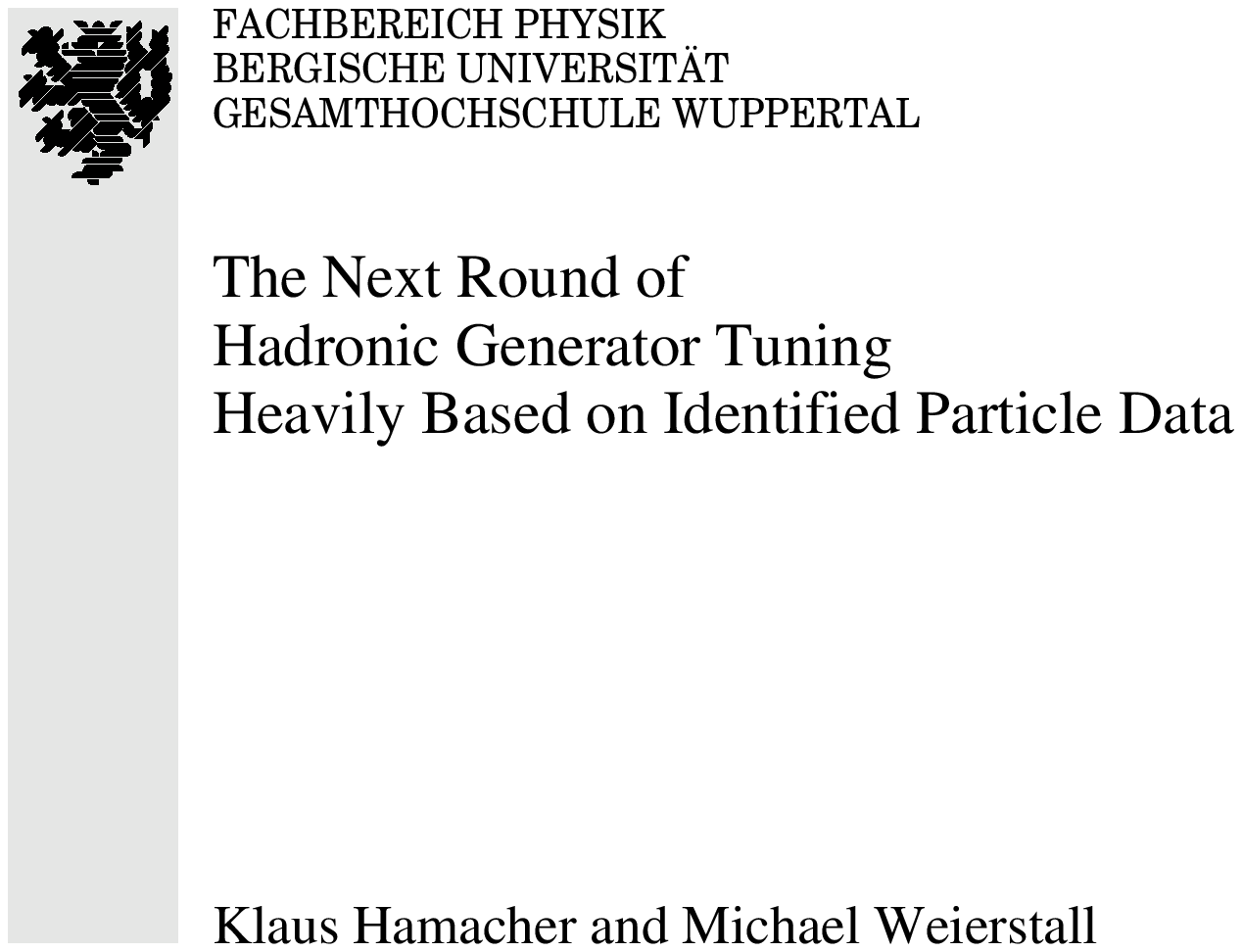,width=21.0cm}}
 \end{minipage}
\vspace{-14cm}

\noindent
\begin{minipage}[h]{13.cm}
{\bf Abstract}\\
{}~\\
Event shape and charged particle inclusive distributions
determined from 750 000 hadronic Z
events measured with the DELPHI detector at LEP
are presented. The statistical and systematic precision of this data allows
for a decisive confrontation with Monte Carlo models of the hadronization
process and a better understanding of the structure of the Z hadronic final
state.

Improved tunings of the JETSET, ARIADNE and HERWIG
parton shower models and the JETSET matrix element model are obtained
by fitting the models to identified particle distributions from all
LEP experiments and the DELPHI data presented.
The description
of the data distributions by the models is critically
reviewed with special importance attributed to identified particles.
{}~\\
{}~\\
{}~\\
{}~\\
Wuppertal, Juni 1995\\
WU B 95-07\\
DELPHI 95-80 PHYS 515\\
\end{minipage}
\end{figure}

\end{titlepage}

\clearpage
\begin{titlepage}
\mbox{}
\end{titlepage}

\pagebreak

\oddsidemargin=0cm
\setcounter{page}{1}

\newcommand{\bc}{\begin{center}}
\newcommand{\ec}{\begin{center}}
%
% ************************************************** Dokument
%
%\psdraft
%
% ************************************************** Der Text

\section{Introduction}
\label{intro}
Precision measurements at LEP using the hadronic final state, such as
determinations of the strong coupling constant $\alpha_s$ from event shapes,
the measurement of the $Z$ mass and width, the forward backward asymmetries
for quarks or at higher energies the $W^\pm$ mass
require a precise modelling of the properties of the corresponding final
states.
Perturbative QCD cannot provide full theoretical insight to
the transition of the primary quarks to observable hadrons, the so-called
fragmentation or hadronization process.
Only part of this transition including large momentum  transfer, mainly the
radiation of hard gluons or the evolution of a parton shower are
calculable perturbatively.
The final formation of hadrons is hidden due to the increase of the strong
coupling constant $\alpha_s$ at small momentum transfer and hence the
failure of perturbation theory.

Guidance to a better understanding of the hadronization process must
therefore come from detailed experimental investigations of the hadronic
final state including attempts to describe this process by phenomenological
models inspired by QCD.
LEP I is a unique unrivaled place to pursue these studies.
The clean well defined initial state in $e^+e^-$ annihilation provides
an excellent testing field, since the event rate at the $Z$ is huge, the
energy is large,
and
the capabilities of the experimental apparatus are much improved with
respect to previous experiments.

This paper attempts to determine parameters for the most frequently used
hadronization models
which lead to an optimal description of the observed hadronic event shapes and
charged particle inclusive distributions, as measured with the DELPHI
experiment at LEP, as well as the available information on identified particles
from all LEP experiments.
The latter allows one to precisely determine more model parameters than
from event shapes only and also to check the internal consistency of the
models.
The performance of these models is compared and critically reviewed.

This paper is organized as follows: Section \ref{exp}
gives a brief overview to the
relevant detector components and describes the experimental procedure
applied to determine event shape and inclusive distributions
and the related systematic errors.
Section \ref{models} discusses the models employed and
the relevant parameters.
Section \ref{fit}
describes the optimization strategy applied to obtain best parameters
for the fragmentation models and justifies the choice of distributions
used in the fit.
The fits are discussed in detail and the resulting optimized parameters
and their errors are presented.
In Section \ref{results}
a confrontation of the models to the event shape,
charge particle spectra and identified particle data is presented.
Finally we summarize in Section \ref{summary}.
The appendices contain the definitions of the variables used throughout this
paper, tables of model parameter settings and tables and plots of
the relevant data distributions and model comparisons.

\section{Detector and Data Analysis}
\label{exp}
The data used in this analysis are from the 1991, 1992 and 1993
data taking with the DELPHI detector at LEP.
For the determination of the event shape distributions we use charged particles
measured in the solenoidal 1.2 T magnetic field of DELPHI
and neutral clusters measured with the electromagnetic or hadronic
calorimeters.
The following detectors are relevant to the analysis
\cite{detector_paper}:
\begin{itemize}
\item[$\bullet$]
The Vertex Detector VD, measuring the $R\varphi$ coordinate
with up to three layers of silicon micro strips at radii between 6.3 cm
and 11 cm.
It covers polar angles $\theta$ between $37^\circ$ and $143^\circ$ .
\item[$\bullet$]
The Inner Detector ID, a 24 layer cylindrical jet chamber ($\theta$ coverage
$17^\circ$ to $163^\circ$ ).
\item[$\bullet$]
The Time Projection Chamber TPC, the principal tracker of DELPHI, with twice
6 sector
plates, 24 pad rows and 192 sense wires each. Inner and outer radius of the TPC
are $30$ cm and $122$ cm, the polar coverage is $20^\circ$ to $160^\circ$.
\item[$\bullet$]
The Outer Detector OD, a five layer drift chamber at $192$ cm radius covering
polar angles between $43^\circ$ and $137^\circ$.
\item[$\bullet$]
Two sets of forward planar drift chambers FCA and FCB with 6 or 12 layers
respectively with a overall polar angle coverage between $11^\circ$ to
$35^\circ$ and $145^\circ$ to $169^\circ$.
\item[$\bullet$]
The High density Projection Chamber HPC, a lead-gas electromagnetic calorimeter
with a very good spatial resolution
inside the DELPHI coil between $208$ cm to $260$ cm radius.
It measures the showers using the TPC principle and
covers polar angles between $43^\circ$ and $137^\circ$.
The precision on the energy can be described by
$\Delta E/E \approx 29\%/\sqrt{(E)}+4\%$.
\item[$\bullet$]
The Forward Electromagnetic Calorimeter FEMC, a lead-glass array in both
endcaps
each consisting of 4500 lead glass blocks with a pointing geometry.
The angular acceptance is between $10^\circ$ and $36.5^\circ$ to the beam.
The precision on the energy is
$\Delta E/E \approx \sqrt{(0.35+5/\sqrt{E})^2 + (6/E)^2}~\%$.
\item[$\bullet$]
The Hadron Calorimeter HAC, a iron-gas hadronic calorimeter outside the coil
consisting of at least 19 layers of streamer tubes and $5$ cm thick iron plates
also used as flux return. The overall angular covarage is $11.2^\circ$
to $168.8^\circ$. The precision on the energy is $120\%/\sqrt{E}$.
\end{itemize}
Depending on the polar angle and the detectors included in the track fit
the precision on the momentum, $p$, for charged particles in hadronic events is
$\Delta p / p^2 = 0.001$ to $0.01 GeV^{-1}$.
Charged particles were
accepted in the analysis if they satisfy the following criteria:
\begin{itemize}
\item[$\bullet$]
$p \geq 200 MeV$
\item[$\bullet$]
$\Delta p /p \leq 1$
\item[$\bullet$]
$20^\circ \leq \theta_{track} \leq 160^\circ$
\item[$\bullet$]
Measured track length $ \geq 50$ cm.
\item[$\bullet$]
Impact parameter with respect to the nominal interaction point within
$2$ cm perpendicular or $5$ cm along the beam.
\end{itemize}
Furthermore we require that charged particles with large momenta
($p \geq 25$ GeV$)$
within the geometrical acceptance of the OD or FCB have indeed been measured
by these detectors as well as by the ID or VD.
This requirement assures a good momentum resolution for high momentum
particles.

Neutral clusters or photons reconstructed from conversions
were accepted if:
\begin{itemize}
\item[$\bullet$]
$1\, GeV \leq E_{cluster}$ for clusters measured with the HAC,
\item[$\bullet$]
$0.5\, GeV \leq E_{cluster/ \gamma}$
for electromagnetic clusters or photons.
\end{itemize}

Because the available statistics is very large, the final experimental error
is dominated by the systematic error. Therefore cuts were applied to the
event kinematics to assure that the major components of the event were
measured in DELPHI with optimal efficiency and resolution, as well as to
minimize secondary interactions in the detector material.
However care has been taken not too strongly bias the measured distributions by
these cuts.
Events were selected if they fulfil the following conditions:
\begin{itemize}
\item[$\bullet$]
there were at least 5 charged particles selected,
\item[$\bullet$]
the total energy of charged particles exceeds 15 GeV and 3 GeV
for each half of the detector, defined by the (x,y) plane,
\item[$\bullet$]
the polar angle of the thrust axis is between $50^\circ$ and $85^\circ$,
\item[$\bullet$]
the momentum imbalance of the event along the beam direction is
$|\sum p_z| / \sqrt{s} \leq 0.15$
\end{itemize}
In total about $750.000$ events satisfy these cuts. The contamination of beam
gas events,
$\gamma\gamma$-events,
 and leptonic events other than $\tau^+\tau^-$, is expected
to be less than 0.1\% and has been neglected.
The influence of $\tau^+\tau^-$ events which have a pronounced 2-jet topology
and contain high momentum particles
has been determined by a simulation study using events generated by the KORALZ
model \cite{koralz} treated by the full simulation of the DELPHI detector
DELSIM \cite{delsim} and the standard data reconstruction chain.
The $\tau^+\tau^-$ contributions
have been subtracted from the measured data according to
the relative abundance of $\tau^+\tau^-$ ($0.16\pm 0.03$\%) and hadronic
events.

We present differential event shape and inclusive single particle
distributions as a function of the physical observables defined in appendix A
normalized to the number of hadronic Z decays.
Three different types of results are given,
cross-sections measured from charged particles only corrected to the initial
charged final state, or to the charged plus neutral final state and measured
charged plus neutral particles corrected to the full final state.
%\ref{variables}
%\begin{center}
%\begin{tabular}{|l|l|}
%\hline
%Name                    & Definition                                        \\
%%\hline
%\multicolumn{2}{|c|}{Single Particle Variables}                             \\
%%\hline
%scaled momentum         & $x_p$            = ${p_{track}}/{p_{beam}}$       \\
%%\hline
%transverse momentum     &                                                   \\
%in the event plane      & $p_\bot^{in}$    = $\vec{p}\cdot\vec{n}_{Major}$  \\
%%\hline
%transverse momentum     &                                                   \\
%out of the event plane  & $p_\bot^{in}$    = $\vec{p}\cdot\vec{n}_{Minor}$  \\
%%\hline
%rapidity $\parallel$  Thrust axis &
%                        $ y_T$             =
%             $\frac{1}{2}\cdot\log\frac{E+p_\parallel}{E-p_\parallel}$      \\
%%\hline
%\multicolumn{2}{|c|}{Event Shape Variables}                                 \\
%%\hline
%Thrust                  & T                =
%$\max_{\vec{n}_{Thrust}}{\sum_{tracks}|\vec{p}_i\cdot\vec{n}_{Thrust}|/\sum_{tracks}|\vec{p}_i|}$
%                                                                            \\
%%\hline
%Major                   & M as T but projecting $\bot$ to
%%$\vec{n}_{Thrust}$\\\hline
%Minor                   & m as M but projecting also $\bot$ to
%%$\vec{n}_{Major}$\\\hline
%$y_{ij}^{Durham}$       &
%\end{tabular}
%\end{center}
%\par
%\label{variables}

The observed data distributions were corrected for kinematic cuts, limited
acceptance, and resolution of the detector as well as effects due to
reinteractions of particles inside the detector material.
This correction has been calculated using simulated events generated by
JETSET 7.3 \cite{jetset} treated by the full simulation and analysis chain
as described above.

For each bin a correction
factor $C_i$ has been calculated as the ratio of the generated
and final distributions.
Particles with a lifetime bigger than  1 ns were considered as stable in
the generated distributions.
The correction for initial state photon radiation has been determined
separately using events generated by JETSET 7.3 PS
with and without initial state
radiation as predicted by DYMU3 \cite{dymu3}.

The simple unfolding by correction factors in general
leads to biases of the final results when the detector smearing is bigger
than the bin width used, and when the model does not describe the
data well \cite{unfolding}.
In our case
the model has been tuned to DELPHI data as described in
\cite{fuerstenau} and in this paper
and is in
good agreement for all distributions considered.
The relevant parameter settings are given in appendix \ref{delphi_par}.
To keep residual
biases small with respect to other systematic errors
we ensure that the bin width of all presented distributions is
at least as big as the detector resolution.

To account for a possible imperfect representation of the DELPHI detector
or secondary processes in the simulation program DELSIM,
the cuts given above have been varied over a wide range, including small polar
angles for the event axis, demanding events with more than 7 tracks etc.
Further cuts have been imposed to exclude the boundaries of the TPC sectors
for high momentum particles where the agreement between data and
simulation is less good.
{}From the stability of the measured distributions a systematic uncertainty
as function
of the physical observables has been deduced as the R.M.S. of the deviation
with respect to the published central value.
As the  systematic error is expected to grow proportional to the deviation of
the overall correction factor from unity an additional relative
systematic uncertainty of
10\% of this deviation has been added quadratically to the above value.
A further systematic error has been
added in quadrature for a few bins where
the results of the individual data sets, corresponding to the different years
of data-taking
were found to be statistically incompatible.
This error has been calculated
such that the $\chi^2$ per degree of freedom for the merging was 1
when this error
was considered in addition.
The individual estimates of the systematic uncertainty show similar trends.
This a
posteriori justifies the methods used, however also indicates correlations
between the error estimates which could lead to an overestimate of the
systematic uncertainty.
The final systematic uncertainty has been
smoothened in dependence of the individual variables.

The uncertainties shown in the graphs in Appendix \ref{plots}
comparing data and models are final experimental uncertainties adding
statistical and systematic uncertainties in quadrature.
With the exception of the point-to-point scatter they should be interpreted
like a statistical uncertainty.
The individual errors are given in the tables in appendix \ref{tables}
together with the cross-sections.

\section{Monte Carlo Models}
\label{models}
In this paper we consider the most frequently used  fragmentation models,
namely
JETSET 7.3/7.4 PS and ME \cite {jetset},
ARIADNE 4.06 \cite{ariadne} and HERWIG 5.8 c \cite{herwig}.
HERWIG and JETSET are complete models describing the parton shower evolution
(or matrix element calculation), the
hadronization of partons into hadrons, and the subsequent decays of short lived
particles. ARIADNE only models the parton shower, the consequent
hadronization and decays are treated by the corresponding JETSET routines.

\subsection{JETSET 7.3~/~7.4 Parton Shower}
\label{jetsetps}
JETSET with the parton shower option is used with the settings as
given in table \ref{jetset_switch}.
\begin{table}[h]
\begin{center}
\begin{tabular}{llllll}
\hline
Variable  & Value &  Variable  & Value  &  Variable  & Value  \\
\hline
MSTJ(11)  & 3 & MSTJ(12)  & 3      & MSTJ(41)  & 2      \\
MSTJ(45)  & 5 & MSTJ(46)  & 3      & MSTJ(51)  & 0      \\
MSTJ(101) & 5 & MSTJ(107) & 0      \\
\end{tabular}
\end{center}
\caption{\label{jetset_switch} Setting of JETSET PS Switches}
\end{table}

The JETSET parton shower algorithm is a coherent leading log algorithm
(LLA) with angular ordering.
The shower evolves in the CM frame of the partons obeying energy momentum
conservation at each step of the shower.
The lowest order 3-jet cross-section is reproduced by rejecting part
of the first branchings of the initial $q\bar{q}$ system
as predicted by the LLA formalism.
Angular ordering of the branchings is explicitely imposed and gluon
helicity effects can be included.
$\alpha_s$ is running with a scale given by the transverse momentum squared of
the branching.
The shower evolution is stopped at a mass scale $Q_0$, then fragmentation
takes over. $Q_0$ and $\alpha_s$ (i.e. $\Lambda_{QCD}$) are parameters of the
parton shower part of JETSET.

The fragmentation is performed using the Lund string scheme which can be
formulated as an iterative procedure.
A string stretches in between the oppositely coloured quark and antiquark
via the gluon colour charges.
Two close by gluons act similary to a single gluon with equal momentum.
Therefore the string model is infrared safe.
The longitudinal momentum fraction $z$ of a hadron
is determined using the Lund symmetric
fragmentation function:
\[
f(z) = \frac{(1-z)}{z}^a \cdot \exp{-\frac{b \cdot m_t^2}{z}}
\]
$m_t^2 = m^2 + p_t^2$ is the transverse mass squared of the hadron.
$a$ and $b$ are parameters of the fragmentation function.
$b$ is universal, $a$ in principle can depend on the quark
flavour.
For heavy quarks we instead use the Peterson fragmentation
function \cite{pet83}:
\[
f(z) = \frac{1} {z \left ( 1 - \frac{1}{z} -\frac{\epsilon_q}{1-z} \right )^2}
\]
with parameters $\epsilon_{b(c)}$ which gives a better description of heavy
quark fragmentation.
The transverse momenta of hadrons are determined from the $p_t$ of its
constituent quarks which in turn is chosen from a tunneling process.
This leads to a exponential $p_t$ distribution, thus a Gaussian $p_t^2$
behaviour.
The relevant parameter is the standard deviation of this distribution,
$\sigma_q$.

The tunneling also determines the quark flavour generated in the string
breakup, leading to a dependence \mbox{$\propto \exp{-m_q^2}$} and thus
negligible heavy quark production in the fragmentation.
Already strangeness production is strongly suppressed:
$P(u\bar{u}):P(d\bar{d}):P(s\bar{s}) = 1:1:\gamma_s \approx 0.3$.
Mesons are produced according to their quark contents
in the six multiplets with smallest mass, i.e. in the
states:
$^1S_0$, $^3S_1$, $^1P_1$, $^3P_0$, $^3P_1$ and $^3P_2$.
Contrary to the formulation implemented in JETSET we define individual
production probabilities for these multiplets and light, strange or (both)
heavy flavours.
Except for the light $^3P_0$ multiplet the probabilities for the
P-multiplets are taken $\propto (1-P(^1S_0)-P(^3S_1)) \cdot (2s+1)$.
In a (2 dimensional) string picture production of particles with
angular momentum (i.e. $P$-states) is suppressed and expected to be
small (10\% \cite{lund_rev}).

Using additional mass relations the tunneling mechanism is also applied to
baryon production (replacing a quark by a diquark).
Parameters related to baryon production are the relative diquark production
rate $P(qq)/P(q)$, an extra strange diquark suppression $P(us)/P(ud)/\gamma_s$
and an extra suppression $P(ud1)/P(ud0)$
of spin 1 diquark relative to spin 0 ones leading to
$s=3/2$ and $s=1/2$ baryons respectively.
Furthermore it turned out to be necessary to include an extra
suppression of leading baryons.
This extra suppression is not used in heavy quark fragmentation and
in the simulation of heavy particle decays by fragmentation because
here it leads to soft baryon spectra and a strong overall suppression
of baryons.

The parameter related to baryon-meson-baryon production, the so-called
Popcorn parameter we leave at its default value,
also because so far experimental
determinations of this parameter are not fully conclusive (see \cite{ak0}
and references therein).
Furthermore we do not include simulation of Bose Einstein interference
(by LUBOEI) although with properly chosen parameters this procedure
results in good representation of the correlation functions for
identical particles as well as in a strongly improved
description of light meson resonance lineshapes.
The energy momentum rescaling performed in this routine strongly
influences angular distributions between particles and multi jet rates.
Inclusion of this routine leads to widely different model parameters
(compare e.g. the resulting parameters given in section \ref{fit}
(tables \ref{terg73d},\ref{terg74d})
and the parameters
of the DELPHI simulation (see table \ref{delphi_tuning} )
which includes Bose Einstein interference).

For JETSET 7.3  and a tuning of ARIADNE 4.06
heavy particles decays have been modified to obtain a better description
of the heavy particle branching fractions.
The modifications act similarly to those performed in JETSET 7.4.

\subsection{JETSET 7.4 Matrix Elements}
The parton shower simulation is replaced in the historically older matrix
element version of JETSET by the exact second order matrix element calculation
which provides up to 4 partons.
Two calculations GKS \cite{GKS} and ERT \cite{ERT} are available in JETSET.
In this paper we consider only the default ERT option because it is expected to
be more exact.
At PETRA/PEP the predicted 4-jet rate turned out to be too small \cite{sj_yb}
for a given 3-jet rate.
This has been connected with higher order terms missing in the second order
calculation.
These terms can be partially accounted for by choosing a suitable
renormalization scale according to the ``optimal perturbation
theory'' description \cite{stevenson} by choosing a new
scale $Q^2 = \mu \cdot s, \mu \leq 1$.
To obtain an optimal description we fit the parameter $\mu$
together with the
$\Lambda_{QCD}$ parameter and the fragmentation parameters as described in the
JETSET PS section \ref{jetsetps}. Formally the scale parameter $\mu$ replaces
the cutoff parameter $Q_0$ in the ME fit.
JETSET 7.4 ME parameter settings are given in table \ref{jetset_me_switch}.
\begin{table}[h]
\begin{center}
\begin{tabular}{llllll}
\hline
Variable  & Value &  Variable  & Value  &  Variable  & Value  \\
\hline
MSTJ(11)  & 3 & MSTJ(12)  & 3      & MSTJ(41)  & 2      \\
MSTJ(45)  & 5 & MSTJ(46)  & 3      & MSTJ(51)  & 0      \\
MSTJ(101) & 2 & MSTJ(107) & 0      & MSTJ(111) & 1      \\
\end{tabular}
\end{center}
\caption{\label{jetset_me_switch} Setting of JETSET ME Switches}
\end{table}

\subsection{ARIADNE 4.06}
ARIADNE is a particulary elegant formulation of a parton shower
based on colour dipoles \cite{leningrad,ariadne}.
The emission of a gluon from a colour dipole, i.e. the initial quark antiquark
pair creates two new dipoles, one stretched from the quark to the gluon and one
from the gluon to the antiquark.
Both in turn can independently radiate further gluons.
This ansatz automatically includes  ordering in angle (or transverse momentum)
as well as azimuthal dependencies in a proper way.
The dipole chain resembles the Lund string picture.
Parameters are the QCD scale parameter $\Lambda_{QCD}$ and the cut off scale
$p_t^{min}$ which corresponds to $Q_0$ for JETSET.
The evolution variable is the transverse momentum squared.
As for JETSET the first order 3 jet cross section is reproduced
in ARIADNE.
The ARIADNE running parameters are set as in table
\ref{ariadne_switch}.
\begin{table}[h]
\begin{center}
\begin{tabular}{llllll}
\hline
Variable  & Value &  Variable  & Value  &  Variable  & Value  \\
\hline
MSTA(1)  & 1     & MSTA(2)  & 1     & MSTA(3)  & 0     \\
MSTA(5)  & 1     & MSTA(12) & 1 \\ \hline
MSTJ(11) & 3     & MSTJ(12) & 3     & MSTJ(41) & 0     \\
MSTJ(45) & 5     & MSTJ(46) & 3     & MSTJ(51) & 0     \\
MSTJ(101)& 5     & MSTJ(105)& 0     & MSTJ(107)& 0      \\
\end{tabular}
\end{center}
\caption{\label{ariadne_switch} Setting of ARIADNE Switches}
\end{table}

\subsection{HERWIG 5.8c}
The evolution of the parton shower in HERWIG is based on the Coherent
Parton Braching formalism, an extension of the LLA. It accounts for the
leading and sub-leading logarithmic terms arising from soft and (or)
collinear gluon emission. HERWIG pays special attention to the simulation of
QCD interference phenomena \cite{herwig2}. Most important parameters of
the parton shower algorithm are $\Lambda_{QCD}$ (QCDLAM), the quark masses
(RMASS(1-6)) and the effective gluon mass (RMASS(13)).
The parton shower in HERWIG 5.8c is matched with the first order 3 jet
cross section.
At the end of the
parton shower evolution gluons are split nonperturbatively into
$q\overline{q}$-pairs.

The hadronization in HERWIG proceeds via the so-called ``cluster algorithm''
based on the preconfinement characteristic of QCD.
The colour charge of a parton is compensated to leading order by a anticolour
object which is close by in phase space.
Combining colour and anticolour objects low mass clusters with no colour
are formed.
Higher mass clusters are further split into two lighter ones.
The splitting is controlled by the parameters CLMAS and CLPOW.
In the decay of a cluster containing a quark from the perturbative phase
the direction of this quark is remembered.
A gaussian smearing is applied controlled by the parameter CLSMR.

Hadrons then are formed in two body cluster
decays according to phase space and spin
factors.
The particle and hadron transverse momentum are thus dynamically produced as a
consequence of the cluster mass spectrum.
Particle production in cluster decays is modified by changing the a priori
weights for the individual hadron types. These are VECWT, TENWT
\footnote{ The $1^{+-}$ and $0^{++}$ meson multiplets are not
included in HERWIG. }
and DECWT
for vector or tensor mesons and decuplet baryons respectively.
The a priory weights for quarks are PWT(6).

Light particle decays are simulated in HERWIG using decay tables.
Particles including heavy quarks decay via quark decay and subsequent
fragmentation.
\newpage

Relevant parameter settings for the tuning of HERWIG are given in table
\ref{herwig_switch}.
\begin{table}[h]
\begin{center}
\begin{tabular}{llllll}
\hline
Variable  & Value &  Variable  & Value  &  Variable  & Value  \\ \hline
IPROC    & 100   & SUDORD   & 1     & CLDIR    & 1 \\
\end{tabular}
\end{center}
\caption{\label{herwig_switch} Setting of HERWIG parameters}
\end{table}

\section{Fit of Monte Carlo Models to Experimental Data}
\label{fit}
Classical optimization strategies like hill climbing methods
in general fail to converge if they are directly applied to the optimization
of a Monte Carlo model because the physical observables predicted by
the models are defined only on a statistical basis (and thus only known
within the statistical errors).
Moreover, this straight forward strategy requires much computer time
and cannot be easily repeated to change the input data to be fitted, or
check the influence of systematic errors of the data etc.
Therefore, similar to previous work \cite{a_fit,t_fit,l_fit} we chose to
approximate the dependence of the physical observables
on the model parameters analytically.
For {\it each bin of each distribution} we use
the quadratic form:
%a Taylor
%approximation up to the quadratic term, i.e.:

\begin{eqnarray}
\label{expans}
MC(\vec{p}_0 + \delta\vec{p})(x)  \approx  f(\vec{p}_0 + \delta\vec{p})(x)
%&=& f(\vec{p}_0)(x)   +
%\sum\limits^{n}_{i=1} \partial_i f(\vec{p}_0)(x)  \delta p_i  +
%\sum\limits^{n}_{i,j=1}\partial_i \partial_j f(\vec{p}_0)(x)  \delta p_i
%%\delta p_j \nonumber \\
&=& a_0^{(1)}(x) + \sum\limits^{n}_{i=1} a^{(2)}_i(x)  \delta p_i +
\sum\limits^{n}_{i,j=1} a^{(3)}_{ij}(x)  \delta p_i \delta p_j
\end{eqnarray}

Here $f(x)$ denotes the predicted distribution of a physical observable $x$,
$\vec{p}_0$ is a central parameter setting in the $n$-dimensional
parameter space and $\delta p_i$ the deviation of parameter i
from this setting.
The
%quadratic
last term
%of the Taylor expansion
includes
correlating terms between the model parameters.
These terms are not present in a linear approximation as used in
\cite{l_fit}.

The optimal $m=\frac{n}{2}\cdot(n+3)+1$ parameters $a^{(k)}$
of the expansion are determined from a fit of ansatz
\ref{expans} to
$l > m$ model reference distributions with different parameter settings.
This fit is equivalent to the solution of a system of linear equations:
\begin{eqnarray}
A \cdot \vec{c} = \vec{f} = \vec{MC}
\label{linear_eq}
%\nonumber
\end{eqnarray}
where column i of the matrix A contains the parameter variations of
model set i:
\begin{eqnarray}
A_{i,1..m} = ( 1, \delta p^{(i)}_1, ~...~ ,\delta p^{(i)}_n,
              \delta p^{(i)2}_1, ~...~ , \delta p^{(i)2}_n,
              \delta p^{(i)}_1 \cdot \delta p^{(i)}_2, ~...~,
              \delta p^{(i)}_n \cdot \delta p^{(i)}_{n-1} )
\nonumber
\end{eqnarray}
$\vec{c}$ is the vector of coefficients $a^{(k)}_{i(j)}(x)$,
and $\vec{MC}$ the
vector of model predictions corresponding to the parameters
$\vec{p_0}+\delta\vec{p}^{(i)}$.
The optimal solution of this, for $l \geq m$ overconstrained
linear system is obtained
using a
standard singular value decomposition method \cite{svd, nag}.

The parameters of the $l$ reference models
(generated with equal statistic) are randomly chosen
in parameter space around the central point $\vec{p}_0$.
Assuming the (a priori chosen)
parameter intervals to be renormalized to $ \pm 1$
it turns out to be
unimportant, except for minor differences in the statistical precision,
whether this volume is a cube or a sphere or when
the points are placed on the surface of the sphere.
Our choice should ensure that the precision of the fitted linear function
is roughly constant within the hypersphere.

Statistical errors of the simulated data sets have been neglected because all
sets are generated with fixed statistics, and the variation of the
distributions
for the different parameter choices is relatively small.
Consequently the
exact statistical uncertainty of the linear approximation is unknown.
However the model statistic is chosen such that
the overall statistical error is small compared to the experimental uncertainty
of the data.

The optimal parameters $p_i$, their
errors $\sigma_i$, and correlation coefficients $\varrho_{ij}$
are then determined from a standard $\chi^2$-fit
using MINUIT
\cite{minuit}
of the analytic approximations
(\ref{expans}) for {\it all distributions and bins} considered to the
corresponding data.
Note that the overall analytic approximation contains $n_{bins}\cdot m$
coefficients $a^{(k)}_{i(j)}(x)$.

If the sensitivity of a distribution to a given model parameter
$p_i$ is negligible compared to the sensitivity to
other parameters (see also next section),
the dependence of the approximation on this parameter has been suppressed
by omitting this parameter from the linear system (\ref{linear_eq}).
This leads to a better convergence of the minimization and more robust
results.

The method has been tested by generating $l+1=51$ simulated sets
of 50.000 events each with 6
independent parameters and simultaneously fitting the parameters of
one set using the other $l$ sets as reference.
Here, the statistical error of the approximation is
negligible compared to the statistical error of the test set.
The pull distributions $(p_i^{fit} - p_i^{true})/\sigma_i$ are approximately
standard normal distributions, i.e. the fitting method is unbiased and
produces correct errors.

\subsection{Choice of Distributions}

The parameters of a fragmentation model have a well-defined physical
meaning.
However some parameters are directly coupled like $a$, $b$, and $\sigma_q$
in the Lund fragmentation function, or the effect of the parameter on
physically observed quantities is obscured by other processes like decays.
%For instance the average momentum of heavy meson depends on the
%parton shower and the fragmentation parameters.
Therefore the choice of distributions to tune the model (parameters) to is not
always evident.
So far among the many possible distributions some have been choosen in an
ad hoc way, and alternative changes of distributions have been used to
estimate systematic errors of parameters
\cite{fuerstenau},\cite{a_fit},\cite{l_fit},\cite{o_fit}.

In practice, to keep the influence of statistical errors as small as possible
it is clear that the models should be fitted to those distributions which
show the strongest dependence on the parameter under consideration and least
correlations with others.
%To choose distributions on a more substantial basis we calculated for
For each
distribution $MC(x)$ we therefore calculated its sensitivity
to a given model parameter, i.e. the quantity:
\begin{eqnarray*}
S_i(x) = \frac{\delta MC(x)}{MC(x)}
\Bigl|_{p_i}
\Bigl/ \frac{\delta p_i}{p_i} \approx
\frac{\partial \ln{MC(x)}}{\partial \ln{p_i}} \Bigr|_{p_i}
\end{eqnarray*}
where $\delta MC(x)$ is the change of the distribution $MC$ when changing
the model parameter $p_i$ by $\delta p_i$ around its central value.
The fraction $\frac{\delta p_i}{p_i}$ gives all parameters the same
normalization.

It is observed, that the sensitivity of single particle inclusive
distributions to all model parameters
%of the parton shower or
%fragmentation process
is in general larger
than for event shape distributions.
Compare for example the sensitivities of inclusive spectra and event shapes
to JETSET parameters (figures \ref{sens3} and \ref{sens1}).

\begin{figure}
\begin{center}
\mbox{\epsfig{file=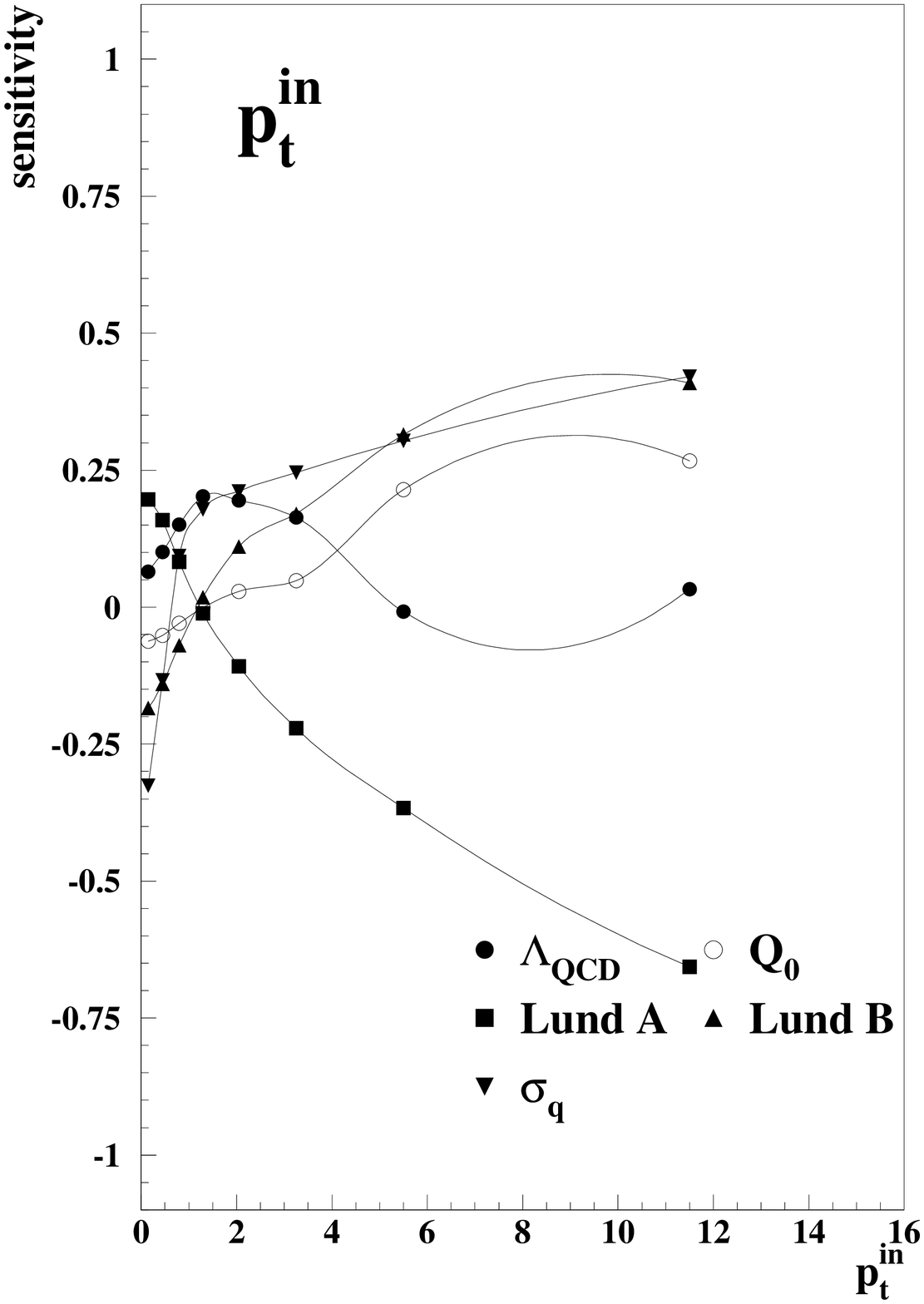,width=5.cm}}   % file=directory
\mbox{\epsfig{file=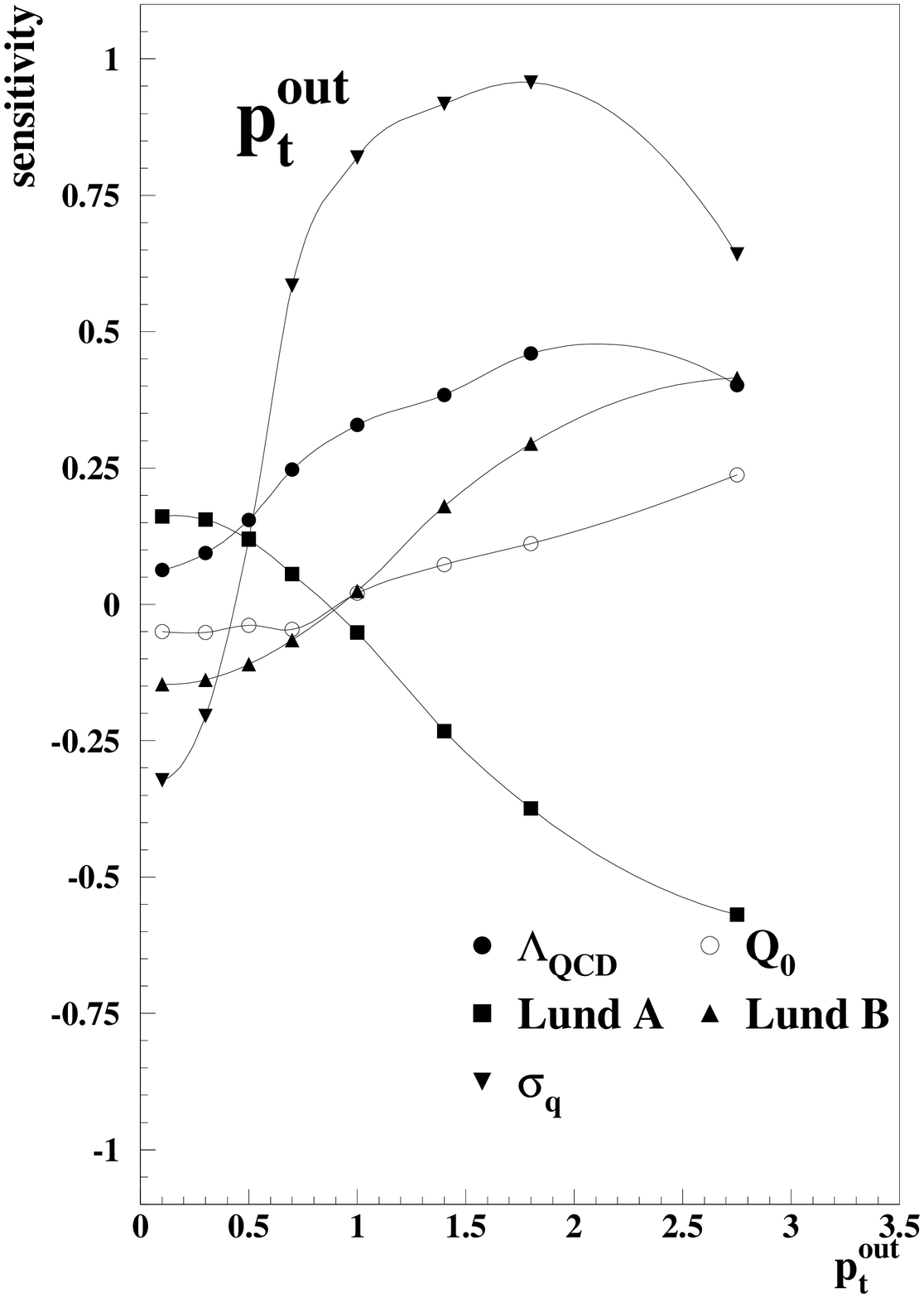,width=5.cm}}   % file=directory
\mbox{\epsfig{file=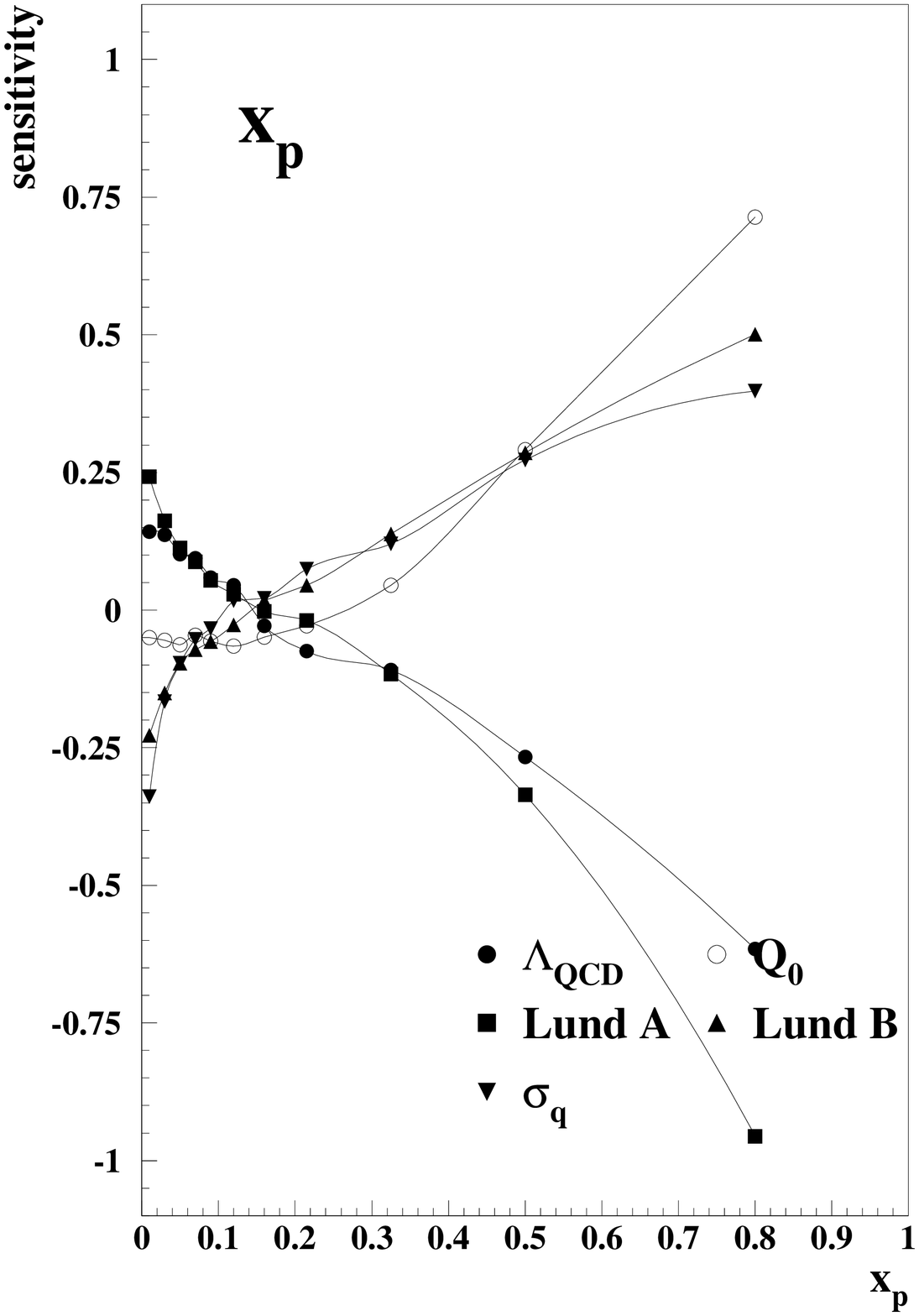,width=5.cm}}   % file=directory
\caption[Sensitivity single track variables]
        {\label{sens3}  Sensitivity single track variables
        }
\end {center}
\end {figure}

It is seen that almost no sensitivity to the parton shower cut off $Q_0$ is
present in the event shape distributions. For this parameter the $x_p$-momentum
spectrum at large $x_p$ is most important.
Correlations among the individual parameters in the inclusive spectra
are very strong as
can be seen by
similar (or opposite) behaviour of the sensitivities for different parameters.
Especially the opposite, almost symmetric
behaviour of the sensitivities on $a$ and $b$ explains
the strong correlation between these parameters and why it is possible
to find good descriptions for many different choices of $a$ and $b$.
As has been expected, $\sigma_Q$ is best determined by the $p_t^{out}$ spectrum
or related quantities.
The variation of the sensitivities are smallest for the rapidity $y$.
Another important quantity is the charged multiplicity which is know to
high precision. Obviously it depends on very many model parameters and
details of particle decays.

Event shape distributions measuring the overall shape like $T$, $S$ or
$M_{high}^2/E^2_{vis}$
mainly depend on $\alpha_s$ i.e. $\Lambda_{QCD}$
(see \ref{sens1})
except in the 2-jet region
where also fragmentation effects are relevant.

\begin{figure}
\begin{center}
\mbox{\epsfig{file=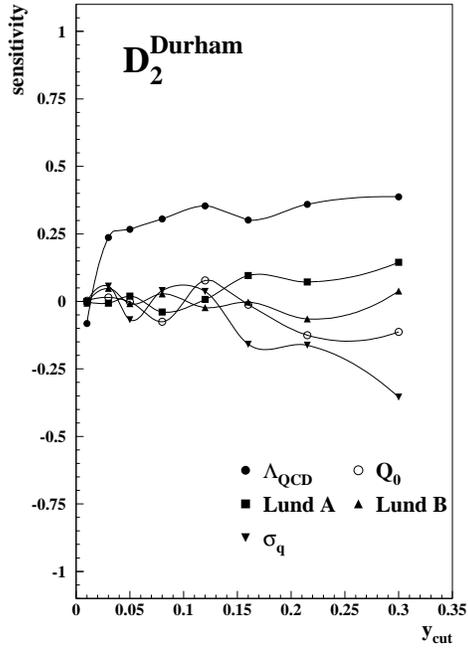,width=7.cm}}   % file=directory
\mbox{\epsfig{file=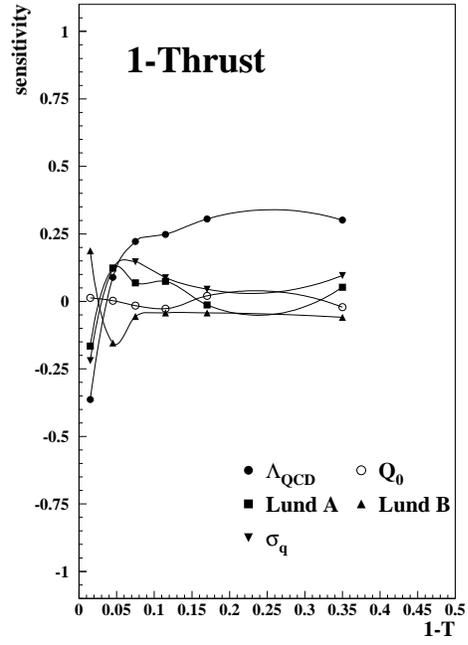,width=7.cm}}   % file=directory
\mbox{\epsfig{file=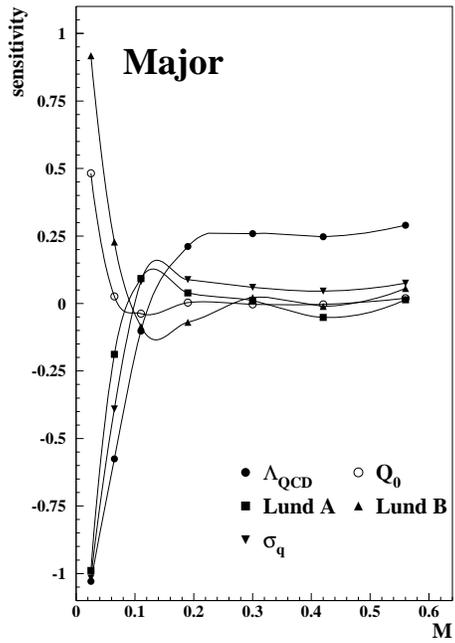,width=7.cm}}   % file=directory
\mbox{\epsfig{file=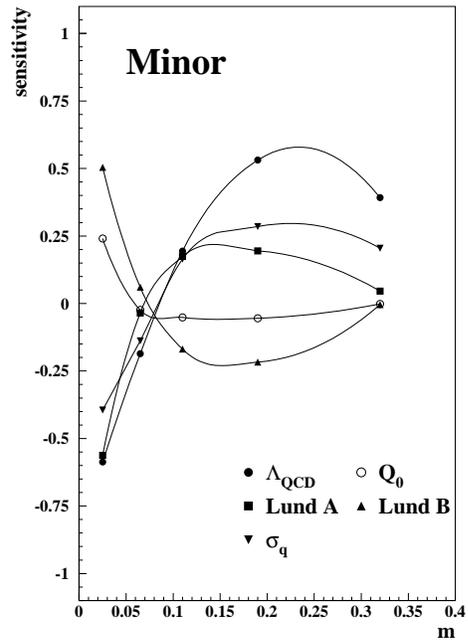,width=7.cm}}   % file=directory

\caption[Sensitivity event shapes variables]
        {\label{sens1} Sensitivity to event shape variables
        }
\end {center}
\end {figure}
%Here
%the experimental error due to resolution and $\tau$-corrections is increased.
The 3-jet rate as measured by $D_2^{Durham}$ or $D_2^{Jade}$ is best
suited to determine $\Lambda_{QCD}$ because the sensitivities on other
parameters
is negligible.
Thus the $\Lambda_{QCD}$-determination can almost be decoupled from
the determination of other parameters.
This also underlines the reliability of $\alpha_s$-determinations from
jet-rates.

Distributions measuring the aplanarity of events like $A$, $m$ and also
$M_{low}^2/E^2_{vis}$
again show increased sensitivities especially on $\alpha_s$
but also strong correlations between the parameters.
This explains why quantities measuring differences of event shape variables
like $O$, $M_{diff}^2/E^2_{vis}$ and the $AEEC$, contrary to popular belief,
depend both on $\Lambda_{QCD}$
and also on many other fragmentation parameters.

For a comprehensive overview tables \ref{sensjt74} and \ref{sensh58} show the
sensitivities to JETSET 7.4 and HERWIG 5.8 model parameters averaged
geometrically over the bins of the distributions.
The averaging dilutes part of the sensitivities (compare e.g. fig. \ref{sens3}
and table \ref{sensjt74}).

{}From the above discussion we conclude that for a determination of the
general parton-shower and fragmentation parameters the models should be fitted
to:
\begin{itemize}
\item[$\bullet$]
the inclusive (charged) particle distributions as function of $x_p$,
$p_t^{in}$ and $p_t^{out}$ with respect to the thrust or sphericity
axis,
\item[$\bullet$]
the differential three jet rate $D_2$ or alternatively $R_3$,
\item[$\bullet$]
a combination of event shape distributions like $T$, $M$, $m$,
 or $S$, $A$, $P$,
or $M_{high}^2/E^2_{vis}$ and $M_{low}^2/E^2_{vis}$.
\end{itemize}
Any combination of event shape distributions or
event axes for the $p_t$-spectra is in principle equivalent and
can be used to estimate the stability of the fit and systematic errors
of the parameters determined or can be viewed as check of
the ``predictive power'' of the models.
However because of the strong correlations among the model parameters
correlations need to be considered.

The action of the model parameters linked directly to
identified particle production (like the strange quark suppression, $\gamma_s$,
or the relative probability to form a qq pair to make a baryon, $P(qq)/P(q)$,
or others) usually follows the physical interpretation more directly.
However identified
particle spectra have also high sensitivity to fragmentation parameters
(see tables \ref{sensjt74} and \ref{sensh58}).
Therefore the concept described above to determine
the optimal dependencies is used.

\subsection{Strategy of the Fit}
\subsubsection{JETSET \& ARIADNE}
First we perform fits of the general fragmentation parameters
( $\Lambda_{QCD}$,$Q_0$ and $p_t^{cut}$ or $\mu$, $a$, $b$ and $\sigma_q$)
to the charged particle inclusive and global event shape distributions.
For each fit 50 simulated sets with 100.000 events are used.
Parameters related to identified particles are set to values
similar to those used for the DELPHI simulation (see table \ref{delphi_tuning})
which are already known to describe identified particle rates and spectra well.

The average scaled momentum $<x_E>$ of charm and beauty particles in the models
only depends
on $\epsilon_{c(b)}$ and $\Lambda_{QCD}$.
Stable hadron spectra only weakly depend on
heavy quark fragmentation parameters.
To further reduce dependencies of stable particle distributions on
heavy quark parameters and at the same time assure correct
$<x_E>$ values for $D^*$ and $B$ mesons
we adopt the following procedure:
the $<x_E>$ values for $D^*$ and $B$ mesons for several choices of
$\Lambda_{QCD}$ and $\epsilon_{c(b)}$ are determined within the model.
Then the dependence of $\epsilon_{c(b)}$ on $<x_E>_{D^* (B)}$ and
$\Lambda_{QCD}$
is parametrized.
This allows to chose for each model set and its given $\Lambda_{QCD}$ value
a corresponding value of $\epsilon_{c(b)}$ such that the model reproduces
the average heavy meson scaled momenta as measured by the LEP experiments
\cite{mean_x_heavy} ($<x_E>_{D^*}=0.504,\,\, <x_E>_{B}=0.701$).

The models were then fit to several combinations of event shape and
semi-inclusive distributions.
The results for $\Lambda_{QCD}$ and $\sigma_q$ were relatively stable, the
variation for $Q_0$ or $p_t^{cut}$
was bigger and there were many different solutions
for $a$ and $b$. The latter is due to the strong correlation between
these two parameters. Therefore from now on we take one central
value of $b$ and treat only $a$ as variable. $a$ is prefered because it
is less directly coupled to $\sigma_q$ in the Lund fragmentation
function.

Parameters relevant only to the production of specific particles were then
adjusted or fitted to the related data. In overview these are:
\begin{itemize}
\item
The extra $\eta$ and $\eta '$  suppressions
were fitted to data from \cite{leta},\cite{aeta}.
\item
The probability for the different B meson multiplets
were adjustet to agree with recent measurements \cite{oB**}, \cite{dB**}.
The corresponding  D mesons probabilities were interpolated between the
B and light meson values.
\item
$P(us)/P(ud)/\gamma_s$,
The strange baryon suppression was
adjusted to the ratio of proton- and $\Lambda^0$-production
\cite{ak0},\cite{ap},\cite{dp},\cite{op}.
\item
$P(ud1)/P(ud0)$ the spin 1 diquark suppression was fitted to the ratio of
$\Sigma(1385)$ and $\Lambda^0$ or proton production
\cite{ak0},\cite{ok0},\cite{ap},\cite{op},\cite{dp}.
\item
The leading baryon suppression was adjusted to the large momentum
tail of the proton and $\Lambda^0$ spectra \cite{dlambda},\cite{ak0}
,\cite{ostr_bary}.
\end{itemize}
Then a simultaneous fit of 10 important parameters (see parameters where
generated ranges are specified in tables
\ref{terg73d} -- \ref{terg74m})
was prepared generating 100 sets of 100.000 events each.
The analytical approximation obtained from these simulated sets was fitted
to different choices of event shape and particle distributions.
For the event shapes we chose the combinations of variables
shown in table \ref{evtfit1}.
%All choices except the first one
%which contains most distributions were considered as equivalent.
The choices for identified particles are shown in table \ref{evtfit2}.
These were used to account for the discrepancies observed in the in the proton
data from different experiments and the imperfect representation of
the $K^{\pm}$ spectra in the models.
\newcommand{\bl} {$\bullet$}
\begin{table}[t]
\begin{center}
\begin{tabular}{lcccccccccccc}\hline
 & \multicolumn{12}{c}{fit choice $S_i$:} \\ \hline
distribution & 0 & 1 & 2 & 3 & 4 & 5 & 6 & 7 & 8 & 9 & 10 & 11\\ \hline
$<N_{ch}>$ &
    &     &     &     &     &     &
\bl & \bl & \bl & \bl & \bl & \bl \\
$x_p$ &
\bl & \bl & \bl & \bl & \bl & \bl &
\bl & \bl & \bl & \bl & \bl & \bl \\
$y_T$ &
\bl & \bl &     & \bl &     & \bl &
\bl & \bl &     & \bl &     & \bl \\
$y_S$ &
\bl &     & \bl &     & \bl & \bl &
\bl &     & \bl &     & \bl & \bl \\
$p^{in}_t,p^{out}_t\,\,\, (T)$ &
\bl & \bl &     & \bl &     & \bl &
\bl & \bl &     & \bl &     & \bl \\
$p^{in}_t,p^{out}_t\,\,\, (S)$ &
\bl &     & \bl &     & \bl & \bl &
\bl &     & \bl &     & \bl & \bl \\
$S,\,\,A,\,\,P$ &
\bl &     & \bl &     & \bl & \bl &
\bl &     & \bl &     & \bl & \bl \\
$T,\,\,M,\,\,m$ &
\bl & \bl &     & \bl &     & \bl &
\bl & \bl &     & \bl &     & \bl \\
$D^{Durham}_2$ &
\bl & \bl &     & \bl &     & \bl &
\bl & \bl &     & \bl &     & \bl \\
$D^{Jade}_2$ &
\bl &     & \bl &     & \bl & \bl &
\bl &     & \bl &     & \bl & \bl \\
$D^{Durham}_3$ &
\bl &     &     & \bl &     & \bl &
\bl &     &     & \bl &     & \bl \\
$D^{Jade}_3$ &
\bl &     &     &     & \bl & \bl &
\bl &     &     &     & \bl & \bl \\
$D^{Durham}_4$ &
    &     &     & \bl &     & \bl &
    &     &     & \bl &     & \bl \\
$D^{Jade}_4$ &
    &     &     &     & \bl & \bl &
    &     &     &     & \bl & \bl \\ \hline
\end{tabular}
\end{center}
\caption[Choices of inclusive and shape distributions for the individual fits]
{\label{evtfit1}
Overview of the combinations of inclusive distributions and shape distributions
fits of the JETSET, ARIADNE and HERWIG models.
Distributions used are marked \\
by \bl .}
\end{table}

\begin{table}
\begin{center}
\begin{tabular}{llccccc}\hline
 & & \multicolumn{5}{c}{fit choice $P_i$:} \\ \hline
\multicolumn{2}{l}{data} & 0 & 1 & 2 & 3 & 4 \\ \hline
$\rho^\circ$ &  DELPHI\cite{drho} &
 \bl & \bl & \bl & \bl & \bl \\
$\omega$ & L3\cite{lomega} &
 \bl & \bl & \bl & \bl & \bl \\
$f_0$, $f_2$ & DELPHI\cite{drho} &
 \bl & \bl & \bl & \bl & \bl \\
$K^0$ & ALEPH\cite{ak0}, OPAL\cite{ok0} &
 \bl & \bl & \bl & \bl & \bl \\
$K^\pm$ & ALEPH\cite{ap}, DELPHI\cite{dp}, OPAL\cite{op}  &
 \bl & \bl & \bl &  &  \\
$K^{\ast 0}$ & OPAL\cite{okstar0} &
 \bl & \bl & \bl & \bl & \bl \\
$K^{\ast \pm}$ & ALEPH\cite{akstar}, DELPHI\cite{drho}, OPAL\cite{okstar0}  &
 \bl & \bl & \bl & \bl & \bl \\
$\Phi$ & DELPHI\cite{dp}, OPAL\cite{op} &
 \bl & \bl & \bl & \bl & \bl \\
$p$ & ALEPH\cite{ap},DELPHI\cite{dp} &
 \bl & \bl & & \bl & \\
$p$ & OPAL\cite{op} &
 \bl & & \bl & & \bl \\ \hline
\end{tabular}
\end{center}
\caption[Choices of identified particle data used for the individual fits
of JETSET and ARIADNE ]
{\label{evtfit2}
Overview of the combinations of identified particle data
used for the JETSET and ARIADNE fits.
Data used are marked by \bl .}
\end{table}

The $\Lambda^0$ data has been excluded from the fit because at small
momenta it is not described by the models. Inclusion of this distribution leads
to ill defined fits and unstable results. The proton data was only used to
determine the diquark suppression parameter $P(qq)/P(q)$ and was excluded
from the $\chi^2$ calculation for the fits of the remaining parameters.
Baryons at intermediate and large momenta are likely, within the models, to be
primary particles. Therefore these data strongly influence the
fragmentation function and the related parameters. Since proton and $\Lambda^0$
spectra appear poorly described by the models and
%especially
because it is even necessary to modify
the fragmentation function by an extra suppression at large momenta
this strong impact on the fit results is felt to be unphysical and therefore
has been excluded.

Separate fits were performed for all combinations of choices for shapes and
identified particles.
To check the stability of the fits the optimization has been started with 6
random start values of fragmentation parameters.
The fits were stable and converged to
the same solution for about 95\% of the cases.
Differences mainly showed up for two strongly correlated parameters.
For those fits which show more than 1 solution we used the one with the better
$\chi^2$.
A full (MINOS \cite{minuit}) error estimate for all parameters was performed.

The central result corresponds to the combination S6/P0.
This combination has been choosen because it contains event shape distributions
linear and quadratic in the particle momenta, the charged multiplicity and
all relevant identified particle information and thus should result in a
comprehensive overall description.
Results are given with the statistical and systematic errors in tables
\ref{terg73d} -- \ref{terg74m}.
The results for the individual parameters are strongly correlated.
The correlation coefficients for the JETSET 7.4 (default decays) and
ARIADNE 4.06 (DELPHI decays) are given representatively in appendix
\ref{correl_tabs}. Beside these statistical correlations further
correlations exist due to the different possible choices of input data.
The systematic error is the R.M.S obtained from all combinations of input data
which turn
out to give parameters smaller respectively bigger than the central fit.

The following observations were made comparing results of the different
input data choices for the JETSET PS fits:\\
As should be expected the results for
$\Lambda_{QCD}$, $Q_0$ and $\sigma_q$
%and $a$
are almost independent
of the choice of identified particle data.
The resulting value of $\Lambda_{QCD}$ does not depend on the algorithm used
(JADE or DURHAM) if only $D_2$ is included in the fit however
if also the higher
jet rates were used the $\Lambda_{QCD}$ value turns out to be somewhat
bigger ($\approx 8\%$) in case of the JADE algorithm.
Values of $\Lambda_{QCD}$ and $Q_0$ are positively correlated. This implies
that the number of final partons is more stable within the models than
might be expected from the error of $Q_0$ alone.
$\Lambda_{QCD}$, and $a$ and $\sigma_q$ are anticorrelated. This is
due to the a compensation of transverse momentum generated in the parton shower
and fragmentation phase of the model.\\
The results for $\gamma_s$ are higher if $K^{\pm}$ and $K^0$ data are included
compared to $K^0$ only. $P(qq)/P(q)$ is larger if the ALEPH proton spectrum
is included.
Production parameters for strange vector and pseudoscalar mesons are
anticorrelated.
If the charged multiplicity is not included and therefore the predicted
multiplicity is smaller than the measured result,
the primary production probabilities
for light pseudoscalar and vector mesons are $P(^1S_0)_{ud} \approx 0.40$ and
$P(^3S_1)_{ud} \approx 0.26$ respectively. If the multiplicity is fixed to
$N_{ch} \approx 20.9$ these values are $0.28$ and $0.29$. This implies
substantial production probabilities for light
p-wave mesons of $0.36$ to $0.43$.

Most of these observations similary also apply in case of ARIADNE.
However here the different choices of $D_2, D_3$ etc. lead to stable results
for
$\Lambda_{QCD}$. There is a tendency to obtain slightly bigger values
from the JADE algorithm (0.245) then from Durham (0.237). This goes
again conform with bigger values for $p_t^{QCD}$ from JADE ($\approx 0.9$)
then from DURHAM  ($\approx 0.6$).

For JETSET ME $\Lambda_{QCD}$ is anticorrelated with the scale $\mu$.
The fragmentation parameters have different values as in the
PS case. Especially $\sigma_q \approx 0.48 $ is much bigger.
This partially compensates missing higher orders in the ME ansatz.
The values of $\gamma_s$ and $P(qq)/P(q)$ are smaller compared to the PS case
and show a decreased dependence on the choice of the related input data.
The probability for p-wave states is much bigger $\approx 0.36$
already for the fits with free multiplicity.

\subsubsection{HERWIG}
HERWIG employs much less parameters than JETSET especially in the hadronization
sector of the model.
Therefore a simultaneous fit of all model parameters which are found to be
important ( compare table \ref{sensh58} )
is easily performed.
These parameters are:
\begin{itemize}
\item
the QCD scale parameter QCDLAM and the gluon mass RMASS(13) as major parameters
of the PS phase,
\item
the cluster fragmentation parameters CLPOW, CLMAX and CLSMR (CLDIR=1)
and
\item
the a priori weights DECWT for decuplet baryons, PWT(3) for strange quarks
and PWT(7) for diquarks.
\end{itemize}
Beside the particle spectra the cluster parameters also strongly influence
the stable charged particle and event shape distributions.
Fits have therefore been performed to combinations of shape distributions
(see table \ref{evtfit1}) and identified particle data
(see table \ref{evtfit3}).
Of special importance is the simultaneous usage of proton and $\Lambda^0$ data.
The average scaled momenta of heavy mesons are important for the
determination of CLPOW and CLMAX.

\begin{table} [t]
\begin{center}
\small
\begin{tabular}{llcccccccccc}\hline
 & & \multicolumn{10}{c}{fit choice $P_i$:} \\ \hline
\multicolumn{2}{l}{data} & 0 & 1 & 2 & 3 & 4 & 5 & 6 & 7 & 8 & 9\\ \hline
$\rho^\circ$ &  DELPHI\cite{drho} &
 \bl & \bl & \bl & \bl & \bl & \bl & \bl & \bl & \bl & \bl \\
$\omega$ & L3\cite{lomega} &
 \bl & \bl & \bl & \bl & \bl & \bl & \bl & \bl & \bl & \bl \\
 $f_2$ & DELPHI\cite{drho} &
 \bl & \bl & \bl & \bl & \bl & \bl & \bl & \bl & \bl & \bl \\
$K^0$ & ALEPH\cite{ak0}, OPAL\cite{ok0} &
 \bl & \bl & \bl & \bl & \bl & \bl & \bl & \bl & \bl & \bl \\
$K^\pm$ & ALEPH\cite{ap}, DELPHI\cite{dp}, OPAL\cite{op}  &
 \bl & \bl & \bl &  &  & \bl & \bl & \bl & & \\
$K^{\ast 0}$ & OPAL\cite{op} &
 \bl & \bl & \bl & \bl & \bl & \bl & \bl & \bl & \bl & \bl \\
$K^{\ast \pm}$ & ALEPH\cite{akstar}, DELPHI\cite{drho}, OPAL\cite{okstar0}  &
 \bl & \bl & \bl & \bl & \bl & \bl & \bl & \bl & \bl & \bl \\
$\Phi$ & DELPHI\cite{dp}, OPAL\cite{op} &
 \bl & \bl & \bl & \bl & \bl & \bl & \bl & \bl & \bl & \bl \\
$\Lambda^0$ & ALEPH\cite{ak0},DELPHI\cite{dlambda} &
 \bl & \bl & & \bl & & \bl & \bl & & \bl  &\\
$\Lambda^0$ & OPAL\cite{okstar0} &
 \bl & & \bl & & \bl & \bl & & \bl & & \bl \\
$p$ & ALEPH\cite{ap},DELPHI\cite{dp} &
 \bl & \bl & & \bl & & \bl & \bl & & \bl  &\\
$p$ & OPAL\cite{op} &
 \bl & & \bl & & \bl & \bl & & \bl & & \bl \\
$\eta$ & ALEPH\cite{aeta} &
\bl & \bl & \bl & \bl & \bl & \bl & \bl & \bl & \bl & \bl \\
$\eta$' & ALEPH\cite{aeta} &
\bl & \bl & \bl & \bl & \bl & \bl & \bl & \bl & \bl & \bl \\
$<x_E>\,\,  D^{\ast\pm}, D^{\ast 0}$ & \cite{mean_x_heavy} &
\bl & \bl & \bl & \bl & \bl & \bl & \bl & \bl & \bl & \bl \\
$<x_E>\,\, B^{0}, B^{\pm}$  &\cite{mean_x_heavy} &
\bl & \bl & \bl & \bl & \bl & \bl & \bl & \bl & \bl & \bl \\
$\Sigma^\pm (1385)$ & DELPHI\cite{dstrange}, OPAL \cite{okstar0} &
\bl & \bl & \bl & \bl & \bl & & & & & \\
$\Xi^-$ & DELPHI\cite{dstrange}, OPAL \cite{okstar0} &
\bl & \bl & \bl & \bl & \bl & & & & & \\
$\Xi_0(1530)$ & DELPHI\cite{dstrange}, OPAL \cite{okstar0} &
\bl & \bl & \bl & \bl & \bl & & & & & \\ \hline
\end{tabular}
\normalsize
\end{center}
\caption[Choices of identified particle data used for
the fits of HERWIG]
{\label{evtfit3}
Overview of the combinations of identified particle data
used for the HERWIG fits. Data used are marked by \bl .}
\end{table}

The different choices of the identified particle data (see table \ref{evtfit3})
are made such that
systematic differences in the data are reflected in the systematic
errors of the parameters.
The central fit result for HERWIG corresponds to the combination $S_6P_5$.

Following observations are made comparing results of the different
input data choices for the HERWIG fits:\\
Values obtained for QCDLAM are in general bigger ($\approx 0.01$) when
fitting DURHAM compared to JADE jet rates.  Contrary to the JETSET case
QCDLAM is smaller (by $\approx 0.005$) when
differential 2-, 3-, and 4-jet rates are fitted compared to 2-jet only.
QCDLAM is insensitive to the inclusion of the charged multiplicity in the fit.
The gluon mass RMASS(13) shows some dependency on the selected identified
particle information and on the multiplicity.
The cluster parameters CLMAX and CLPOW depend on the identified particle
spectra only in case that the multiplicity is not included in the fit.
CLPOW is higher for JADE than for DURHAM jet rates. CLSMR depends
only on the charged particle and event shape information. The inclusion
of the baryon decuplet data tends to spoil the description of
the octet sector.
\clearpage

\section{Comparison of Models to Data}%//
\label{results}

The fits of the individual fragmentation models are compared to DELPHI
data in figures
\ref{bildxpp} - \ref{bilddjb}
shown in appendix \ref{plots}.
The lower insets of the plots depict the relative deviation of the models
to the data. Also shown as shaded area in these insets is the total
experimental error obtained by adding quadratically the systematic and
statistical error in each bin. Except for the point-to-point scatter the error
should be interpreted like a statistical ''$1 \sigma$'' uncertainty.
For most shape distributions we show on the left side the data as measured
from charged particles only, on the right measured from charged and neutral
particles.
The distributions are corrected to the corresponding final states.

We restrict to comparisons to the following models:
\begin{itemize}
\item
JETSET 7.3 DELPHI decays labeled JT 7.3 PS
\item
JETSET 7.4 default decays labeled JT 7.4 PS
\item
ARIADNE 4.06 DELPHI decays labeled AR 4.06
\item
HERWIG 5.8 c default decays labeled H 5.8C
\item
JETSET 7.4 ME default decays labeled JT 7.4 ME
\end{itemize}
The different decay treatments lead only to negligible differences for
the stable charged particle and event shape distributions.

{\bf Inclusive Charged Particle Spectra}\\
All models describe well the general trends of the data. Almost no
discrepencies show up from a direct data model comparison. More
quantitatively the comparison of model and data (lower insets) shows:\\
The $x_p$-spectrum (fig. \ref{bildxpp})
for $x_p < 0.4$ is almost perfectly described by the
ARIADNE and JETSET PS models.
At large $x_p$ these models slightly underestimate the data.
%Within these models it is impossible to reproduce both the high momentum tail
%of the $x_p$-spectrum and the average charged multiplicity simultaneously.
This trend is reduced if the multiplicity is left free in the fit.
HERWIG and JETSET ME exhibit a slight wave structure when compared to the data.
This structure is also reflected in the rapidity distributions (see fig.
\ref{bildyts}).
\\
Although the $p^{in}_t$-distribution (fig. \ref{bildptin})
is in agreement to data for all models
the tail of the $p^{out}_t$-distribution (fig. \ref{bildptout})
 ($p^{out}_t > 1$) disagrees with
the data for all models. This tail is almost unaffected by fragmentation
but is sensitive to the partonic part of the models.
For the ME model a discrepancy  might be expected because
of missing higher order terms in the second order matrix element calculation.
The failure of the parton shower models can possibly be traced back to missing
large angle terms in the LLA basic to the models.
Accepting these reasons for the failure of the models it should be
expected that a matching of the second order calculation and the LLA formalism
should lead to an improved description of $p^{out}_t$
and related distributions.

{\bf Shape Distributions}\\
The general shape distributions $1-T$, $S$, $C$ and $B_{sum}$
 (figs. \ref{bildthr},\ref{bildsph},\ref{bildcp},\ref{bildtjb})
measured from charged particles
are described
within the small experimental errors (typically 2-3\%) by all parton shower
models.
HERWIG for large values of these observables tends to lie slightly above the
data.
The agreement to the corresponding
charged plus neutral distributions is still satisfactory,
although these distributions have not been used in the fit.
This illustrates furher the quality of the models and of the data.
The description of distributions sensitive to
transverse momenta in the event plane like
$M$, $B_{max}$ or $M_h^2/E^2_{vis}$
\mbox{(figs. \ref{bildmaj},\ref{bildwjb},\ref{bildhjm})}
is only slightly less good (typically better than 5\%).
Turning to distributions sensitive to
transverse momenta out of the event plane $m$, $O$, $A$, $M_l^2/E^2_{vis}$,
or $B_{min}$  \mbox{(figs.
\ref{bildmin},\ref{bildob},\ref{bildapl},\ref{bildljm},
\ref{bildnjb})}
the following pattern is observed:
ARIADNE generally describes the data well,
HERWIG tends to overestimate the distributions and
JETSET PS is usually below the data for higher values of the observables.
The latter is to be expected from the underestimation of the
$p_t^{out}$-distribution by all PS models.
A similar pattern is also observed for the jet rates.
The differential 2-jet rate $D_2$ (figs. \ref{bild2dd},\ref{bild2dj})
is well described by all models.
ARIADNE describes well also the higher jet rates (figs.
\ref{bild3dd}--\ref{bild4dj}), HERWIG
overestimates and JETSET PS underestimates them.
This behaviour is consistently observed for the JADE as well as for the
DURHAM algorithm and for the charged and charged plus neutral data.\\
The description of the shape distributions by the JETSET ME model is less
perfect than for the PS models. There is a shortcome for the extreme
2 jet region, the multijet rates
and observables sensitive to radiation out of the event plane.
In general however the description by the JETSET ME is quite satisfactory.

{\bf Identified Particle Rates}\\
Table \ref{part_rates} compares the particle rates predicted from the models
with the current measured LEP averages.
\scriptsize
\begin{table}
\begin{center}
\begin{tabular}
{lr@{.}lr@{.}lr@{.}lr@{.}lr@{.}lr@{.}l@{ $\pm$ }r@{.}l}\hline
 &         \multicolumn{2}{r}{\tiny  JETSET 7.3 PS}&
           \multicolumn{2}{r}{\tiny JETSET 7.4 PS}&
           \multicolumn{2}{r}{\tiny ARIADNE 4.06 }&
           \multicolumn{2}{r}{\tiny JETSET 7.4 ME}&
           \multicolumn{2}{r}{\tiny HERWIG 5.8 C}&
\multicolumn{4}{c}{\tiny
LEP\cite{pdg},\cite{ada},\cite{multipl},\cite{deltapp}} \\ \hline
\multicolumn{8}{l}{Charged Particles} \\
$<N_{ch}>$          & 20&87 & 20&81 & 20&80 & 20&86 & 20&94 & 20&95 &  0&21
  \\ \hline
\multicolumn{8}{l}{Pseudoscalar Mesons} \\
$\pi^{\pm}$          & 17&19 & 17&09 & 17&13 & 17&36 & 17&66 & 17&1  &  0&4
   \\
$\pi^{0}$            &  9&85 &  9&83 &  9&82 & 10&03 &  9&81 &  9&9  &  0&08
    \\
$K^{\pm}$           &  2&20 &  2&23 &  2&19 &  2&15 &  2&11 &  2&42  &  0&13
    \\
$K^{0}$             &  2&13 &  2&17 &  2&12 &  2&10 &  2&08 &  2&12  &  0&06
    \\
$\eta$               &  1&07 &  1&10 &  1&09 &  1&16 &  1&02 &  0&73 &  0&07
    \\
$\eta$'(958)         &  0&10 &  0&09 &  0&10 &  0&10 &  0&14 &  0&17 &  0&05
    \\
$D^{+}$             &  0&19 &  0&20 &  0&20 &  0&20 &  0&24 &  0&20  &  0&03
    \\
$D^{0}$             &  0&46 &  0&49 &  0&48 &  0&49 &  0&53 &  0&40  &  0&06
    \\
$B^{\pm}$,$B^{0}$   &  0&36 &  0&36 &  0&36 &  0&36 &  0&36 &  0&34  &  0&06
\\ \hline
\multicolumn{8}{l}{Scalar Mesons} \\
$f_0(980)$           &  0&17 &  0&16 &  0&17 &  0&16 & \multicolumn{2}{l}{~}
&  0&14   &  0&06  \\ \hline
\multicolumn{8}{l}{Vector Mesons} \\
$\rho^{\circ}(770)$  &  1&29 &  1&27 &  1&26 &  1&29 &  1&43 &  1&40 &  0&1
 \\
$K^{\ast\pm}(892)$ &  0&78 &  0&77 &  0&79 &  0&77 &  0&74 &  0&78   &  0&08
     \\
$K^{\ast 0}(892)$  &  0&80 &  0&77 &  0&81 &  0&78 &  0&74 &  0&77   &  0&09
\\
$\phi(1020)$         &  0&109&  0&107&  0&107 &  0&102 &  0&099&  0&086 & 0&018
    \\
$D^{\ast\pm}(2010)$ &  0&18 &  0&22 &  0&19 &  0&22 &  0&22 &  0&17     & 0&02
\\
$D^{\ast 0}(2007)$  &  0&20 &  0&22 &  0&20 &  0&22 &  0&23             &
 \\  \hline
\multicolumn{8}{l}{Tensor Mesons} \\
$f_2(1270)$           &  0&29 &  0&29 &  0&29 &  0&30 &  0&26 &  0&31    & 0&12
  \\ \hline
\multicolumn{8}{l}{Baryons} \\
$p$                  &  0&97 &  0&97 &  0&96 &  0&90 &  0&78 &  0&92     & 0&11
  \\
$\Lambda^0$            &  0&361&  0&349&  0&365&  0&309&  0&368&  0&348    &
0&013   \\
$\Xi^-$              &  0&0288&  0&0300&  0&0300&  0&0256&  0&0493 &  0&0238  &
 0&0024   \\
$\Delta^{++}(1232)$  &  0&158&  0&160&  0&136&  0&158&  0&154&  0&077
&0&018\\
$\Sigma^{\pm}(1385)$ &   0&037&  0&036&  0&032&  0&033&  0&065&  0&0380       &
 0&0062    \\
$\Xi^0(1530)$        &  0&0073&  0&0069&  0&0063 &  0&0060&  0&0249 &  0&0063 &
 0&0014    \\
$\Omega^-$           &  0&0013&  0&0019&  0&0021&  0&0010&  0&0077&  0&0051   &
 0&0013    \\
$\Lambda_b^0$        &  0&032&  0&033& 0&032 &  0&029& 0&007&  0&031          &
 0&016     \\ \hline
\end{tabular}
\end{center}
\caption[Particle Production Rates]
{\label{part_rates}
{The Production Rates for the different Generators compared to LEP data}}
\end{table}
\normalsize
After inclusion of the charged multiplicity in the fits it is well
described by the models. Neglecting this constraint ARIADNE and JETSET PS
predict the multiplicity too low ( 20.2 -- 20.4 ) and JETSET ME too high
\mbox{( 22.7 )}.
HERWIG is correct without constraint.

All meson rates with the exception of the $K^{\pm}$ and the $\eta$
rate are well described within errors (2$\sigma$).
The measured $\eta$ momentum spectrum is however well described.
We therefore suspect a too small error quoted in the extrapolation needed
to determine the total $\eta$ rate.
The  $K^{\pm}$ rate is sensitive to heavy quark decays (see below).
The octet baryons also agree reasonably.
Only HERWIG overestimates the $\Xi^-$ rate by about a factor 2.
Some discrepancies show up in the decuplet baryon sector.

{\bf Meson Momentum Spectra}\\
In figures \ref{bildk0},\ref{bildkpm} we compare the models  to the identified
kaon
spectra as function of $\xi_p=\log{\frac{1}{x_p}}$ from different LEP
experiments.
For a more quantitative comparison which also allows to judge the
agreement among the different experiments we show on the right the
relative deviation between the individual data sets and each model as
lines.
%Within $\approx 15\%$ $K^{\pm}$ and $K^0$ are simultaneously described
%by all models.
In the fragmentation region at large $\xi_p$ (i.e. small $x_p$) where
the $K^{\pm}$ data is more precise and therefore dominates the fit the
$K^0$'s are overestimated by $\approx 10-20\%$ by all models.
At more central momenta all kaons are well described.
$K^{\pm}$ are underestimated in the range ($0.8 < \xi_p < 2.5$).
K's from heavy particle decays tend to contribute mainly in this momentum
range.

Wrong branching fractions of b-hadrons in the models
presumably  cause this discrepancy.
Indeed a recent DELPHI measurement of the inclusive particle
production in b-events supports this interpretation \cite{b-inclusive}.
The difference of $K^{\pm}$ and $K^0$ production was found to be
$0.60 \pm 0.50$.
The corresponding value in the models is only $\approx 0.30$ for
JETSET and ARIADNE and $0.04$ for HERWIG.
This difference causes the relatively large systematic error on
$\gamma_s$.
An improvement is possible as soon as K measurements for
identified b and light quark events become available.
Finally at very small $\xi_p$ (i.e. large $x_p$)
there is an indication that the
models underestimate slightly the K production.

Fragmentation functions of vector mesons, which are likely to be primary
particles, are compared to in figs.
\ref{bildkstar0},\ref{bildkstarpm},\ref{bildphi}.
Within the large errors of these resonance measurements all models
describe the spectra very well.
There is only a tendency to predict a somewhat harder fragmentation
than measured.
This is most evident for the $\Phi (1020)$ spectrum.\\

It is important to note that the strange pseudoscalar and vector
meson data shown  already imply large production rates of higher mass
resonances with strangeness.
The large $\xi_p$ part of the $K^{0, \pm}$ spectra can only be
described if large resonance rates are assumed in the models.
%This is caused by the higher mass of stable K's with respect to
%$\pi$'s.
Due to the stronger boost (because of their high mass)
kaons from resonance decays acquire in average higher momenta
(i.e. smaller $\xi_p$).
As the strange vector meson rates are relatively moderate higher mass
strange resonances must therefore be present.
This reflects in high production probabilities, of the order of 25\%,
for p-wave
resonances with strangeness  in JETSET and ARIADNE.
HERWIG dynamically predicts similar rates.

Fig. \ref{bildf0} depicts the the good agreement of the model predictions
to the measurements of the $f_0(980)$ scalar and $f_2(1270)$
tensor meson resonance.
With large production probabilities for p-wave resonances all models
describe the measured data well.
Note that high production of p-wave states ($>10\%$) is not
expected in the string picture \cite{lund_rev}.

{\bf Baryon Momentum Spectra}\\
The proton spectra of OPAL and ALEPH show a severe discrepancy at
high momentum (or small $\xi_p$) (see fig. \ref{bildproton}).
The central model fits interpolate between these experimental results.
Including only the ALEPH or OPAL data in the fit it turns out that
the OPAL proton spectrum cannot be described by JETSET PS and ARIADNE.
The ALEPH data can be reasonably well described.
On the other hand the JETSET ME model gives a fair description of the OPAL
proton data.
The rate predicted by HERWIG is too high at small $\xi_p$ and too low at small
$\xi_p$.
At small $\xi_p$ the behaviour of JETSET PS and ARIADNE is similar to HERWIG
if the extra baryon suppression is not used.
The need for an extra suppression of baryon production may also indicate
the presence of nucleon resonances.
Although the mass splitting
between angular excited baryons and the related ground states is smaller
than in the mesons case
these states are so far not foreseen in JETSET or HERWIG.
Decays of these states would lead to a softening of the spectra
of stable baryons
and would also modify the predicted correlations between
baryon antibaryon pairs.

The experimental situation for the $\Lambda^0$ spectrum also shows
some discrepancies (see fig. \ref{bildlambda}).
JETSET PS and ARIADNE describe or slightly overestimate the small $\xi_p$
region. HERWIG here is again too high.
At large $\xi_p$ where the measurements agree all models underestimate
the $\Lambda^0$ production by $\approx 15\%$.
This region is of special interest because here particle production is expected
to be influenced by coherence phenomena.

Fig. \ref{bildbaryons} shows in a comprehensive overview a comparison
of the octet and decuplet baryon production and the model predictions.
The discrepancies discussed in the last paragraphs are hidden here
due to the large scales. JETSET and ARIADNE describe the gross features
of the octet and decuplet baryon production well.
The $\Delta^{++}$ rate \cite{deltapp}
is however overestimated by about a factor 2.
HERWIG predicts
too hard baryon fragmentation. The relative production rates of the different
multiplet states are less well predicted by HERWIG.

\section{Summary}
\label{summary}

{}From 750.000 $e^+e^- \rightarrow Z \rightarrow hadrons$ events measured by
the DELPHI experiment
precise fully corrected semi-inclusive charged particle and event
shape distributions have been determined.

A systematic, quantitative study has been undertaken to
determine the optimal choice
of distributions to tune fragmentation models to.
Semi-inclusive charged particle and identified particle distributions
constrain the hadronization part of the models whereas
3-jet rate and most event shape distributions mainly control the
parton shower parameters ( especially $\Lambda_{QCD}$) in the models.

Optimal parameters for the ARIADNE 4.06, HERWIG 5.8c and
JETSET 7.3 and 7.4 parton shower models and for the JETSET 7.4 matrix element
model have been determined.
The models were fitted to the measured event properties and inclusive
data and to identified particle data measured by the LEP experiments.
The fit algorithm employed allowed for a simultaneous fit of up to 10 model
parameters.
Statistical and systematic errors
as well as correlations of the model parameters have been determined.

All models reasonably describe the inclusive and the event shape
distributions.
The data measured form charged particles and charged plus neutral
particles when compared with the corresponding model predictions
yield consistent results.\\
All models underestimate the tail of the $p_t^{out}$ distribution by more than
$25\%$.
With this exception the best overall description of event shapes
is provided by the ARIADNE 4.06 model.
HERWIG 5.8c tends to overestimate and JETSET 7.3/7.4 to underestimate
the production of 4 and more jet events.
Correspondingly the tails of
shape distributions sensitive to particle production out of the event plane
are overestimated (underestimated) by HERWIG (JETSET).
The matrix element model JETSET 7.4 ME with optimized scale also provides
reasonable predictions. It however shows the expected discrepancies
due to missing higher orders in the extreme 2-jet and multijet regions.

Identified meson spectra are fairly well described by all models.
It has been found that strong production of p-wave resonances ($25-40\%$) has
to be
considered. This is not expected in a string fragmentation picture.

The gross features of baryon production are described by JETSET and ARIADNE.
Some discrepancies however show up in the momentum spectra and in the relative
abundance of the individual multiplet states.
HERWIG shows stronger discrepancies, especially the predicted
fragmentation functions are too hard.
In JETSET and ARIADNE a similar tendency has been corrected by an extra
leading baryon suppression.

{}~

{\bf\Large Acknowledgement}\\

We would like to thank L. L\"onnblad, I. Knowles, M. Seymour and T. Sj\"ostrand
for useful discussions.

\clearpage

\appendix
\section{Definition of Variables}
\label{vars}
Throughout this paper we use the following definitions of event shape
and inclusive particle variables:
\subsection{Inclusive Single Particle Variables}
\begin{description}
\setlength{\leftmargin}{0.5cm}
\item[Scaled Momentum, $x_{p}$]~\\
The Scaled Momentum $x_{p}$ is the absolute momentum $|\vec{p}|$ of a
particle scaled to the beam momentum.
\item[Transverse Momenta, $p_t^{in}, p_t^{out}$] ~\\
We distinguish the transverse momentum of a particle in the event
plane $p_t^{in}=\vec{p}\cdot\vec{n}_{Major}$ and out of the event plane
$p_t^{out}=\vec{p}\cdot\vec{n}_{Minor}$. Alternatively we use the axes
as defined by the eigenvectors of the quadratic momentum tensor.

\item[Rapidity $y$] ~\\
The rapidity is given by:
\[
y=\frac{1}{2}\cdot\log\frac{E-p_{\parallel}}{E+p_{\parallel}}
\]
where $p_{\parallel}$ is a particles momentum parallel to $\vec{n}_{Thrust}$
or $\vec{n}_{Sphericity}$.
\end{description}
\subsection{Event Shapes Variables}
\begin{description}
\leftmargin0.05cm
\item[Thrust $T$, Major $M$, Minor $m$, Oblateness $O$] ~\\
Thrust $T$ \cite{thrust} and Thrust-axis are defined by:
\[
\nonumber
T = \max_{\vec{n}_{Thrust}}
\frac{\sum\limits_{i=1}^{N_{particle}} \left| \vec{p}_{i} \cdot
\vec{n}_{Thrust}
\right| } {\sum\limits_{i=1}^{N_{particle}} \left| \vec{p}_{i} \right| }
\]
$\vec{p}$ is the 3-momentum of a particle, $\vec{n}_{Thrust}$ is a
unit-vector along the Thrust-axis.
Major and Minor are defined via the same expression, however replacing
$\vec{n}_{Thrust}$
by
$\vec{n}_{Major} \perp \vec{n}_{Thrust}$
or
$ \vec{n}_{Minor} \, = \, \vec{n}_{Major} \times \vec{n}_{Thrust}$
respectively. The oblateness is $O=M-m$.
\item[Sphericity $S$, Aplanarity $A$, Planarity $P$ ] ~\\
Ordering the eigenvalues $\lambda$ of the quadratic  momentum tensor:
\[
M^{\alpha\beta}=\sum\limits_{i=1}^{N_{particle}} p^{\alpha}_{i} p^{\beta}_{j}
{}~~~~~~~~~~~( \alpha,\beta = 1,2,3 )
\]
\[
\lambda_{1}\ge\lambda_{2} \ge\lambda_{3}
{}~~~~~~~~~~~\lambda_{1}+\lambda_{2}+\lambda_{3}=1
\]
The Spericity is $S=\frac{3}{2} (\lambda_{2}+\lambda_{3})$, the Aplanarity
$A=\frac{3}{2} \lambda_{3}$ and the Planarity is
$P=\frac{2}{3}(S-2A)$ \cite{sphericity}.
The Sphericity-axis is parallel to the eigenvector corresponding to
$\lambda_1$.
As the momenta enter quadratically the Sphericity-axis is influenced
more strongly by large momentum particles than the Thrust-axis.
\item[$C$- and $D$-Parameter] ~\\
C- and D-Parameter are defined through the eigenvalues $\lambda$
of the linear momentum
tensor \cite{c_d}:
\[
\Theta^{\alpha\beta} =
\frac{1}
{\sum\limits_{k=1}^{N_{particle}} \left|\vec{p}_{k}\right|}
\cdot
\sum\limits_{i=1}^{N_{particle}}
\frac{p^{i}_{k}p^{j}_{k}}{\left|\vec{p}_{k}\right|}
\]
\[
C=3\cdot(\lambda_{1}\lambda_{2}+\lambda_{2}\lambda_{3}+
\lambda_{3}\lambda_{1})~~~~~~~~~~
D=27\cdot\lambda_{1}\lambda_{2}\lambda_{3}
\]

\item[Jet Masses $M^2_{high}/E^{2}_{vis}, M^2_{low}/E^{2}_{vis},
M^2_{diff}/E^{2}_{vis} $]~\\
Particles are ordered in the two hemispheres of an event
separated by the plane normal to $\vec{n}_{Thrust}$.
\[
\frac{M^{2}_{high}}{E^{2}_{vis}} = \frac{1}{E^{2}_{vis}} \cdot
\max \left( \left(
\sum\limits_{ \vec{p}_{k} \cdot \vec{n}_{Thrust}>0}  p_{k}
\right)^{2} \, , \,\, \left(
\sum\limits_{ \vec{p}_{k} \cdot \vec{n}_{Thrust}<0}  p_{k}
\right)^{2}\right)
\]
For $\frac{M^2_{low}}{E^{2}_{vis}}$ the maximum in the above formula is
replaced by the minimum and
$\frac{M^2_{diff}}{E^{2}_{vis}} =
\frac{M^2_{high}}{E^{2}_{vis}}-\frac{M^2_{low}}{E^{2}_{vis}}$.
This definition differs from the original one by Clavelli \cite{clavelli}
but is easier to calculate and therefore prefered by many
experiments.
\item[Jet Broadening $B_{max}, B_{min}, B_{sum}, B_{diff}$]~\\
Similarly to the definition of the jetmasses the transverse momenta
with respect to the Thrust-axis are added up \cite{bbw}, i.e.:
\[
B_{\pm} = \frac{
\sum
\limits_{ \pm \vec{p}_{i} \cdot \vec{n}_{Thrust} > 0}
\left | \right. \vec{p}_{i} \times \vec{n}_{Thrust} \left  | \right.
}
{2 \sum\limits_i \left | \vec{p}_i  \right |}
\]
Then:
\[
B_{max}=\max{B_+, B_-}~~~~B_{min}=\min{B_+, B_-}
{}~~~~B_{sum}=B_+ + B_-~~~~B_{diff}=|B_+ - B_-|
\]

%\item[Fox-Wolfram-Moments $H_k$]~\\
%Fox-Wolfram moments are event shape measures independent of
%an axis definition \cite{fox}:
%\[
%H_l = \sum\limits_{i,j}^{N_{particle}}
%\frac{|\vec{p}_i||\vec{p}_j|}{E_{vis}^2}\cdot P_l(cos{\chi_{ij}})
%\]
%$\chi_{ij}$ is the angle between the particles $i$ and $j$.
%$H_1=0$ and $H_0 \approx 1$ due to energy momentum conservation. We present
%the moments normalized to $H_0$, i.e. $H_{k0}=H_k/H_0$.
\item[Differential Jet Rates $D_i(y)$] ~\\
Jets are reconstructed using cluster finding algorithms
of the JADE typ \cite{jade}. For each
event and each pair of particles i and j
the scaled invariant mass or transverse momentum
$y_{ij}$ for the JADE or Durham algorithm respectively
are evaluated:
\[
y_{ij}^{Jade} =
 2E_{i}E_{j}/E_{vis}^{2}\cdot(1-\cos\theta_{ij}) ~\\
\]
\[
y_{ij}^{Durham} =
 2\min(E_{i}^{2},E_{j}^{2})/E_{vis}^{2}\cdot(1-\cos\theta_{ij})       \\
\]
where $E_i$, $E_j$ are the energies and
$\theta_{ij}$ the angle between the momentum vectors of the two particles.
The particle pair with the lowest value $y_{ij}$ is selected and replaced
by a pseudo-particle with four momentum $(p_i + p_j)$, hereby reducing the
multiplicity by one. In successive steps the procedure is repeated until the
scaled invariant masses of all pairs of (pseudo-)particles are
larger than a given resolution  $y$. The remaining (pseudo-)particles
are called jets.
The differential 2-jet rate $D_{2}$ is derived from
the 2-jet rate $R_2=\frac{N_{2-jets}}{N}$  \cite{smolik}:
\[
D_{2}(y)   =  \frac {R_{2}(y + \Delta y)- R_{2}(y)}{\Delta y}
\]
Higher differential rates follow from the recursion:
\[
D_{n}(y)   =  \frac {R_{n}(y + \Delta y)- R_{n}(y)}{\Delta y} + D_{n-1}(y)
\]
\item[Energy-Energy-Correlation $EEC$ and Asymmetry $AEEC$] ~\\
The Energy-Energy-Correlation $EEC$ is the histogram of angles $\chi_{ij}$
between all particles weighted by their scaled energies \cite{eec}:
\[
EEC(cos{\chi}) = \frac{1}{N}\frac{1}{\Delta\cos{\chi}} \cdot
\sum_{events}^N\sum_{i,j} \frac{E_i}{E_{vis}}\frac{E_j}{E_{vis}}
\Theta(\Delta\cos{\chi} - |cos{\chi}-cos{\chi_{ij}}| )
\]
$cos{\chi}$ and $\Delta\cos{\chi}$ are the low edge and width
of a bin and $\Theta$ is the step function.
For $cos{\chi}\geq 0$ the asymmetry $AEEC$ of the $EEC$ then is:
\[
AEEC(cos{\chi}) = EEC(-cos{\chi}) - EEC(cos{\chi})
\]
\end{description}

\clearpage

\section{Parameter Settings of the DELPHI Monte Carlo}
\label{delphi_par}
The DELPHI tuning of JETSET has been obtained by tuning the model to charged
particle data from the 1991 and 1992 data taking.
Care has been taken to describe especially well the observables relevant
to standard precision analyses which are the charged multiplicity, the momentum
spectrum, 2-jet rate and Thrust and Sphericity distribution.
Also included is the simulation of Bose Einstein interference (by LUBOEI)
to obtain a correct description of two particle correlations and light
resonance line shapes.
The BE para\-meters are taken from the DELPHI measurement \cite{bepaper}
Particle spectra have been adjusted and partially
fitted to available data \cite{ada}.
Heavy particle decays have been adjusted.
When using adjusted decay tables this is quoted as DELPHI decays.
\begin{table}[h]
\begin{tabular}{lcc}
\hline
Variable & Dec. 93 & Sept. 94 \\
\hline
MSTJ(11) & 3 & 3  \\
MSTJ(12) & 3 & 3  \\
MSTJ(41) & 2 & 2 \\
MSTJ(45) & 5 & 5  \\
MSTJ(46) & 3 & 3  \\
MSTJ(51) & 2 & 2  \\
MSTJ(52) & 7 & 7  \\
MSTJ(101)& 5 & 5 \\
MSTJ(107) & 0 & 0 \\
\hline
PARJ(1)  & 0.10  & 0.10  \\
PARJ(2)  & 0.28  & 0.28  \\
PARJ(3)  & 0.55  & 0.55  \\
PARJ(4)  & 0.07  & 0.07  \\
PARJ(5)  & 0.5   & 0.5   \\
PARJ(11) & 0.55  &       \\
PARJ(12) & 0.55  &       \\
PARJ(13) & 0.75  &       \\
PARJ(14) & 0.090 &       \\
PARJ(15) & 0.070 &       \\
PARJ(16) & 0.085 &       \\
PARJ(17) & 0.140 &       \\
PARJ(19) & 0.5   & 0.5   \\
PARJ(21)  & 0.417   & 0.428 \\
PARJ(25)  & 0.7     & 0.7  \\
PARJ(26)  & 0.2     & 0.2  \\
PARJ(41)  & 0.5     & 0.354  \\
PARJ(42)  & 0.701   & 0.523 \\
PARJ(54)  & -0.0631 & -0.0305 \\
PARJ(55)  & -0.00414& -0.00233 \\
\hline
\end{tabular}
\hfill
\begin{tabular}{lcc}
\hline
Variable & Dec. 93 & Sept. 94 \\
\hline
PARJ(81)  & 0.297   & 0.346\\
PARJ(82)  & 1.732   & 2.25\\
PARJ(92)  & 1       & 1  \\
PARJ(93)  & 0.394   & 0.394  \\
$P(^1S_0)_{ud}$  &      &   .423  \\
$P(^3S_1)_{ud}$  &      &   .275  \\
$P(^1P_1)_{ud}$  &      &   .067  \\
$P(^3P_0)_{ud}$  &      &   .056  \\
$P(^3P_1)_{ud}$  &      &   .067  \\
$P(^3P_2)_{ud}$  &      &   .112  \\
$P(^1S_0)_{s }$  &      &   .388  \\
$P(^3S_1)_{s }$  &      &   .296  \\
$P(^1P_1)_{s }$  &      &   .079  \\
$P(^3P_0)_{s }$  &      &   .026  \\
$P(^3P_1)_{s }$  &      &   .079  \\
$P(^3P_2)_{s }$  &      &   .132  \\
$P(^1S_0)_{c }$  &      &   .250  \\
$P(^3S_1)_{c }$  &      &   .400  \\
$P(^1P_1)_{c }$  &      &   .087  \\
$P(^3P_0)_{c }$  &      &   .030  \\
$P(^3P_1)_{c }$  &      &   .087  \\
$P(^3P_2)_{c }$  &      &   .146  \\
$P(^1S_0)_{b }$  &      &   .1625 \\
$P(^3S_1)_{b }$  &      &   .4875 \\
$P(^1P_1)_{b }$  &      &   .087  \\
$P(^3P_0)_{b }$  &      &   .030  \\
$P(^3P_1)_{b }$  &      &   .087  \\
$P(^3P_2)_{b }$  &      &   .146  \\
\hline
\end{tabular}
\caption{DELPHI parameter setting of JETSET 7.3 PS (December 1993 tuning used
for modelling detector effects for 1993 data and
September 1994 tuning used for modelling detector effects for 1994 data )}
\label{delphi_tuning}
\end{table}

\clearpage
%\newpage

\section{ Tables of Sensitivities }
\scriptsize

\scriptsize

\begin{table}[h]
\begin{center}
\begin{tabular}
{lcccccccccc}\hline
 Parameter &
 \multicolumn{2}{c}{JETSET 7.3 PS}&
           \multicolumn{2}{c}{JETSET 7.4 PS}&
           \multicolumn{2}{c}{ARIADNE 4.06 }&
           \multicolumn{2}{c}{JETSET 7.4 ME}\\
 & {\tiny{$<x_E> D^{\ast}$}} &  {\tiny{$<x_E> B$}} &
  {\tiny{$<x_E> D^{\ast}$}} &  {\tiny{$<x_E> B$}} &
  {\tiny{$<x_E> D^{\ast}$}} &  {\tiny{$<x_E> B$}} &
  {\tiny{$<x_E> D^{\ast}$}} &  {\tiny{$<x_E> B$}} \\
$\Lambda_{QCD}$  & 32& 30& 29& 26&  23& 18& 22& 23\\
$\epsilon_C$     &  9&   &  9&   &  10&   &  9&   \\
$\epsilon_B$     &   &  8&   &  8&    &  9&   & 10\\ \hline
\end{tabular}
\end{center}
\caption[Sensitivity (x100) of $<x_E>$ for $D^\ast$- and B-mesons]
 {\label{sensxe} Sensitivity (x100) of $<x_E>$ for $D^\ast$- and B-mesons}
\end{table}

\begin{table}[h]
\begin{center}
\begin{tabular}
{lccccccccccc}\hline
\multicolumn{8}{c}{~} &
\multicolumn{2}{c}{u/d quarks} &
\multicolumn{2}{c}{s quarks} \\
Property &
   \multicolumn{1}{c}{$a$} &
   \multicolumn{1}{c}{$b$} &
   \multicolumn{1}{c}{$\sigma_q$} &
   \multicolumn{1}{c}{$\Lambda_{QCD}$} &
   \multicolumn{1}{c}{$Q_0$} &
   \multicolumn{1}{c}{$\gamma_s$} &
   \multicolumn{1}{c}{{\tiny $\frac{P(qq)}{P(q)}$}} &
   \multicolumn{1}{c}{{\tiny $P(^1S_0)$}} &
   \multicolumn{1}{c}{{\tiny $P(^3S_1)$}} &
   \multicolumn{1}{c}{{\tiny $P(^1S_0)$}} &
   \multicolumn{1}{c}{{\tiny $P(^3S_1)$}} \\
  \hline
$S      $&                     1&  2&  5& 10&  1&  4&  2&  1&  1&  0&  0\\
$A      $&                     2&  6& 12& 20&  1&  9& 12&  1&  2&  4&  1\\
$P      $&                     1&  2&  2&  8&  1&  2&  1&  1&  1&  0&  1\\
$1-T    $&                     2&  4&  6& 10&  1&  2&  2&  0&  1&  0&  0\\
$M      $&                     6& 13& 11& 19&  6&  4&  4&  1&  1&  2&  0\\
$m      $&                     5& 13& 11& 18&  4&  1&  5&  1&  1&  1&  1\\
$O      $&                     2&  5&  6&  6&  1&  3&  2&  1&  0&  0&  1\\
$D_2^D  $&                     1&  1&  6&  9&  2&  6&  2&  0&  2&  1&  1\\
$D_3^D  $&                     1&  2&  9& 22&  6& 12& 11&  3&  4&  1&  1\\
$D_4^D  $&                     4&  9& 16& 45&  5& 16& 17&  9&  6&  5&  3\\
$D_2^J  $&                     1&  1&  2&  8&  2&  3&  2&  0&  1&  0&  1\\
$D_3^J  $&                     2&  4&  6& 19&  1&  2&  1&  2&  0&  0&  1\\
$D_4^J  $&                     3&  9& 11& 20&  3&  7&  4&  3&  1&  1&  1\\
$C      $&                     2&  5&  5& 10&  1&  3&  2&  1&  1&  1&  1\\
$D      $&                     2&  6&  6& 15&  2&  4&  5&  1&  0&  1&  1\\
$M^2_h/E_{vis.}^2 $&          2&  5&  3&  9&  1&  4&  1&  1&  0&  1&  1\\
$M^2_l/E_{vis.}^2 $&           5& 11&  2& 23&  3&  4&  4&  1&  1&  1&  1\\
$M^2_d/E_{vis.}^2 $&           1&  4&  4&  6&  0&  6&  2&  1&  2&  2&  0\\
$B_{max} $&                    3&  8&  7& 14&  2&  2&  2&  1&  0&  1&  1\\
$B_{min} $&                    3&  7&  7& 20&  4&  2&  6&  3&  1&  2&  2\\
$B_{sum} $&                    5& 10&  8& 15&  4&  5&  4&  1&  1&  1&  0\\
$B_{diff.} $&                  1&  4&  5&  5&  1&  5&  3&  1&  2&  2&  1\\
\hline
$p_t^{in} $&                   5&  8&  9&  5&  3&  3&  3&  2&  2&  2&  1\\
$p_t^{out}$&                   4&  8& 27& 11&  2&  2&  2&  4&  2&  2&  0\\
$y_T    $&                     2&  5&  6&  8&  2&  1&  2&  1&  1&  1&  0\\
$x_p    $&                     3&  6&  7&  6&  4&  1&  1&  2&  1&  1&  0\\
\hline
$p_t^{out} vs. x_p $&          3&  7&  7&  9&  7& 11&  6&  3&  2&  2&  0\\
$p_t^{trans.} vs. x_p  $&      3&  7&  7&  9&  7& 11&  6&  3&  2&  2&  0\\
$EEC    $&                     1&  1&  1&  4&  0&  1&  0&  0&  0&  0&  0\\
$AEEC   $&                     1&  6& 16&  9& 13& 32& 17&  6&  2&  7&  3\\
\hline
$\rho^{\circ}      $&          4&  7&  9&  8&  4&  4&  3&  3&  9&  1&  1\\
$K^0 / K^{\pm}  $&             4&  7&  9&  6&  4& 13&  4&  1&  0&  6&  2\\
$K^{\ast 0}/ K^{\ast \pm} $&   4&  8& 10&  7&  4& 14&  4&  0&  1&  8& 10\\
$\phi          $&              5&  8&  7&  9&  4& 35&  9&  3&  3&  5& 23\\
$p             $&              5& 12& 13&  5&  6&  3& 22&  3&  1&  1&  3\\
$\Lambda^0              $&     5& 11& 12& 12&  4& 17& 21&  2&  2&  2&  2\\
$\Sigma^{\pm}(1193)    $&      4& 15& 18& 11&  8&  9& 21&  4&  4&  5&  5\\
\hline
\end{tabular}
\end{center}
\caption[Sensitivies for JETSET 7.4 PS]
 {\label{sensjt74} Sensitivities (x100) for JETSET 7.4 PS with standard decays.
ARIADNE (with $Q_0$ in place of $p_t^{QCD}$) and JETSET 7.3 PS
sensitivities are similar.
The identified particle sensitivities have been calculated for the $\xi_p$
distributions.
The quoted values have  significant statistical errors.}
\end{table}

\begin{table}
\begin{center}
\begin{tabular}
{lccccccccc}\hline
Property &
   \multicolumn{1}{c}{{\tiny $QCDLAM$}} &
   \multicolumn{1}{c}{{\tiny $RMASS(13)$}} &
   \multicolumn{1}{c}{{\tiny $CLMAX$}} &
   \multicolumn{1}{c}{{\tiny $CLPOW$}} &
   \multicolumn{1}{c}{{\tiny $CLSMR$}} &
   \multicolumn{1}{c}{{\tiny $PWT(3)$}} &
   \multicolumn{1}{c}{{\tiny $PWT(7)$}} &
   \multicolumn{1}{c}{{\tiny $DECWT$}} \\ \hline
$S      $&                    11&  3&  2&  3&  1&  1&  0&  1\\
$A      $&                    25&  3&  7&  5&  2&  2&  1&  3\\
$P      $&                     9&  2&  2&  2&  0&  0&  1&  1\\
$1-T    $&                    12&  6&  1&  1&  0&  1&  0&  1\\
$M      $&                    13&  7&  3&  7&  1&  1&  1&  2\\
$m      $&                    20&  5&  3&  4&  1&  1&  1&  2\\
$O      $&                     8&  1&  2&  1&  0&  1&  1&  1\\
$D_2^D  $&                    12&  2&  3&  1&  1&  1&  1&  0\\
$D_3^D  $&                    27&  6&  4&  3&  1&  1&  1&  2\\
$D_4^D  $&                    39&  5&  8&  6&  3&  5&  3&  2\\
$D_2^J  $&                    11&  2&  2&  1&  0&  0&  1&  1\\
$D_3^J  $&                    21&  5&  3&  3&  1&  2&  1&  1\\
$D_4^J  $&                    24&  6&  7&  5&  1&  1&  1&  2\\
$C      $&                    12&  6&  1&  1&  0&  1&  0&  1\\
$D      $&                    20&  5&  2&  1&  1&  1&  1&  2\\
$M^2_h/E_{vis.}^2 $&         10&  7&  4&  2&  1&  1&  1&  1\\
$M^2_l/E_{vis.}^2 $&          17& 14&  5&  2&  1&  2&  1&  1\\
$M^2_d/E_{vis.}^2 $ &         10&  3&  4&  3&  1&  0&  1&  1\\
$B_{max} $&                   13&  8&  3&  4&  1&  1&  0&  1\\
$B_{min} $&                   19&  7&  3&  2&  1&  1&  1&  2\\
$B_{sum} $&                   12&  7&  1&  4&  1&  0&  1&  2\\
$B_{diff.} $&                  8&  1&  2&  1&  0&  1&  1&  1\\\hline
$p_t^{in} $&                   6&  4&  5& 11&  3&  1&  0&  1\\
$p_t^{out}$&                  15&  5& 14&  7&  2&  1&  0&  2\\
$y_T    $&                     6&  4&  7& 10&  2&  1&  1&  1\\
$x_p    $&                     2&  3&  4&  9&  3&  1&  1&  0\\\hline
$p_t^{out} vs. x_p $&          1&  3&  3& 50&  5&  2&  2&  5\\
$p_t^{trans.} vs. x_p  $&      1&  3&  3& 50&  5&  2&  2&  5\\
$EEC    $&                     5&  1&  1&  0&  0&  0&  0&  0\\
$AEEC   $&                     7&  3&  3&  4&  1&  1&  2&  4\\\hline
$\rho^{\circ}      $&          4&  6&  7& 12&  3&  4&  2&  2\\
$K^0 / K^{\pm}  $&             6& 11& 13& 23&  6& 13&  4&  2\\
$K^{\ast 0}/ K^{\ast \pm} $&        8& 14&  8& 15&  4& 14&  5&  3\\
$\phi          $&             15& 24& 10& 21&  6& 33& 10&  3\\
$p             $&              5&  7& 48& 20&  7&  2&  3& 15\\
$\Lambda^0             $&     9& 13& 65& 27&  8& 18&  8& 11\\
$\Sigma^{\pm}(1193)    $&     10& 18& 46& 16&  7& 13&  7&  7\\ \hline
\end{tabular}
\end{center}
\caption[Sensitivies for HERWIG 5.8c]
 {\label{sensh58} Sensitivities (x100) for HERWIG 5.8c with standard decays.
The identified particle sensitivities have been calculated for the $\xi_p$
distributions.
The quoted values have  significant statistical errors.}
\end{table}
\normalsize
\clearpage
\normalsize

\section{ Tables of Inclusive Charge Particle and Event Shape Distributions }
\label{tables}
The cross section has been determined from charged particles only and
from charged and neutral particles. The tables show first the measurement
from charged particles corrected to the charged final state, then corrected
to the full final state. The third column contains the measurement from
charged plus neutral particles corrected to the full final state.
For inclusive charged particle distributions the relevant event axes
have been evaluated for the corresponding final states.
The cross section given in the table is
followed by the statistical and  systematic errors.

\subsection{ Tables of Inclusive Charged Particle Distributions }
\scriptsize

\begin{table}[h]
\begin{center}
\begin{tabular}
 {r@{ - }rr@{.}l@{ $\pm$ }r@{.}l@{ $\pm$ }r@{.}lr@{.}l@{ $\pm$ }r@{.}l@{ $\pm$
}r@{.}l
r@{.}l@{$\pm$ }r@{.}l@{ $\pm$ }r@{.}l }\hline
\multicolumn{10}{l}{ $p^{in}_{t}$ with respect to the Thrust axis }&
\multicolumn{10}{l}{  $\frac{1}{N} \; \frac{dN}{dp^{in}_{t}}$} \\ \hline
\multicolumn{2}{c}{  } &
  \multicolumn{6}{c}{charged particles} &
  \multicolumn{6}{c}{charged particles} &
  \multicolumn{6}{c}{charged plus neutral particles} \\
\multicolumn{2}{c}{Interval/GeV} &
  \multicolumn{6}{c}{corrected to } &
  \multicolumn{6}{c}{corrected to } &
  \multicolumn{6}{c}{corrected to } \\
\multicolumn{2}{c}{  } &
  \multicolumn{6}{c}{charged particles} &
  \multicolumn{6}{c}{charged plus neutral particles} &
  \multicolumn{6}{c}{charged plus neutral particles} \\ \hline
 0.000 & 0.100 & 44&667 & 0&036 & 1&653  & 44&740 & 0&036 & 1&685  & 46&663 &
0&037 & 1&758  \\
 0.100 & 0.200 & 39&331 & 0&033 & 1&062  & 39&475 & 0&033 & 1&078  & 39&823 &
0&033 & 1&092  \\
 0.200 & 0.300 & 29&429 & 0&028 & 0&603  & 29&491 & 0&028 & 0&606  & 29&351 &
0&028 & 0&608  \\
 0.300 & 0.400 & 21&326 & 0&024 & 0&353  & 21&338 & 0&024 & 0&352  & 21&034 &
0&024 & 0&350  \\
\hline
\end{tabular}
\end{center}
\caption[$p^{in}_{t}$ with respect to the Thrust axis ]{\label{pint}
{Transverse momentum  $p^{in}_{t}$ with respect to the Thrust axis}}
\end{table}

\begin{table}[h]
\begin{center}
\begin{tabular}
 {r@{ - }rr@{.}l@{ $\pm$ }r@{.}l@{ $\pm$ }r@{.}lr@{.}l@{ $\pm$ }r@{.}l@{ $\pm$
}r@{.}l
r@{.}l@{$\pm$ }r@{.}l@{ $\pm$ }r@{.}l }\hline
\multicolumn{10}{l}{ $p^{out}_{t}$ with respect to the Thrust axis  }&
\multicolumn{10}{l}{$\frac{1}{N} \; \frac{dN}{dp^{out}_{t}}$} \\ \hline
\multicolumn{2}{c}{  } &
  \multicolumn{6}{c}{charged particles} &
  \multicolumn{6}{c}{charged particles} &
  \multicolumn{6}{c}{charged plus neutral particles} \\
\multicolumn{2}{c}{Interval/GeV} &
  \multicolumn{6}{c}{corrected to } &
  \multicolumn{6}{c}{corrected to } &
  \multicolumn{6}{c}{corrected to } \\
\multicolumn{2}{c}{  } &
  \multicolumn{6}{c}{charged particles} &
  \multicolumn{6}{c}{charged plus neutral particles} &
  \multicolumn{6}{c}{charged plus neutral particles} \\ \hline
 0.000 & 0.100 & 67&512 & 0&043 & 1&956  & 67&673 & 0&044 & 1&818  & 66&160 &
0&043 & 1&822  \\
 0.100 & 0.200 & 50&517 & 0&037 & 1&024  & 50&698 & 0&037 & 1&166  & 49&794 &
0&037 & 1&149  \\
 0.200 & 0.300 & 33&618 & 0&030 & 0&593  & 33&803 & 0&030 & 0&692  & 33&544 &
0&030 & 0&678  \\
 0.300 & 0.400 & 21&228 & 0&024 & 0&368  & 21&278 & 0&024 & 0&403  & 21&407 &
0&024 & 0&397  \\
\hline
\end{tabular}
\end{center}
\caption[$p^{out}_{t}$ with respect to the Thrust axis ]{\label{poutt}
{Transverse momentum  $p^{out}_{t}$ with respect to the Thrust axis}}
\end{table}

\newpage

\begin{table}[h]
\begin{center}
\begin{tabular}
 {r@{ - }rr@{.}l@{ $\pm$ }r@{.}l@{ $\pm$ }r@{.}lr@{.}l@{ $\pm$ }r@{.}l@{ $\pm$
}r@{.}l
r@{.}l@{$\pm$ }r@{.}l@{ $\pm$ }r@{.}l }\hline
\multicolumn{10}{l}{ $p^{in}_{t}$ with respect to the Sphericity axis }&
\multicolumn{10}{l}{ $\frac{1}{N} \; \frac{dN}{dp^{in}_{t}}$} \\ \hline
\multicolumn{2}{c}{  } &
  \multicolumn{6}{c}{charged particles} &
  \multicolumn{6}{c}{charged particles} &
  \multicolumn{6}{c}{charged plus neutral particles} \\
\multicolumn{2}{c}{Interval}/GeV &
  \multicolumn{6}{c}{corrected to } &
  \multicolumn{6}{c}{corrected to } &
  \multicolumn{6}{c}{corrected to } \\
\multicolumn{2}{c}{  } &
  \multicolumn{6}{c}{charged particles} &
  \multicolumn{6}{c}{charged plus neutral particles} &
  \multicolumn{6}{c}{charged plus neutral particles} \\ \hline
 0.000 & 0.100 & 49&063 & 0&038 & 1&068  & 48&833 & 0&038 & 1&696  & 49&206 &
0&038 & 1&672  \\
 0.100 & 0.200 & 38&338 & 0&032 & 0&819  & 38&301 & 0&033 & 1&265  & 38&461 &
0&033 & 0&984  \\
 0.200 & 0.300 & 28&259 & 0&027 & 0&593  & 28&231 & 0&027 & 0&886  & 28&203 &
0&027 & 0&571  \\
 0.300 & 0.400 & 20&492 & 0&023 & 0&422  & 20&554 & 0&023 & 0&613  & 20&391 &
0&023 & 0&349  \\
\hline
\end{tabular}
\end{center}
\caption[$p^{in}_{t}$ with respect to the Sphericity axis ]{\label{pins}
{Transverse momentum  $p^{in}_{t}$ with respect to the Sphericity axis }}
\end{table}

\begin{table}[h]
\begin{center}
\begin{tabular}
 {r@{ - }rr@{.}l@{ $\pm$ }r@{.}l@{ $\pm$ }r@{.}lr@{.}l@{ $\pm$ }r@{.}l@{ $\pm$
}r@{.}l
r@{.}l@{$\pm$ }r@{.}l@{ $\pm$ }r@{.}l }\hline
\multicolumn{10}{l}{ $p^{out}_{t}$ with respect to the Sphericity axis }&
\multicolumn{10}{l}{ $\frac{1}{N} \; \frac{dN}{dp^{out}_{t}}$} \\ \hline
\multicolumn{2}{c}{  } &
  \multicolumn{6}{c}{charged particles} &
  \multicolumn{6}{c}{charged particles} &
  \multicolumn{6}{c}{charged plus neutral particles} \\
\multicolumn{2}{c}{Interval/GeV} &
  \multicolumn{6}{c}{corrected to } &
  \multicolumn{6}{c}{corrected to } &
  \multicolumn{6}{c}{corrected to } \\
\multicolumn{2}{c}{  } &
  \multicolumn{6}{c}{charged particles} &
  \multicolumn{6}{c}{charged plus neutral particles} &
  \multicolumn{6}{c}{charged plus neutral particles} \\ \hline
 0.000 & 0.100 & 68&848 & 0&044 & 1&931  & 69&511 & 0&045 & 1&992  & 66&825 &
0&043 & 1&506  \\
 0.100 & 0.200 & 51&541 & 0&037 & 1&039  & 51&918 & 0&038 & 1&141  & 50&556 &
0&037 & 1&102  \\
 0.200 & 0.300 & 34&351 & 0&030 & 0&607  & 34&419 & 0&030 & 0&644  & 34&241 &
0&030 & 0&726  \\
 0.300 & 0.400 & 21&343 & 0&024 & 0&373  & 21&274 & 0&024 & 0&372  & 21&708 &
0&024 & 0&451  \\
\hline
\end{tabular}
\end{center}
\caption[$p^{out}_{t}$ with respect to the Sphericity axis ]{\label{pouts}
{Transverse momentum  $p^{out}_{t}$ with respect to the Sphericity axis}}
\end{table}

%\clearpage

\begin{table}[p]
\begin{center}
\begin{tabular}
 {r@{ - }rr@{.}l@{ $\pm$ }r@{.}l@{ $\pm$ }r@{.}lr@{.}l@{ $\pm$ }r@{.}l@{ $\pm$
}r@{.}l
r@{.}l@{$\pm$ }r@{.}l@{ $\pm$ }r@{.}l }\hline
\multicolumn{10}{l}{ Rapidity $y_{T}$ with respect to the Thrust axis }&
\multicolumn{10}{l}{$\frac{1}{N} \; \frac{dN}{dy_{T}}$} \\ \hline
\multicolumn{2}{c}{  } &
  \multicolumn{6}{c}{charged particles} &
  \multicolumn{6}{c}{charged particles} &
  \multicolumn{6}{c}{charged plus neutral particles} \\
\multicolumn{2}{c}{Interval/GeV} &
  \multicolumn{6}{c}{corrected to } &
  \multicolumn{6}{c}{corrected to } &
  \multicolumn{6}{c}{corrected to } \\
\multicolumn{2}{c}{  } &
  \multicolumn{6}{c}{charged particles} &
  \multicolumn{6}{c}{charged plus neutral particles} &
  \multicolumn{6}{c}{charged plus neutral particles} \\ \hline
 0.000 & 0.250 & 5&7889 & 0&0093 & 0&5509  & 5&7590 & 0&0093 & 0&5296  & 5&9517
& 0&0095 & 0&5628  \\
 0.250 & 0.500 & 6&3749 & 0&0094 & 0&4306  & 6&4046 & 0&0095 & 0&4411  & 6&4291
& 0&0095 & 0&4417\\
 0.500 & 0.750 & 6&6630 & 0&0093 & 0&3197  & 6&6834 & 0&0094 & 0&3399  & 6&6831
& 0&0094 & 0&3319  \\
 0.750 & 1.000 & 6&7572 & 0&0089 & 0&2316  & 6&8119 & 0&0090 & 0&2522  & 6&7763
& 0&0089 & 0&2429  \\
\hline
\end{tabular}
\end{center}
\caption[$y_{T}$ Thrust axis ]{\label{yt} {Rapidity $y_{T}$ with respect to the
Thrust axis}}
\end{table}

\begin{table}[p]
\begin{center}
\begin{tabular}
 {r@{ - }rr@{.}l@{ $\pm$ }r@{.}l@{ $\pm$ }r@{.}lr@{.}l@{ $\pm$ }r@{.}l@{ $\pm$
}r@{.}l
r@{.}l@{$\pm$ }r@{.}l@{ $\pm$ }r@{.}l }\hline
\multicolumn{10}{l}{ Rapidity $y_S $ with respect to the Sphericity axis  }&
\multicolumn{10}{l}{ $\frac{1}{N} \; \frac{dN}{dy_S}$} \\ \hline
\multicolumn{2}{c}{  } &
  \multicolumn{6}{c}{charged particles} &
  \multicolumn{6}{c}{charged particles} &
  \multicolumn{6}{c}{charged plus neutral particles} \\
\multicolumn{2}{c}{Interval} &
  \multicolumn{6}{c}{corrected to } &
  \multicolumn{6}{c}{corrected to } &
  \multicolumn{6}{c}{corrected to } \\
\multicolumn{2}{c}{  } &
  \multicolumn{6}{c}{charged particles} &
  \multicolumn{6}{c}{charged plus neutral particles} &
  \multicolumn{6}{c}{charged plus neutral particles} \\ \hline
 0.000 & 0.250 & 6&5422 & 0&0096 & 0&5337  & 6&5778 & 0&0097 & 0&5082  & 6&5680
& 0&0097 & 0&5323  \\
 0.250 & 0.500 & 6&5743 & 0&0096 & 0&4132  & 6&6009 & 0&0096 & 0&4157  & 6&5901
& 0&0096 & 0&4246  \\
 0.500 & 0.750 & 6&5997 & 0&0092 & 0&3151  & 6&6327 & 0&0093 & 0&3348  & 6&6094
& 0&0093 & 0&3329  \\
 0.750 & 1.000 & 6&5896 & 0&0087 & 0&2357  & 6&6311 & 0&0088 & 0&2630  & 6&6152
& 0&0088 & 0&2554  \\
\hline
\end{tabular}
\end{center}
\caption[$y_S $ with respect to the Sphericity axis ]{\label{ys}
{Rapidity $y_S$ with respect to the Sphericity axis}}
\end{table}
%\clearpage

\twocolumn
\begin{table}[p]
\begin{center}
\begin{tabular}
 {r@{ - }rr@{.}l@{ $\pm$ }r@{.}l@{ $\pm$ }r@{.}l}\hline
\multicolumn{5}{l}{Scaled momentum $x_p$ }& \multicolumn{3}{l}{$\frac{1}{N} \;
\frac{dN}{dx_p}$} \\ \hline
\multicolumn{2}{c}{  } &
  \multicolumn{6}{c}{charged particles} \\
\multicolumn{2}{c}{Interval} &
  \multicolumn{6}{c}{corrected to } \\
\multicolumn{2}{c}{  } &
  \multicolumn{6}{c}{charged particles} \\ \hline
 0.000 & 0.010 & 389&18 & 0&37 & 28&68  \\
 0.010 & 0.020 & 407&47 & 0&32 & 10&44  \\
 0.020 & 0.030 & 266&10 & 0&26 & 3&67  \\
 0.030 & 0.040 & 185&45 & 0&22 & 1&85  \\
 0.040 & 0.050 & 137&73 & 0&19 & 1&38  \\
 0.050 & 0.060 & 106&07 & 0&16 & 1&08  \\
 0.060 & 0.070 &  84&57 & 0&15 & 0&95  \\
 0.070 & 0.080 &  68&87 & 0&13 & 0&85  \\
 0.080 & 0.090 &  57&09 & 0&12 & 0&78  \\
 0.090 & 0.100 &  47&79 & 0&11 & 0&71  \\
 0.100 & 0.120 &  37&756 & 0&071 & 0&619  \\
 0.120 & 0.140 &  28&006 & 0&061 & 0&501  \\
 0.140 & 0.160 &  21&330 & 0&054 & 0&409  \\
 0.160 & 0.180 &  16&660 & 0&048 & 0&337  \\
 0.180 & 0.200 &  13&043 & 0&042 & 0&273  \\
 0.200 & 0.250 &   9&019 & 0&022 & 0&195  \\
 0.250 & 0.300 &   5&464 & 0&018 & 0&124  \\
 0.300 & 0.400 &   2&8036 & 0&0088 & 0&0683  \\
 0.400 & 0.500 &   1&1938 & 0&0057 & 0&0327  \\
 0.500 & 0.600 &   0&5322 & 0&0040 & 0&0169  \\
 0.600 & 0.700 &   0&2287 & 0&0027 & 0&0084  \\
 0.700 & 0.800 &   0&0936 & 0&0016 & 0&0039  \\
 0.800 & 1.000 &   0&0201 & 0&0005 & 0&0023  \\ \hline
\end{tabular}
\end{center}
\caption[$x_p$]{\label{xp} {Scaled momentum $x_p$}}
\end{table}

\begin{table}[p]
\begin{center}
\begin{tabular}
 {r@{ - }rr@{.}l@{ $\pm$ }r@{.}l@{ $\pm$ }r@{.}l}\hline
\multicolumn{5}{l}{ $\xi_p=\ln(\frac{1}{x_p}$)    }& \multicolumn{3}{l}{
$\frac{1}{N} \; \frac{dN}{d\xi_p}$} \\ \hline
\multicolumn{2}{c}{  } &
  \multicolumn{6}{c}{charged particles} \\
\multicolumn{2}{c}{Interval} &
  \multicolumn{6}{c}{corrected to } \\
\multicolumn{2}{c}{  } &
  \multicolumn{6}{c}{charged particles} \\ \hline
 0.000 & 0.200 & 0&0151 & 0&0004 & 0&0022  \\
 0.200 & 0.400 & 0&0774 & 0&0011 & 0&0051  \\
 0.400 & 0.600 & 0&1987 & 0&0018 & 0&0083  \\
 0.600 & 0.800 & 0&3970 & 0&0023 & 0&0116  \\
 0.800 & 1.000 & 0&6805 & 0&0031 & 0&0161  \\
 1.000 & 1.200 & 1&0520 & 0&0038 & 0&0226  \\
 1.200 & 1.400 & 1&5128 & 0&0046 & 0&0311  \\
 1.400 & 1.600 & 2&0270 & 0&0053 & 0&0398  \\
 1.600 & 1.800 & 2&5997 & 0&0060 & 0&0481  \\
 1.800 & 2.000 & 3&1993 & 0&0066 & 0&0553  \\
 2.000 & 2.200 & 3&8069 & 0&0071 & 0&0610  \\
 2.200 & 2.400 & 4&3851 & 0&0076 & 0&0651  \\
 2.400 & 2.600 & 4&9235 & 0&0080 & 0&0680  \\
 2.600 & 2.800 & 5&4160 & 0&0083 & 0&0702  \\
 2.800 & 3.000 & 5&8079 & 0&0086 & 0&0716  \\
 3.000 & 3.200 & 6&1608 & 0&0088 & 0&0734  \\
 3.200 & 3.400 & 6&4026 & 0&0090 & 0&0753  \\
 3.400 & 3.600 & 6&5371 & 0&0091 & 0&0779  \\
 3.600 & 3.800 & 6&5822 & 0&0091 & 0&0819  \\
 3.800 & 4.000 & 6&4738 & 0&0090 & 0&0869  \\
 4.000 & 4.200 & 6&2383 & 0&0089 & 0&0933  \\
 4.200 & 4.400 & 5&8490 & 0&0087 & 0&1003  \\
 4.400 & 4.600 & 5&3231 & 0&0084 & 0&1068  \\
 4.600 & 4.800 & 4&6603 & 0&0079 & 0&1098  \\
 4.800 & 5.000 & 3&9029 & 0&0073 & 0&1064  \\
 5.000 & 5.200 & 3&1150 & 0&0066 & 0&0952  \\
 5.200 & 5.400 & 2&3781 & 0&0058 & 0&0777  \\ \hline
\end{tabular}
\end{center}
\caption[$\xi_p$ ]{\label{logxp}
{$\xi_p=\ln(\frac{1}{x_p})$ }}
\end{table}

\begin{table}[p]
\begin{center}
\begin{tabular}
 {r@{ - }rr@{.}l@{ $\pm$ }r@{.}l@{ $\pm$ }r@{.}l}\hline
\multicolumn{5}{l}{$x_p$ }& \multicolumn{3}{l}{$<p^{out}_{t}>/GeV$} \\ \hline
\multicolumn{2}{c}{  } &
  \multicolumn{6}{c}{charged particles} \\
\multicolumn{2}{c}{Interval} &
  \multicolumn{6}{c}{corrected to } \\
\multicolumn{2}{c}{  } &
  \multicolumn{6}{c}{charged particles} \\ \hline
 0.000 & 0.010 & 0&11124 & 0&00012 & 0&00097  \\
 0.010 & 0.020 & 0&19506 & 0&00014 & 0&00113  \\
 0.020 & 0.030 & 0&23694 & 0&00019 & 0&00118  \\
 0.030 & 0.040 & 0&25986 & 0&00023 & 0&00130  \\
 0.040 & 0.050 & 0&27485 & 0&00028 & 0&00137  \\
 0.050 & 0.060 & 0&28678 & 0&00032 & 0&00143  \\
 0.060 & 0.070 & 0&29812 & 0&00037 & 0&00149  \\
 0.070 & 0.080 & 0&30653 & 0&00041 & 0&00153  \\
 0.080 & 0.090 & 0&31324 & 0&00046 & 0&00157  \\
 0.090 & 0.100 & 0&31969 & 0&00051 & 0&00160  \\
 0.100 & 0.120 & 0&32899 & 0&00041 & 0&00164  \\
 0.120 & 0.140 & 0&34140 & 0&00049 & 0&00171  \\
 0.140 & 0.160 & 0&35107 & 0&00058 & 0&00176  \\
 0.160 & 0.180 & 0&35833 & 0&00066 & 0&00179  \\
 0.180 & 0.200 & 0&37165 & 0&00076 & 0&00186  \\
 0.200 & 0.250 & 0&38189 & 0&00059 & 0&00191  \\
 0.250 & 0.300 & 0&40047 & 0&00078 & 0&00200  \\
 0.300 & 0.400 & 0&41720 & 0&00079 & 0&00223  \\
 0.400 & 0.500 & 0&4348 & 0&0012 & 0&0031  \\
 0.500 & 0.600 & 0&4436 & 0&0019 & 0&0044  \\
 0.600 & 0.700 & 0&4412 & 0&0031 & 0&0062  \\
 0.700 & 0.800 & 0&4525 & 0&0046 & 0&0093  \\
 0.800 & 1.000 & 0&3778 & 0&0051 & 0&0112  \\ \hline
\end{tabular}
\end{center}
\caption[$<p^{out}_{t}>$ vs. $x_p$ ]{\label{poutxp}
{$<p^{out}_{t}>$ vs. $x_p$ }}
\end{table}

\begin{table}[p]
\begin{center}
\begin{tabular}
 {r@{ - }rr@{.}l@{ $\pm$ }r@{.}l@{ $\pm$ }r@{.}l}\hline
\multicolumn{5}{l}{$x_p$ }& \multicolumn{3}{l}{$<p_{t}>/GeV$} \\ \hline
\multicolumn{2}{c}{  } &
  \multicolumn{6}{c}{charged particles} \\
\multicolumn{2}{c}{Interval} &
  \multicolumn{6}{c}{corrected to } \\
\multicolumn{2}{c}{  } &
  \multicolumn{6}{c}{charged particles} \\
 0.000 & 0.010 & 0&18791 & 0&00016 & 0&00171  \\
 0.010 & 0.020 & 0&36450 & 0&00019 & 0&00214  \\
 0.020 & 0.030 & 0&48985 & 0&00027 & 0&00245  \\
 0.030 & 0.040 & 0&57682 & 0&00035 & 0&00288  \\
 0.040 & 0.050 & 0&64402 & 0&00043 & 0&00322  \\
 0.050 & 0.060 & 0&69974 & 0&00051 & 0&00350  \\
 0.060 & 0.070 & 0&75195 & 0&00059 & 0&00376  \\
 0.070 & 0.080 & 0&79370 & 0&00067 & 0&00397  \\
 0.080 & 0.090 & 0&83170 & 0&00075 & 0&00416  \\
 0.090 & 0.100 & 0&85958 & 0&00083 & 0&00430  \\
 0.100 & 0.120 & 0&90697 & 0&00069 & 0&00453  \\
 0.120 & 0.140 & 0&96796 & 0&00083 & 0&00484  \\
 0.140 & 0.160 & 1&01237 & 0&00098 & 0&00506  \\
 0.160 & 0.180 & 1&0555 & 0&0011 & 0&0053  \\
 0.180 & 0.200 & 1&0903 & 0&0013 & 0&0055  \\
 0.200 & 0.250 & 1&1594 & 0&0010 & 0&0061  \\
 0.250 & 0.300 & 1&2363 & 0&0014 & 0&0076  \\
 0.300 & 0.400 & 1&3226 & 0&0014 & 0&0100  \\
 0.400 & 0.500 & 1&4043 & 0&0022 & 0&0135  \\
 0.500 & 0.600 & 1&4327 & 0&0035 & 0&0183  \\
 0.600 & 0.700 & 1&4210 & 0&0056 & 0&0246  \\
 0.700 & 0.800 & 1&4134 & 0&0084 & 0&0336  \\
 0.800 & 1.000 & 1&226 & 0&011 & 0&040  \\ \hline
\end{tabular}
\end{center}
\caption[$<p_{t}>$ vs. $x_p$ ]{\label{ptransxp}
$<p_{t}>$ vs. $x_p$}
\end{table}

\begin{table}[p]
\begin{center}
\begin{tabular}
 {r@{ - }rr@{.}l@{ $\pm$ }r@{.}l@{ $\pm$ }r@{.}l}\hline
\multicolumn{5}{l}{ Energy Energy Correlation }& \multicolumn{3}{l}{
$EEC$}\\ \hline
\multicolumn{2}{c}{  } &
  \multicolumn{6}{c}{charged particles} \\
\multicolumn{2}{c}{Interval} &
  \multicolumn{6}{c}{corrected to } \\
\multicolumn{2}{c}{$\cos \chi$  } &
  \multicolumn{6}{c}{charged particles} \\ \hline
 -1.000 & -0.960 &    3&0850 & 0&0085 & 0&0154  \\
 -0.960 & -0.920 &    0&7097 & 0&0040 & 0&0035  \\
 -0.920 & -0.880 &    0&3794 & 0&0029 & 0&0019  \\
 -0.880 & -0.840 &    0&2517 & 0&0024 & 0&0013  \\
\hline
\end{tabular}
\end{center}
\caption[EEC]{\label{EEC}
{Energy Energy Correlation $EEC$} }
\end{table}

\begin{table}[t]
\begin{center}
\begin{tabular} {r@{ - }rr@{.}l@{ $\pm$ }r@{.}l@{ $\pm$ }r@{.}l}\hline
\multicolumn{8}{l}{ Asymmetry of the Energy Energy Correlation   }\\
\multicolumn{5}{l}{ ~}&
\multicolumn{3}{l}{$AEEC$ } \\ \hline
\multicolumn{2}{c}{  } &
  \multicolumn{6}{c}{charged particles} \\
\multicolumn{2}{c}{Interval} &
  \multicolumn{6}{c}{corrected to } \\
\multicolumn{2}{c}{{$\cos \chi$}  } &
  \multicolumn{6}{c}{charged particles} \\ \hline
 -1.000 & -0.960 &   0&18959 & 0&00051 & 0&00924  \\
 -0.960 & -0.920 &   0&22480 & 0&00150 & 0&00220  \\
 -0.920 & -0.880 &   0&12560 & 0&00120 & 0&00130  \\
 -0.880 & -0.840 &   0&08023 & 0&00092 & 0&00080  \\
 -0.840 & -0.800 &   0&05496 & 0&00073 & 0&00055  \\
 -0.800 & -0.760 &   0&04122 & 0&00062 & 0&00049  \\
 -0.760 & -0.720 &   0&03054 & 0&00050 & 0&00046  \\
 -0.720 & -0.680 &   0&02383 & 0&00042 & 0&00044  \\
 -0.680 & -0.640 &   0&01929 & 0&00036 & 0&00043  \\
 -0.640 & -0.600 &   0&01613 & 0&00032 & 0&00042  \\
 -0.600 & -0.560 &   0&01391 & 0&00029 & 0&00041  \\
 -0.560 & -0.520 &   0&01182 & 0&00026 & 0&00079  \\
 -0.520 & -0.480 &   0&00938 & 0&00021 & 0&00036  \\
 -0.480 & -0.440 &   0&00803 & 0&00019 & 0&00035  \\
 -0.440 & -0.400 &   0&00696 & 0&00017 & 0&00033  \\
 -0.400 & -0.360 &   0&00546 & 0&00014 & 0&00029  \\
 -0.360 & -0.320 &   0&00459 & 0&00012 & 0&00027  \\
 -0.320 & -0.280 &   0&00408 & 0&00011 & 0&00026  \\
 -0.280 & -0.240 &   0&00360 & 0&00010 & 0&00025  \\
 -0.240 & -0.200 &   0&00268 & 0&00008 & 0&00020  \\
 -0.200 & -0.160 &   0&00212 & 0&00006 & 0&00017  \\
 -0.160 & -0.120 &   0&00148 & 0&00005 & 0&00013  \\
 -0.120 & -0.080 &   0&00104 & 0&00003 & 0&00010  \\
 -0.080 & -0.040 &   0&00069 & 0&00003 & 0&00010  \\
 -0.040 &  0.000 &   0&00013 & 0&00002 & 0&00005  \\ \hline
\end{tabular}
\end{center}
\caption[AEEC]{\label{AEEC}
{Asymmetry of the Energy Energy Correlation $AEEC$} }
\end{table}

\onecolumn
\clearpage
\normalsize

\subsection{ Tables of Event Shape Distributions }
\scriptsize

\begin{table}[h]

\begin{center}
\begin{tabular}
 {r@{ - }rr@{.}l@{ $\pm$ }r@{.}l@{ $\pm$ }r@{.}lr@{.}l@{ $\pm$ }r@{.}l@{
$\pm$ }r@{.}lr@{.}l@{$\pm$ }r@{.}l@{ $\pm$ }r@{.}l }\hline
\multicolumn{10}{l}{ Sphericity} & \multicolumn{10}{l}{  $\frac{1}{N} \;
\frac{dN}{dS}$} \\ \hline
\multicolumn{2}{c}{  } &
  \multicolumn{6}{c}{charged particles} &
  \multicolumn{6}{c}{charged particles} &
  \multicolumn{6}{c}{charged plus neutral particles} \\
\multicolumn{2}{c}{Interval} &
  \multicolumn{6}{c}{corrected to } &
  \multicolumn{6}{c}{corrected to } &
  \multicolumn{6}{c}{corrected to } \\
\multicolumn{2}{c}{  } &
  \multicolumn{6}{c}{charged particles} &
  \multicolumn{6}{c}{charged plus neutral particles} &
  \multicolumn{6}{c}{charged plus neutral particles} \\ \hline
 0.000 & 0.010 & 17&598 & 0&068 & 0&176  & 17&956 & 0&072 & 0&458  & 16&198 &
0&067 & 0&208  \\
 0.010 & 0.020 & 18&570 & 0&068 & 0&186  & 19&635 & 0&071 & 0&267  & 20&008 &
0&072 & 0&246  \\
 0.020 & 0.030 & 12&129 & 0&053 & 0&121  & 12&249 & 0&054 & 0&122  & 12&896 &
0&056 & 0&153  \\
 0.030 & 0.040 &  8&225 & 0&044 & 0&082  & 7&985 & 0&042 & 0&080  & 8&237 &
0&043 & 0&094  \\
\hline
\end{tabular}
\end{center}
\caption[Sphericity]{\label{SPH} {Sphericity $S$}}
\end{table}

%\clearpage

\begin{table}[h]
\begin{center}
\begin{tabular} {r@{ - }rr@{.}l@{ $\pm$ }r@{.}l@{ $\pm$ }r@{.}lr@{.}l@{ $\pm$
}r@{.}l@{ $\pm$ }r@{.}l
 r@{.}l@{$\pm$ }r@{.}l@{ $\pm$ }r@{.}l }\hline
\multicolumn{10}{l}{ Aplanarity } & \multicolumn{10}{l}{  $\frac{1}{N} \;
\frac{dN}{dA}$} \\ \hline
\multicolumn{2}{c}{  } &
  \multicolumn{6}{c}{charged particles} &
  \multicolumn{6}{c}{charged particles} &
  \multicolumn{6}{c}{charged plus neutral particles} \\
\multicolumn{2}{c}{Interval} &
  \multicolumn{6}{c}{corrected to } &
  \multicolumn{6}{c}{corrected to } &
  \multicolumn{6}{c}{corrected to } \\
\multicolumn{2}{c}{  } &
  \multicolumn{6}{c}{charged particles} &
  \multicolumn{6}{c}{charged plus neutral particles} &
  \multicolumn{6}{c}{charged plus neutral particles} \\ \hline
 0.000 & 0.005 & 81&41 & 0&20 & 0&81  & 81&83 & 0&21 & 1&03  & 75&10 & 0&19 &
0&75  \\
 0.005 & 0.010 & 49&12 & 0&15 & 0&49  & 51&34 & 0&16 & 0&62  & 55&31 & 0&17 &
0&55  \\
 0.010 & 0.015 & 24&05 & 0&11 & 0&24  & 24&15 & 0&10 & 0&33  & 26&03 & 0&11 &
0&28  \\
 0.015 & 0.020 & 13&668 & 0&079 & 0&137  & 13&242 & 0&076 & 0&224  & 13&927 &
0&079 & 0&176  \\
\hline
\end{tabular}
\end{center}
\caption[Aplanarity]{\label{Apl} {Aplanarity $A$}}
\end{table}

\begin{table}[h]
\begin{center}
\begin{tabular}
 {r@{ - }rr@{.}l@{ $\pm$ }r@{.}l@{ $\pm$ }r@{.}lr@{.}l@{ $\pm$ }r@{.}l@{ $\pm$
}r@{.}l
r@{.}l@{$\pm$ }r@{.}l@{ $\pm$ }r@{.}l }\hline
\multicolumn{10}{l}{ Planarity} & \multicolumn{10}{l}{  $\frac{1}{N} \;
\frac{dN}{dP}$} \\ \hline
\multicolumn{2}{c}{  } &
  \multicolumn{6}{c}{charged particles} &
  \multicolumn{6}{c}{charged particles} &
  \multicolumn{6}{c}{charged plus neutral particles} \\
\multicolumn{2}{c}{Interval} &
  \multicolumn{6}{c}{corrected to } &
  \multicolumn{6}{c}{corrected to } &
  \multicolumn{6}{c}{corrected to } \\
\multicolumn{2}{c}{  } &
  \multicolumn{6}{c}{charged particles} &
  \multicolumn{6}{c}{charged plus neutral particles} &
  \multicolumn{6}{c}{charged plus neutral particles} \\ \hline
 0.000 & 0.005 & 64&81 & 0&18 & 0&65  & 68&07 & 0&19 & 0&82  & 68&69 & 0&19 &
0&74  \\
 0.005 & 0.010 & 32&31 & 0&12 & 0&32  & 31&44 & 0&12 & 0&39  & 31&66 & 0&12 &
0&35  \\
 0.010 & 0.015 & 18&177 & 0&092 & 0&182  & 17&196 & 0&088 & 0&215  & 17&091 &
0&088 & 0&188  \\
 0.015 & 0.020 & 12&001 & 0&075 & 0&120  & 11&362 & 0&072 & 0&145  & 11&370 &
0&072 & 0&127  \\
\hline
\end{tabular}
\end{center}
\caption[Planarity ]{\label{Pla} {Planarity $P$}}
\end{table}

%\clearpage

\begin{table}[h]
\begin{center}
\begin{tabular}
 {r@{ - }rr@{.}l@{ $\pm$ }r@{.}l@{ $\pm$ }r@{.}lr@{.}l@{ $\pm$ }r@{.}l@{ $\pm$
}r@{.}l
r@{.}l@{$\pm$ }r@{.}l@{ $\pm$ }r@{.}l }\hline
\multicolumn{10}{l}{ 1-Thrust} & \multicolumn{10}{l}{  $\frac{1}{N} \;
\frac{dN}{d(1-T)}$} \\ \hline
\multicolumn{2}{c}{  } &
  \multicolumn{6}{c}{charged particles} &
  \multicolumn{6}{c}{charged particles} &
  \multicolumn{6}{c}{charged plus neutral particles} \\
\multicolumn{2}{c}{Interval} &
  \multicolumn{6}{c}{corrected to } &
  \multicolumn{6}{c}{corrected to } &
  \multicolumn{6}{c}{corrected to } \\
\multicolumn{2}{c}{  } &
  \multicolumn{6}{c}{charged particles} &
  \multicolumn{6}{c}{charged plus neutral particles} &
  \multicolumn{6}{c}{charged plus neutral particles} \\ \hline
 0.000 & 0.010 &  3&386 & 0&028 & 0&132  & 1&366 & 0&025 & 0&125  & 1&030 &
0&019 & 0&076  \\
 0.010 & 0.020 & 13&901 & 0&057 & 0&249  & 10&185 & 0&048 & 0&446  & 10&951 &
0&051 & 0&527  \\
 0.020 & 0.030 & 15&733 & 0&061 & 0&157  & 16&184 & 0&062 & 0&288  & 17&645 &
0&066 & 0&547  \\
 0.030 & 0.040 & 12&535 & 0&055 & 0&125  & 13&572 & 0&059 & 0&173  & 14&192 &
0&061 & 0&292  \\
\hline
\end{tabular}
\end{center}
\caption[1-Thrust]{\label{Thr} {1-Thrust $(1-T)$}}
\end{table}

\begin{table}[h]
\begin{center}
\begin{tabular}
 {r@{ - }rr@{.}l@{ $\pm$ }r@{.}l@{ $\pm$ }r@{.}lr@{.}l@{ $\pm$ }r@{.}l@{ $\pm$
}r@{.}l
r@{.}l@{$\pm$ }r@{.}l@{ $\pm$ }r@{.}l }\hline
\multicolumn{10}{l}{ Major}  & \multicolumn{10}{l}{ $\frac{1}{N} \;
\frac{dN}{dM}$} \\ \hline
\multicolumn{2}{c}{  } &
  \multicolumn{6}{c}{charged particles} &
  \multicolumn{6}{c}{charged particles} &
  \multicolumn{6}{c}{charged plus neutral particles} \\
\multicolumn{2}{c}{Interval} &
  \multicolumn{6}{c}{corrected to } &
  \multicolumn{6}{c}{corrected to } &
  \multicolumn{6}{c}{corrected to } \\
\multicolumn{2}{c}{  } &
  \multicolumn{6}{c}{charged particles} &
  \multicolumn{6}{c}{charged plus neutral particles} &
  \multicolumn{6}{c}{charged plus neutral particles} \\ \hline
 0.000 & 0.020 &   0&00207 & 0&00087 & 0&00313  & 0&0022 & 0&0023 & 0&0016  &
0&00040 & 0&00090 & 0&00005  \\
 0.020 & 0.040 &   0&2165 & 0&0053 & 0&0114  & 0&1274 & 0&0058 & 0&0094  &
0&0590 & 0&0030 & 0&0058  \\
 0.040 & 0.050 &   1&300 & 0&018 & 0&013  & 0&895 & 0&017 & 0&024  & 0&642 &
0&013 & 0&028  \\
 0.050 & 0.060 &   2&734 & 0&027 & 0&027  & 2&337 & 0&026 & 0&059  & 2&178 &
0&024 & 0&086  \\
\hline
\end{tabular}
\end{center}
\caption[Major]{\label{Maj} {Major $M$}}
\end{table}
%\clearpage

\begin{table}[h]
\begin{center}
\begin{tabular}
 {r@{ - }rr@{.}l@{ $\pm$ }r@{.}l@{ $\pm$ }r@{.}lr@{.}l@{ $\pm$ }r@{.}l@{ $\pm$
}r@{.}l
r@{.}l@{$\pm$ }r@{.}l@{ $\pm$ }r@{.}l }\hline
\multicolumn{10}{l}{ Minor }& \multicolumn{10}{l}{$\frac{1}{N} \;
\frac{dN}{dm}$} \\ \hline
\multicolumn{2}{c}{  } &
  \multicolumn{6}{c}{charged particles} &
  \multicolumn{6}{c}{charged particles} &
  \multicolumn{6}{c}{charged plus neutral particles} \\
\multicolumn{2}{c}{Interval} &
  \multicolumn{6}{c}{corrected to } &
  \multicolumn{6}{c}{corrected to } &
  \multicolumn{6}{c}{corrected to } \\
\multicolumn{2}{c}{  } &
  \multicolumn{6}{c}{charged particles} &
  \multicolumn{6}{c}{charged plus neutral particles} &
  \multicolumn{6}{c}{charged plus neutral particles} \\ \hline
 0.000 & 0.020 &  0&1449 & 0&0043 & 0&0107  & 0&0689 & 0&0067 & 0&0117  &
0&0156 & 0&0017 & 0&0036  \\
 0.020 & 0.040 &  3&109 & 0&019 & 0&031  & 1&968 & 0&019 & 0&100  & 1&236 &
0&013 & 0&066  \\
 0.040 & 0.050 &  7&968 & 0&044 & 0&080  & 6&346 & 0&041 & 0&071  & 5&706 &
0&037 & 0&073  \\
 0.050 & 0.060 & 10&353 & 0&050 & 0&104  & 9&491 & 0&048 & 0&116  & 9&714 &
0&048 & 0&125  \\
\hline
\end{tabular}
\end{center}
\caption[Minor]{\label{Min} {Minor $m$}}
\end{table}

\begin{table}[h]
\begin{center}
\begin{tabular}
 {r@{ - }rr@{.}l@{ $\pm$ }r@{.}l@{ $\pm$ }r@{.}lr@{.}l@{ $\pm$ }r@{.}l@{ $\pm$
}r@{.}l
r@{.}l@{$\pm$ }r@{.}l@{ $\pm$ }r@{.}l }\hline
\multicolumn{10}{l}{ Oblateness }& \multicolumn{10}{l}{ $\frac{1}{N} \;
\frac{dN}{dO}$} \\ \hline
\multicolumn{2}{c}{  } &
  \multicolumn{6}{c}{charged particles} &
  \multicolumn{6}{c}{charged particles} &
  \multicolumn{6}{c}{charged plus neutral particles} \\
\multicolumn{2}{c}{Interval} &
  \multicolumn{6}{c}{corrected to } &
  \multicolumn{6}{c}{corrected to } &
  \multicolumn{6}{c}{corrected to } \\
\multicolumn{2}{c}{  } &
  \multicolumn{6}{c}{charged particles} &
  \multicolumn{6}{c}{charged plus neutral particles} &
  \multicolumn{6}{c}{charged plus neutral particles} \\ \hline
 0.000 & 0.020 &  7&415 & 0&031 & 0&106  & 8&103 & 0&033 & 0&193  & 9&357 &
0&036 & 0&178  \\
 0.020 & 0.040 & 10&840 & 0&037 & 0&108  & 10&809 & 0&036 & 0&204  & 11&508 &
0&038 & 0&140  \\
 0.040 & 0.060 &  7&800 & 0&030 & 0&078  & 7&294 & 0&029 & 0&109  & 7&215 &
0&029 & 0&072  \\
 0.060 & 0.080 &  5&273 & 0&025 & 0&053  & 5&030 & 0&024 & 0&060  & 4&736 &
0&023 & 0&047  \\
\hline
\end{tabular}
\end{center}
\caption[Oblateness]{\label{Obl} {Oblateness $O$}}
\end{table}

%\newpage

\begin{table}[h]
\begin{center}
\begin{tabular}
 {r@{ - }rr@{.}l@{ $\pm$ }r@{.}l@{ $\pm$ }r@{.}lr@{.}l@{ $\pm$ }r@{.}l@{ $\pm$
}r@{.}l
r@{.}l@{$\pm$ }r@{.}l@{ $\pm$ }r@{.}l }\hline
\multicolumn{10}{l}{Heavy Jet Mass }&
\multicolumn{10}{l}{  $\frac{1}{N} \;
\frac{dN}{dM^{2}_{high}/E_{vis}^{2}}$} \\ \hline
\multicolumn{2}{c}{  } &
  \multicolumn{6}{c}{charged particles} &
  \multicolumn{6}{c}{charged particles} &
  \multicolumn{6}{c}{charged plus neutral particles} \\
\multicolumn{2}{c}{Interval} &
  \multicolumn{6}{c}{corrected to } &
  \multicolumn{6}{c}{corrected to } &
  \multicolumn{6}{c}{corrected to } \\
\multicolumn{2}{c}{  } &
  \multicolumn{6}{c}{charged particles} &
  \multicolumn{6}{c}{charged plus neutral particles} &
  \multicolumn{6}{c}{charged plus neutral particles} \\ \hline
 0.000 & 0.010 &  4&450 & 0&032 & 0&260  & 2&035 & 0&028 & 0&180  & 1&994 &
0&027 & 0&166  \\
 0.010 & 0.020 & 17&310 & 0&062 & 0&173  & 16&492 & 0&059 & 0&165  & 18&580 &
0&065 & 0&709  \\
 0.020 & 0.030 & 17&290 & 0&066 & 0&267  & 20&056 & 0&074 & 0&611  & 20&678 &
0&076 & 0&729  \\
 0.030 & 0.040 & 12&996 & 0&058 & 0&206  & 13&772 & 0&061 & 0&377  & 13&377 &
0&060 & 0&412  \\
\hline
\end{tabular}
\end{center}
\caption[Heavy Jet Mass ]{\label{Mh} {Heavy Jet Mass $M^{2}_{high}/E_{vis}^{2}
$}}
\end{table}

\begin{table}[h]
\begin{center}
\begin{tabular}
 {r@{ - }rr@{.}l@{ $\pm$ }r@{.}l@{ $\pm$ }r@{.}lr@{.}l@{ $\pm$ }r@{.}l@{ $\pm$
}r@{.}lr@{.}l@{$\pm$ }r@{.}l@{ $\pm$ }r@{.}l }\hline
\multicolumn{10}{l}{Light Jet Mass }& \multicolumn{10}{l}{ $\frac{1}{N}
\;\frac{dN}{dM^{2}_{low}/E_{vis}^{2} }$} \\ \hline
\multicolumn{2}{c}{  } &
  \multicolumn{6}{c}{charged particles} &
  \multicolumn{6}{c}{charged particles} &
  \multicolumn{6}{c}{charged plus neutral particles} \\
\multicolumn{2}{c}{Interval} &
  \multicolumn{6}{c}{corrected to } &
  \multicolumn{6}{c}{corrected to } &
  \multicolumn{6}{c}{corrected to } \\
\multicolumn{2}{c}{  } &
  \multicolumn{6}{c}{charged particles} &
  \multicolumn{6}{c}{charged plus neutral particles} &
  \multicolumn{6}{c}{charged plus neutral particles} \\ \hline
 0.000 & 0.010 & 39&362 & 0&093 & 1&220  & 24&708 & 0&077 & 1&403  & 23&414 &
0&074 & 1&595 \\
 0.010 & 0.020 & 31&177 & 0&092 & 1&074  & 35&91  & 0&10 & 2&35  & 39&12 & 0&11
& 2&65  \\
 0.020 & 0.030 & 13&733 & 0&065 & 0&523  & 17&370 & 0&079 & 1&196  & 18&080 &
0&081 & 1&215  \\
 0.030 & 0.040 &  6&486 & 0&044 & 0&270  & 7&628 & 0&051 & 0&516  & 7&704 &
0&052 & 0&514  \\
 0.040 & 0.050 &  3&358 & 0&032 & 0&147  & 3&829 & 0&036 & 0&251  & 3&922 &
0&036 & 0&260  \\
 0.050 & 0.060 &  1&879 & 0&024 & 0&082  & 2&085 & 0&026 & 0&132  & 2&128 &
0&026 & 0&140  \\
 0.060 & 0.080 &  0&901 & 0&012 & 0&040  & 0&989 & 0&013 & 0&063  & 1&013 &
0&013 & 0&066  \\
 0.080 & 0.100 &  0&3212 & 0&0072 & 0&0175  & 0&3326 & 0&0073 & 0&0234  &
0&3748 & 0&0079 & 0&0241  \\
 0.100 & 0.120 &  0&1087 & 0&0044 & 0&0082  & 0&1135 & 0&0044 & 0&0096  &
0&1412 & 0&0050 & 0&0089  \\ \hline
\end{tabular}
\end{center}
\caption[Light Jet Mass ]{\label{Ml} {Light Jet Mass  $M^{2}_{low}/E_{vis}^{2}
$}}
\end{table}

\begin{table}[h]
\begin{center}
\begin{tabular}
 {r@{ - }rr@{.}l@{ $\pm$ }r@{.}l@{ $\pm$ }r@{.}lr@{.}l@{ $\pm$ }r@{.}l@{ $\pm$
}r@{.}lr@{.}l@{$\pm$ }r@{.}l@{ $\pm$ }r@{.}l }\hline
\multicolumn{10}{l}{Difference of the Jet Masses }&
\multicolumn{10}{l}{$\frac{1}{N}
\;\frac{dN}{dM^{2}_{diff}/E_{vis}^{2}}$} \\ \hline
\multicolumn{2}{c}{  } &
  \multicolumn{6}{c}{charged particles} &
  \multicolumn{6}{c}{charged particles} &
  \multicolumn{6}{c}{charged plus neutral particles} \\
\multicolumn{2}{c}{Interval} &
  \multicolumn{6}{c}{corrected to } &
  \multicolumn{6}{c}{corrected to } &
  \multicolumn{6}{c}{corrected to } \\
\multicolumn{2}{c}{  } &
  \multicolumn{6}{c}{charged particles} &
  \multicolumn{6}{c}{charged plus neutral particles} &
  \multicolumn{6}{c}{charged plus neutral particles} \\ \hline
 0.000 & 0.010 & 29&229 & 0&082 & 0&292  & 31&955 & 0&086 & 0&332  & 35&393 &
0&092 & 0&354  \\
 0.010 & 0.020 & 19&648 & 0&068 & 0&196  & 20&327 & 0&071 & 0&203  & 20&745 &
0&071 & 0&207  \\
 0.020 & 0.030 & 12&462 & 0&055 & 0&125  & 11&909 & 0&054 & 0&119  & 11&426 &
0&052 & 0&114  \\
 0.030 & 0.040 &  8&332 & 0&045 & 0&083  & 7&654 & 0&043 & 0&077  & 7&170 &
0&041 & 0&072  \\
\hline
\end{tabular}
\end{center}
\caption[Difference of the Jet Masses ]
{\label{Md} {Difference of the Jet Masses $M^{2}_{diff}/E_{vis}^{2} $}}
\end{table}

\clearpage

\begin{table}[h]
\begin{center}
\begin{tabular}
 {r@{ - }rr@{.}l@{ $\pm$ }r@{.}l@{ $\pm$ }r@{.}lr@{.}l@{ $\pm$ }r@{.}l@{ $\pm$
}r@{.}lr@{.}l@{$\pm$ }r@{.}l@{ $\pm$ }r@{.}l }\hline
\multicolumn{10}{l}{ Diff. Jet Rate $D_2$ Durham  Algorithm  }&
\multicolumn{10}{l}{$\frac{1}{N} \; \frac{dN}{dD_2^D}$} \\ \hline
\multicolumn{2}{c}{  } &
  \multicolumn{6}{c}{charged particles} &
  \multicolumn{6}{c}{charged particles} &
  \multicolumn{6}{c}{charged plus neutral particles} \\
\multicolumn{2}{c}{Interval} &
  \multicolumn{6}{c}{corrected to } &
  \multicolumn{6}{c}{corrected to } &
  \multicolumn{6}{c}{corrected to } \\
\multicolumn{2}{c}{  } &
  \multicolumn{6}{c}{charged particles} &
  \multicolumn{6}{c}{charged plus neutral particles} &
  \multicolumn{6}{c}{charged plus neutral particles} \\ \hline
 0.000 & 0.010 &  63&52 & 0&12 & 0&64  & 62&64 & 0&12 & 0&63  & 63&81 & 0&12 &
0&64  \\
 0.010 & 0.020 &  12&230 & 0&054 & 0&122  & 12&326 & 0&054 & 0&126  & 11&589 &
0&052 & 0&116  \\
 0.020 & 0.030 &   6&128 & 0&038 & 0&061  & 6&324 & 0&039 & 0&080  & 6&040 &
0&038 & 0&060  \\
 0.030 & 0.040 &   3&828 & 0&030 & 0&038  & 3&993 & 0&031 & 0&058  & 3&823 &
0&030 & 0&038  \\
\hline
\end{tabular}
\end{center}
\caption[Differential Jet Rate $D_2$ Durham  Algorithm  ]{\label{D2D}
{Differential Jet Rate $D_2$ Durham  Algorithm  $D_2^D$ }}
\end{table}

\begin{table}[h]
\begin{center}
\begin{tabular}
 {r@{ - }rr@{.}l@{ $\pm$ }r@{.}l@{ $\pm$ }r@{.}lr@{.}l@{ $\pm$ }r@{.}l@{ $\pm$
}r@{.}l
r@{.}l@{$\pm$ }r@{.}l@{ $\pm$ }r@{.}l }\hline
\multicolumn{10}{l}{ Diff. Jet Rate $D_3$ Durham  Algorithm   }&
\multicolumn{10}{l}{ $\frac{1}{N} \; \frac{dN}{dD_3^D}$} \\ \hline
\multicolumn{2}{c}{  } &
  \multicolumn{6}{c}{charged particles} &
  \multicolumn{6}{c}{charged particles} &
  \multicolumn{6}{c}{charged plus neutral particles} \\
\multicolumn{2}{c}{Interval} &
  \multicolumn{6}{c}{corrected to } &
  \multicolumn{6}{c}{corrected to } &
  \multicolumn{6}{c}{corrected to } \\
\multicolumn{2}{c}{  } &
  \multicolumn{6}{c}{charged particles} &
  \multicolumn{6}{c}{charged plus neutral particles} &
  \multicolumn{6}{c}{charged plus neutral particles} \\ \hline
 0.000 & 0.002 & 318&65 & 0&62 & 3&19  & 318&41 & 0&62 & 3&18  & 333&97 & 0&64
& 3&34  \\
 0.002 & 0.004 &  91&82 & 0&33 & 0&92  & 90&83 & 0&32 & 1&07  & 85&51 & 0&31 &
0&86  \\
 0.004 & 0.006 &  33&87 & 0&20 & 0&34  & 33&23 & 0&20 & 0&49  & 29&49 & 0&18 &
0&29  \\
 0.006 & 0.008 &  16&86 & 0&14 & 0&17  & 16&99 & 0&14 & 0&31  & 14&78 & 0&13 &
0&15  \\
\hline
\end{tabular}
\end{center}
\caption[Differential Jet Rate $D_3$ Durham  Algorithm  ]{\label{D3D}
{Differential Jet Rate $D_3$ Durham  Algorithm  $D_3^D$}}
\end{table}

\begin{table}[h]
\begin{center}
\begin{tabular}
 {r@{ - }rr@{.}l@{ $\pm$ }r@{.}l@{ $\pm$ }r@{.}lr@{.}l@{ $\pm$ }r@{.}l@{ $\pm$
}r@{.}l
r@{.}l@{$\pm$ }r@{.}l@{ $\pm$ }r@{.}l }\hline
\multicolumn{10}{l}{ Diff. Jet Rate $D_4$ Durham  Algorithm  }&
\multicolumn{10}{l}{$\frac{1}{N} \; \frac{dN}{dD_4^D}$} \\ \hline
\multicolumn{2}{c}{  } &
  \multicolumn{6}{c}{charged particles} &
  \multicolumn{6}{c}{charged particles} &
  \multicolumn{6}{c}{charged plus neutral particles} \\
\multicolumn{2}{c}{Interval} &
  \multicolumn{6}{c}{corrected to } &
  \multicolumn{6}{c}{corrected to } &
  \multicolumn{6}{c}{corrected to } \\
\multicolumn{2}{c}{  } &
  \multicolumn{6}{c}{charged particles} &
  \multicolumn{6}{c}{charged plus neutral particles} &
  \multicolumn{6}{c}{charged plus neutral particles} \\ \hline
 0.000 & 0.001 & 688&3 & 1&3 & 6&9  & 684&2 & 1&3 & 6&8  & 711&0 & 1&3 & 7&1
\\
 0.001 & 0.002 & 190&40 & 0&67 & 1&90  & 193&30 & 0&67 & 1&93  & 184&32 & 0&66
& 1&84  \\
 0.002 & 0.003 &  60&98 & 0&37 & 0&61  & 61&50 & 0&38 & 2&31  & 53&35 & 0&35 &
1&06  \\
 0.003 & 0.004 &  25&69 & 0&24 & 0&26  & 25&32 & 0&25 & 0&98  & 21&18 & 0&22 &
0&53  \\
\hline
\end{tabular}
\end{center}
\caption[Differential Jet Rate $D_4$ Durham  Algorithm  ]{\label{D4D}
{Differential Jet Rate $D_4$ Durham  Algorithm  $D_4^D$}}
\end{table}

\begin{table}[h]
\begin{center}
\begin{tabular}
 {r@{ - }rr@{.}l@{ $\pm$ }r@{.}l@{ $\pm$ }r@{.}lr@{.}l@{ $\pm$ }r@{.}l@{ $\pm$
}r@{.}l
r@{.}l@{$\pm$ }r@{.}l@{ $\pm$ }r@{.}l }\hline
\multicolumn{10}{l}{ Diff. Jet Rate $D_2$ Jade  Algorithm  }&
\multicolumn{10}{l}{ $\frac{1}{N} \; \frac{dN}{dD^J_2}$} \\ \hline
\multicolumn{2}{c}{  } &
  \multicolumn{6}{c}{charged particles} &
  \multicolumn{6}{c}{charged particles} &
  \multicolumn{6}{c}{charged plus neutral particles} \\
\multicolumn{2}{c}{Interval} &
  \multicolumn{6}{c}{corrected to } &
  \multicolumn{6}{c}{corrected to } &
  \multicolumn{6}{c}{corrected to } \\
\multicolumn{2}{c}{  } &
  \multicolumn{6}{c}{charged particles} &
  \multicolumn{6}{c}{charged plus neutral particles} &
  \multicolumn{6}{c}{charged plus neutral particles} \\ \hline
 0.000 & 0.010 & 26&014 & 0&077 & 0&260  & 24&267 & 0&073 & 0&442  & 26&317 &
0&078 & 0&289  \\
 0.010 & 0.020 & 18&243 & 0&065 & 0&182  & 18&064 & 0&066 & 0&269  & 17&670 &
0&064 & 0&197  \\
 0.020 & 0.030 & 11&945 & 0&054 & 0&119  & 12&277 & 0&056 & 0&157  & 11&622 &
0&054 & 0&134  \\
 0.030 & 0.040 &  8&378 & 0&046 & 0&084  & 8&677 & 0&047 & 0&103  & 8&177 &
0&045 & 0&099  \\
\hline
\end{tabular}
\end{center}
\caption[Differential Jet Rate $D_2$ Jade  Algorithm  ]{\label{D2J}
{Differential Jet Rate $D_2$ Jade  Algorithm  $D^J_2$}}
\end{table}

\begin{table}[h]
\begin{center}
\begin{tabular}
 {r@{ - }rr@{.}l@{ $\pm$ }r@{.}l@{ $\pm$ }r@{.}lr@{.}l@{ $\pm$ }r@{.}l@{ $\pm$
}r@{.}l
r@{.}l@{$\pm$ }r@{.}l@{ $\pm$ }r@{.}l }\hline
\multicolumn{10}{l}{ Diff. Jet Rate $D_3$ Jade  Algorithm  }&
\multicolumn{10}{l}{$\frac{1}{N} \;\frac{dN}{dD^J_3}$} \\ \hline
\multicolumn{2}{c}{  } &
  \multicolumn{6}{c}{charged particles} &
  \multicolumn{6}{c}{charged particles} &
  \multicolumn{6}{c}{charged plus neutral particles} \\
\multicolumn{2}{c}{Interval} &
  \multicolumn{6}{c}{corrected to } &
  \multicolumn{6}{c}{corrected to } &
  \multicolumn{6}{c}{corrected to } \\
\multicolumn{2}{c}{  } &
  \multicolumn{6}{c}{charged particles} &
  \multicolumn{6}{c}{charged plus neutral particles} &
  \multicolumn{6}{c}{charged plus neutral particles} \\ \hline
 0.000 & 0.002 & 107&87 & 0&36 & 1&08  & 103&55 & 0&35 & 1&04  & 113&89 & 0&37
& 1&14  \\
 0.002 & 0.004 & 121&68 & 0&38 & 1&22  & 122&74 & 0&37 & 1&41  & 130&22 & 0&39
& 1&30  \\
 0.004 & 0.006 &  74&79 & 0&29 & 0&75  & 73&19 & 0&29 & 0&95  & 71&16 & 0&28 &
0&80  \\
 0.006 & 0.008 &  46&84 & 0&23 & 0&47  & 47&04 & 0&24 & 0&69  & 43&53 & 0&22 &
0&51  \\
\hline
\end{tabular}
\end{center}
\caption[Differential Jet Rate $D_3$ Jade  Algorithm  ]{\label{D3J}
{Differential Jet Rate $D_3$ Jade  Algorithm  $D^J_3$}}
\end{table}

\begin{table}[h]
\begin{center}
\begin{tabular}
 {r@{ - }rr@{.}l@{ $\pm$ }r@{.}l@{ $\pm$ }r@{.}lr@{.}l@{ $\pm$ }r@{.}l@{ $\pm$
}r@{.}l
r@{.}l@{$\pm$ }r@{.}l@{ $\pm$ }r@{.}l }\hline
\multicolumn{10}{l}{ Diff. Jet Rate $D_4$ Jade  Algorithm  }&
\multicolumn{10}{l}{ $\frac{1}{N} \; \frac{dN}{dD^J_4}$} \\ \hline
\multicolumn{2}{c}{  } &
  \multicolumn{6}{c}{charged particles} &
  \multicolumn{6}{c}{charged particles} &
  \multicolumn{6}{c}{charged plus neutral particles} \\
\multicolumn{2}{c}{Interval} &
  \multicolumn{6}{c}{corrected to } &
  \multicolumn{6}{c}{corrected to } &
  \multicolumn{6}{c}{corrected to } \\
\multicolumn{2}{c}{  } &
  \multicolumn{6}{c}{charged particles} &
  \multicolumn{6}{c}{charged plus neutral particles} &
  \multicolumn{6}{c}{charged plus neutral particles} \\ \hline
 0.000 & 0.001 & 248&15 & 0&77 & 2&48  & 227&18 & 0&75 & 3&97  & 240&63 & 0&78
& 4&58  \\
 0.001 & 0.002 & 286&76 & 0&82 & 2&87  & 289&93 & 0&82 & 5&54  & 309&26 & 0&86
& 4&96  \\
 0.002 & 0.003 & 167&03 & 0&63 & 1&67  & 173&10 & 0&64 & 3&59  & 171&69 & 0&64
& 2&39  \\
 0.003 & 0.004 &  97&00 & 0&48 & 1&04  & 99&50 & 0&49 & 2&22  & 92&65 & 0&47 &
1&18  \\
\hline
\end{tabular}
\end{center}
\caption[Differential Jet Rate $D_4$ Jade  Algorithm  ]{\label{D4J}
{Differential Jet Rate $D_4$ Jade  Algorithm  $D^J_4$ }}
\end{table}

%\clearpage

\begin{table}[h]
\begin{center}
\begin{tabular}
 {r@{ - }rr@{.}l@{ $\pm$ }r@{.}l@{ $\pm$ }r@{.}lr@{.}l@{ $\pm$ }r@{.}l@{ $\pm$
}r@{.}l
r@{.}l@{$\pm$ }r@{.}l@{ $\pm$ }r@{.}l }\hline
\multicolumn{10}{l}{Wide Jet Broadening }& \multicolumn{10}{l}{ $\frac{1}{N} \;
\frac{dN}{dB_{W}}$} \\ \hline
\multicolumn{2}{c}{  } &
  \multicolumn{6}{c}{charged particles} &
  \multicolumn{6}{c}{charged particles} &
  \multicolumn{6}{c}{charged plus neutral particles} \\
\multicolumn{2}{c}{Interval} &
  \multicolumn{6}{c}{corrected to } &
  \multicolumn{6}{c}{corrected to } &
  \multicolumn{6}{c}{corrected to } \\
\multicolumn{2}{c}{  } &
  \multicolumn{6}{c}{charged particles} &
  \multicolumn{6}{c}{charged plus neutral particles} &
  \multicolumn{6}{c}{charged plus neutral particles} \\ \hline
 0.010 & 0.020 &  2&027 & 0&023 & 0&020  & 1&396 & 0&023 & 0&014  & 0&6707 &
0&0096 & 0&1077  \\
 0.020 & 0.030 &  9&923 & 0&051 & 0&232  & 9&815 & 0&051 & 0&747  & 7&538 &
0&038 & 0&809  \\
 0.030 & 0.040 & 14&758 & 0&061 & 0&245  & 15&520 & 0&062 & 0&819  & 14&690 &
0&058 & 0&745  \\
 0.040 & 0.050 & 13&374 & 0&056 & 0&140  & 12&966 & 0&054 & 0&413  & 13&942 &
0&057 & 0&592  \\
\hline
\end{tabular}
\end{center}
\caption[Wide Jet Broadening ]{\label{bmax} {Wide Jet Broadening $B_{W}$ }}
\end{table}

\begin{table}[h]
\begin{center}
\begin{tabular}
 {r@{ - }rr@{.}l@{ $\pm$ }r@{.}l@{ $\pm$ }r@{.}lr@{.}l@{ $\pm$ }r@{.}l@{ $\pm$
}r@{.}l
r@{.}l@{$\pm$ }r@{.}l@{ $\pm$ }r@{.}l }\hline
\multicolumn{10}{l}{Small Jet Broadening}& \multicolumn{10}{l}{ $\frac{1}{N} \;
\frac{dN}{dB_{min}}$} \\ \hline
\multicolumn{2}{c}{  } &
  \multicolumn{6}{c}{charged particles} &
  \multicolumn{6}{c}{charged particles} &
  \multicolumn{6}{c}{charged plus neutral particles} \\
\multicolumn{2}{c}{Interval} &
  \multicolumn{6}{c}{corrected to } &
  \multicolumn{6}{c}{corrected to } &
  \multicolumn{6}{c}{corrected to } \\
\multicolumn{2}{c}{  } &
  \multicolumn{6}{c}{charged particles} &
  \multicolumn{6}{c}{charged plus neutral particles} &
  \multicolumn{6}{c}{charged plus neutral particles} \\ \hline
 0.000 & 0.010 &  2&502 & 0&025 & 0&025  & 1&552 & 0&026 & 0&115  & 0&645 &
0&010 & 0&096  \\
 0.010 & 0.020 & 18&823 & 0&068 & 0&188  & 15&484 & 0&064 & 0&243  & 11&169 &
0&045 & 1&006  \\
 0.020 & 0.030 & 28&683 & 0&084 & 0&287  & 29&092 & 0&084 & 0&291  & 28&908 &
0&082 & 1&823  \\
 0.030 & 0.040 & 20&703 & 0&071 & 0&207  & 22&082 & 0&073 & 0&275  & 25&972 &
0&083 & 1&478  \\
 0.040 & 0.050 & 11&654 & 0&052 & 0&117  & 11&600 & 0&052 & 0&139  & 14&119 &
0&061 & 0&860  \\
 0.050 & 0.060 &  6&558 & 0&039 & 0&066  & 6&423 & 0&038 & 0&089  & 7&500 &
0&044 & 0&494  \\
 0.060 & 0.080 &  3&191 & 0&019 & 0&032  & 3&205 & 0&019 & 0&052  & 3&405 &
0&021 & 0&233  \\
 0.080 & 0.100 &  1&295 & 0&012 & 0&026  & 1&376 & 0&013 & 0&034  & 1&320 &
0&013 & 0&089  \\
 0.100 & 0.120 &  0&5561 & 0&0078 & 0&0185  & 0&5970 & 0&0086 & 0&0222  &
0&5448 & 0&0082 & 0&0328  \\
 0.120 & 0.150 &  0&1921 & 0&0037 & 0&0054  & 0&2114 & 0&0042 & 0&0070  &
0&1916 & 0&0040 & 0&0104  \\
 0.150 & 0.180 &  0&0319 & 0&0015 & 0&0012  & 0&0356 & 0&0017 & 0&0016  &
0&0366 & 0&0017 & 0&0034  \\ \hline
\end{tabular}
\end{center}
\caption[Small Jet Broadening ]{\label{bmin} {Small Jet Broadening $B_{min}$ }}
\end{table}
\clearpage
\begin{table}
\begin{center}
\begin{tabular}
 {r@{ - }rr@{.}l@{ $\pm$ }r@{.}l@{ $\pm$ }r@{.}lr@{.}l@{ $\pm$ }r@{.}l@{ $\pm$
}r@{.}l
r@{.}l@{$\pm$ }r@{.}l@{ $\pm$ }r@{.}l }\hline
\multicolumn{10}{l}{Total Jet Broadening }& \multicolumn{10}{l}{ $\frac{1}{N}
\; \frac{dN}{dB_{sum}}$} \\ \hline
\multicolumn{2}{c}{  } &
  \multicolumn{6}{c}{charged particles} &
  \multicolumn{6}{c}{charged particles} &
  \multicolumn{6}{c}{charged plus neutral particles} \\
\multicolumn{2}{c}{Interval} &
  \multicolumn{6}{c}{corrected to } &
  \multicolumn{6}{c}{corrected to } &
  \multicolumn{6}{c}{corrected to } \\
\multicolumn{2}{c}{  } &
  \multicolumn{6}{c}{charged particles} &
  \multicolumn{6}{c}{charged plus neutral particles} &
  \multicolumn{6}{c}{charged plus neutral particles} \\ \hline
 0.020 & 0.030 &  0&956 & 0&016 & 0&010  & 0&550 & 0&017 & 0&030  & 0&2030 &
0&0055 & 0&0383  \\
 0.030 & 0.040 &  3&765 & 0&031 & 0&038  & 2&804 & 0&029 & 0&033  & 1&628 &
0&015 & 0&183  \\
 0.040 & 0.050 &  7&117 & 0&043 & 0&071  & 6&646 & 0&042 & 0&066  & 4&999 &
0&031 & 0&463  \\
 0.050 & 0.060 &  9&301 & 0&048 & 0&093  & 9&342 & 0&048 & 0&093  & 8&190 &
0&041 & 0&644  \\
\hline
\end{tabular}
\end{center}
\caption[Total Jet Broadening ]{\label{bsum} {Total Jet Broadening $B_{sum}$ }}
\end{table}

\begin{table}[h]
\begin{center}
\begin{tabular}
 {r@{ - }rr@{.}l@{ $\pm$ }r@{.}l@{ $\pm$ }r@{.}lr@{.}l@{ $\pm$ }r@{.}l@{ $\pm$
}r@{.}l
r@{.}l@{$\pm$ }r@{.}l@{ $\pm$ }r@{.}l }\hline
\multicolumn{10}{l}{ Difference of the Jet Broadenings }& \multicolumn{10}{l}{
$\frac{1}{N} \; \frac{dN}{dB_{diff}}$} \\ \hline
\multicolumn{2}{c}{  } &
  \multicolumn{6}{c}{charged particles} &
  \multicolumn{6}{c}{charged particles} &
  \multicolumn{6}{c}{charged plus neutral particles} \\
\multicolumn{2}{c}{Interval} &
  \multicolumn{6}{c}{corrected to } &
  \multicolumn{6}{c}{corrected to } &
  \multicolumn{6}{c}{corrected to } \\
\multicolumn{2}{c}{  } &
  \multicolumn{6}{c}{charged particles} &
  \multicolumn{6}{c}{charged plus neutral particles} &
  \multicolumn{6}{c}{charged plus neutral particles} \\ \hline
 0.000 & 0.010 & 24&778 & 0&078 & 0&248  & 26&513 & 0&082 & 0&265  & 26&630 &
0&081 & 0&459  \\
 0.010 & 0.020 & 18&957 & 0&068 & 0&190  & 18&385 & 0&066 & 0&184  & 18&684 &
0&066 & 0&292  \\
 0.020 & 0.030 & 13&050 & 0&056 & 0&131  & 12&151 & 0&053 & 0&122  & 12&343 &
0&054 & 0&186  \\
 0.030 & 0.040 &  9&129 & 0&046 & 0&091  & 8&722 & 0&045 & 0&087  & 8&819 &
0&046 & 0&134  \\
\hline
\end{tabular}
\end{center}
\caption[Difference of the Jet Broadenings ]{\label{bdiff} {Difference of
the Jet Broadenings $B_{diff.}$ }}
\end{table}
%\clearpage

\begin{table}[h]
\begin{center}
\begin{tabular}
 {r@{ - }rr@{.}l@{ $\pm$ }r@{.}l@{ $\pm$ }r@{.}lr@{.}l@{ $\pm$ }r@{.}l@{ $\pm$
}r@{.}l
r@{.}l@{$\pm$ }r@{.}l@{ $\pm$ }r@{.}l }\hline
\multicolumn{10}{l}{ C-Parameter }& \multicolumn{10}{l}{ $\frac{1}{N} \;
\frac{dN}{dC}$} \\ \hline
\multicolumn{2}{c}{  } &
  \multicolumn{6}{c}{charged particles} &
  \multicolumn{6}{c}{charged particles} &
  \multicolumn{6}{c}{charged plus neutral particles} \\
\multicolumn{2}{c}{Interval} &
  \multicolumn{6}{c}{corrected to } &
  \multicolumn{6}{c}{corrected to } &
  \multicolumn{6}{c}{corrected to } \\
\multicolumn{2}{c}{  } &
  \multicolumn{6}{c}{charged particles} &
  \multicolumn{6}{c}{charged plus neutral particles} &
  \multicolumn{6}{c}{charged plus neutral particles} \\ \hline
 0.000 & 0.040 & 0&3806 & 0&0047 & 0&0140  & 0&1375 & 0&0047 & 0&0138  & 0&0881
& 0&0030 & 0&0067  \\
 0.040 & 0.080 & 2&388 & 0&012 & 0&050  & 1&5046 & 0&0098 & 0&0858  & 1&5383 &
0&0100 & 0&0831  \\
 0.080 & 0.120 & 3&653 & 0&015 & 0&042  & 3&514 & 0&014 & 0&108  & 3&909 &
0&016 & 0&142  \\
 0.120 & 0.160 & 3&323 & 0&014 & 0&033  & 3&570 & 0&015 & 0&058  & 3&833 &
0&016 & 0&088  \\
\hline
\end{tabular}
\end{center}
\caption[C-Parameter ]{\label{CP} {C-Parameter $C$}}
\end{table}

\begin{table}[h]
\begin{center}
\begin{tabular}
 {r@{ - }rr@{.}l@{ $\pm$ }r@{.}l@{ $\pm$ }r@{.}lr@{.}l@{ $\pm$ }r@{.}l@{ $\pm$
}r@{.}l
r@{.}l@{$\pm$ }r@{.}l@{ $\pm$ }r@{.}l }\hline
\multicolumn{10}{l}{ D-Parameter }& \multicolumn{10}{l}{$\frac{1}{N} \;
\frac{dN}{dD}$} \\ \hline
\multicolumn{2}{c}{  } &
  \multicolumn{6}{c}{charged particles} &
  \multicolumn{6}{c}{charged particles} &
  \multicolumn{6}{c}{charged plus neutral particles} \\
\multicolumn{2}{c}{Interval} &
  \multicolumn{6}{c}{corrected to } &
  \multicolumn{6}{c}{corrected to } &
  \multicolumn{6}{c}{corrected to } \\
\multicolumn{2}{c}{  } &
  \multicolumn{6}{c}{charged particles} &
  \multicolumn{6}{c}{charged plus neutral particles} &
  \multicolumn{6}{c}{charged plus neutral particles} \\ \hline
 0.000 & 0.008 & 29&731 & 0&093 & 0&308  & 22&567 & 0&083 & 0&645  & 22&228 &
0&082 & 0&868  \\
 0.008 & 0.016 & 19&922 & 0&077 & 0&199  & 20&963 & 0&080 & 0&401  & 22&766 &
0&085 & 0&440  \\
 0.016 & 0.030 & 11&026 & 0&044 & 0&110  & 11&522 & 0&045 & 0&169  & 12&107 &
0&047 & 0&150  \\
 0.030 & 0.044 &  6&513 & 0&034 & 0&065  & 6&763 & 0&035 & 0&093  & 6&879 &
0&035 & 0&079  \\
\hline
\end{tabular}
\end{center}
\caption[D-Parameter ]{\label{DP} {D-Parameter $D$} }
\end{table}
\clearpage

%\clearpage
\normalsize

\section{Tables of Results of the Fits}
\scriptsize
\label{fit_tab}

\footnotesize
\begin{table}[h]
\begin{center}
\begin{tabular}{l r@{.}l r@{.}l@{ - }r@{.}l r@{.}l @{$\pm$} r@{.}l @{ $^+_-$}
r@{.}l} \hline
Parameter & \multicolumn{2}{c}{Default} &
 \multicolumn{4}{c}{Range gen.}   &\multicolumn{6}{c}{Fit Result} \\
  \multicolumn{7}{c}{~} &  \multicolumn{2}{c}{Val.} &
                           \multicolumn{2}{c}{stat.}&
                           \multicolumn{2}{c}{sys.} \\ \hline
QCDLAM  &  0&18   &  0&155  &  0&205  &
                  0&163 & 0&001 &\multicolumn{2}{l}{$^{0.004}_{0.005}$}  \\
RMASS(13)  &  0&75   &  0&64  &  0&96  &
                  0&65 & 0&01 &\multicolumn{2}{l}{$^{0.02}_{0.01}$}  \\
CLMAX   &  3&35   &  2&35  &  4&35  &
                  3&48 & 0&04 &\multicolumn{2}{l}{$^{0.20}_{0.13}$}  \\
CLPOW   &  2&0   &  1&0  &  2&0  &
                  1&49 & 0&04 &\multicolumn{2}{l}{$^{0.16}_{0.08}$}  \\
CLSMR   &  0&   &  0&0  &  1&0  &
                  0&36  & 0&04 &\multicolumn{2}{l}{$^{0.06}_{0.01}$}  \\
DECWT   &  1&0   &  0&0  &  1&0  &
                  0&77  & 0&08 &\multicolumn{2}{l}{$^{0.28}_{0.24}$}  \\
PWT(3)  &  1&0   &  0&0  &  1&0  &
                  0&83 & 0&02 &\multicolumn{2}{l}{$^{0.03}_{0.09}$}  \\
PWT(7)  &  1&0   &  0&0  &  1&0  &
                  0&74 & 0&09 &\multicolumn{2}{l}{$^{0.37}_{0.25}$}  \\ \hline
\end{tabular}
\end{center}
\caption[Parameter setting and fit results for HERWIG 5.8 C]
{\label{th58c} Parameter setting and fit results for HERWIG 5.8 C}
\end{table}

\begin{table}[h]
\begin{center}
\begin{tabular}{cl r@{.}l r@{.}l@{ - }r@{.}l r@{.}l @{$\pm$} r@{.}l @{ $^+_-$}
r@{.}l} \hline
Parameter & Name & \multicolumn{2}{c}{Default} &
 \multicolumn{4}{c}{Range gen.}   &\multicolumn{6}{c}{Fit Result} \\
  \multicolumn{8}{c}{~} &  \multicolumn{2}{c}{Val.} &
                           \multicolumn{2}{c}{stat.}&
                           \multicolumn{2}{c}{sys.} \\ \hline
$\lambda_{QCD}$   &  PARJ(81)       &  0&4   &  0&25  &  0&35  &
                  0&297 & 0&005 &\multicolumn{2}{l}{$^{0.012}_{0.007}$}  \\
$Q_0    $  &  PARJ(82)       &  1&0   &  1&0   &  2&0   &
                  1&34  & 0&08 & \multicolumn{2}{l}{$^{0.15}_{0.09}$} \\ \hline
$a$  &  PARJ(41)       &  0&5   &  0&1 & 0&5  &
                  0&409 &  0&020 &  \multicolumn{2}{l}{$^{0.009}_{0.013}$}\\
$b$  &  PARJ(42)  &  0&9 &\multicolumn{4}{c}{0.844}
                               &\multicolumn{6}{c}{optimized}\\ \hline
$\sigma_q$      &  PARJ(21)  &  0&35  &  0&36  & 0&44  &
                  0&401 & 0&006 & \multicolumn{2}{l}{$^{0.003}_{0.013}$}\\
\hline
$P(^1S_0)_{ud}$ &     -      &  0&5   &  0&3 & 0&5   &
                  0&272 &  0&030 & \multicolumn{2}{l}{$^{0.108}_{0.008}$}\\
\hline
\end{tabular}
\end{center}
\caption[Parameter setting and fit results for JETSET 7.3 PS]
{\label{terg73d} Parameter setting and fit results for JETSET 7.3 PS with
DELPHI decays}
\end{table}

\begin{table}
\begin{center}
\begin{tabular}{cl r@{.}l r@{.}l@{ - }r@{.}l r@{.}l @{$\pm$} r@{.}l @{ $^+_-$}
r@{.}l} \hline
Parameter & Name & \multicolumn{2}{c}{Default} &
 \multicolumn{4}{c}{Range gen.}   &\multicolumn{6}{c}{Fit Result} \\
   \multicolumn{8}{c}{~} &  \multicolumn{2}{c}{Val.} &
                           \multicolumn{2}{c}{stat.}&
                           \multicolumn{2}{c}{sys.} \\ \hline
$\lambda_{QCD}$   &  PARJ(81)       &  0&4   &  0&25  &  0&35  &
                  0&297 & 0&004 &\multicolumn{2}{l}{$^{0.007}_{0.008}$}  \\
$Q_0    $  &  PARJ(82)       &  1&0   &  1&0   &  2&0   &
                  1&56  & 0&11 & \multicolumn{2}{l}{$^{0.21}_{0.15}$} \\ \hline
$a$  &  PARJ(41)       &  0&5   &  0&1 & 0&5  &
                  0&417 &  0&022 &  \multicolumn{2}{l}{$^{0.011}_{0.015}$}\\
$b$  &  PARJ(42)  &  0&9 &\multicolumn{4}{c}{0.850}
                               &\multicolumn{6}{c}{optimized}\\ \hline
$\sigma_q$      &  PARJ(21)  &  0&35  &  0&36  & 0&44  &
                  0&408 & 0&005 & \multicolumn{2}{l}{$^{0.004}_{0.004}$}\\
\hline
$P(^1S_0)_{ud}$ &     -      &  0&5   &  0&3 & 0&5   &
                  0&297 &  0&021 & \multicolumn{2}{l}{$^{0.102}_{0.011}$}\\
\hline
\end{tabular}
\end{center}
\caption[Parameter setting and fit results for JETSET 7.4 PS]
{\label{terg74d} Parameter setting and fit results for JETSET 7.4 PS with
default decays}
\end{table}

\begin{table}
\begin{center}
\begin{tabular}{cl r@{.}l r@{.}l@{ - }r@{.}l r@{.}l @{$\pm$} r@{.}l @{ $^+_-$}
r@{.}l} \hline
Parameter & Name & \multicolumn{2}{c}{Default} &
 \multicolumn{4}{c}{Range gen.}   &\multicolumn{6}{c}{Fit Result} \\
  \multicolumn{8}{c}{~} &  \multicolumn{2}{c}{Val.} &
                           \multicolumn{2}{c}{stat.}&
                           \multicolumn{2}{c}{sys.} \\ \hline
$\lambda_{QCD}$   &  PARA(1)       &  0&22   &  0&20  &  0&30  &
                  0&237 & 0&003 &\multicolumn{2}{l}{$^{0.006}_{0.002}$}  \\
$p_t^{QCD}$&  PARA(3)       &  0&6   &  0&4   &  1&2   &
                  0&64  & 0&08 & \multicolumn{2}{l}{$^{0.20}_{0.08}$} \\ \hline
$a$  &  PARJ(41)       &  0&5   &  0&1 & 0&5  &
                  0&391 &  0&047 &  \multicolumn{2}{l}{$^{0.039}_{0.046}$}\\
$b$  &  PARJ(42)  &  0&9 &\multicolumn{4}{c}{0.850}
                               &\multicolumn{6}{c}{optimized}\\ \hline
$\sigma_q$      &  PARJ(21)  &  0&35  &  0&36  & 0&44  &
                  0&390 & 0&005 & \multicolumn{2}{l}{$^{0.003}_{0.007}$}\\
\hline
$P(^1S_0)_{ud}$ &     -      &  0&5   &  0&3 & 0&5   &
                  0&276 &  0&017 & \multicolumn{2}{l}{$^{0.103}_{0.012}$}\\
\hline
\end{tabular}
\end{center}
\caption[Parameter setting and fit results for ARIADNE 4.06]
{\label{tar46d} Parameter setting and fit results for ARIADNE 4.06 with
DELPHI decays}
\end{table}

\begin{table}
\begin{center}
\begin{tabular}{cl r@{.}l r@{.}l@{ - }r@{.}l r@{.}l @{$\pm$} r@{.}l @{ $^+_-$}
r@{.}l} \hline
Parameter & name & \multicolumn{2}{c}{Default} &
 \multicolumn{4}{c}{gen. range}   &\multicolumn{6}{c}{Results} \\
  \multicolumn{8}{c}{~} &  \multicolumn{2}{c}{Value} &
                           \multicolumn{2}{c}{stat.}&
                           \multicolumn{2}{c}{sys.} \\ \hline
$\lambda_{QCD}$   &  PARA(1)       &  0&22   &  0&20  &  0&30  &
                  0&237 & 0&004 &\multicolumn{2}{l}{$^{0.005}_{0.004}$}  \\
$p_t^{QCD}$&  PARA(3)       &  0&6   &  0&4   &  1&2   &
                  0&68  & 0&06 & \multicolumn{2}{l}{$^{0.11}_{0.11}$} \\ \hline
$a$  &  PARJ(41)       &  0&5   &  0&1 & 0&5  &
                  0&388 &  0&030 &  \multicolumn{2}{l}{$^{0.036}_{0.112}$}\\
$b$  &  PARJ(42)  &  0&9 &\multicolumn{4}{c}{0.850}
                               &\multicolumn{6}{c}{optimized}\\ \hline
$\sigma_q$      &  PARJ(21)  &  0&35  &  0&36  & 0&44  &
                  0&387 & 0&008 & \multicolumn{2}{l}{$^{0.012}_{0.016}$}\\
\hline
$P(^1S_0)_{ud}$ &     -      &  0&5   &  0&3 & 0&5   &
                  0&326 &  0&017 & \multicolumn{2}{l}{$^{0.058}_{0.023}$}\\
\hline
\end{tabular}
\end{center}
\caption[Parameter setting and fit results for ARIADNE 4.06]
{\label{tar46dd} Parameter setting and fit results for ARIADNE 4.06 with
default decays}
\end{table}

\begin{table}[h]
\begin{center}
\begin{tabular}{cl r@{.}l r@{.}l@{ - }r@{.}l r@{.}l @{$\pm$} r@{.}l @{ $^+_-$}
r@{.}l} \hline
Parameter & Name & \multicolumn{2}{c}{Default} &
 \multicolumn{4}{c}{Range gen.}   &\multicolumn{6}{c}{Fit Result} \\
  \multicolumn{8}{c}{~} &  \multicolumn{2}{c}{Val.} &
                           \multicolumn{2}{c}{stat.}&
                           \multicolumn{2}{c}{sys.} \\ \hline
$\lambda_{QCD}$   &  PARJ(122)       &  0&22   &  0&20  &  0&30  &
                  0&163 & 0&002 &\multicolumn{2}{l}{$^{0.001}_{0.003}$}  \\
$Scale$ $\mu$ &  PARJ(129)       &  0&002   &  0&0005   &  0&0025   &
                  0&00250  & 0&00006 &
\multicolumn{2}{l}{$^{0.00013}_{0.00018}$} \\ \hline
$a$  &  PARJ(41)       &  0&5   &  0&1 & 0&5  &
                  0&903 &  0&014 &  \multicolumn{2}{l}{$^{0.038}_{0.014}$}\\
$b$ &  PARJ(42)  &  0&9 &\multicolumn{4}{c}{0.525}
                               &\multicolumn{6}{c}{optimized}\\ \hline
$\sigma_q$      &  PARJ(21)  &  0&35  &  0&36  & 0&44  &
                  0&477 & 0&004 & \multicolumn{2}{l}{$^{0.005}_{0.006}$}\\
\hline
$P(^1S_0)_{ud}$ &     -      &  0&5   &  0&3 & 0&5   &
                  0&327 &  0&023 & \multicolumn{2}{l}{$^{0.051}_{0.021}$}\\
\hline
\end{tabular}
\end{center}
\caption[Parameter setting and fit results for JETSET 7.4 ME]
{\label{terg74m} Parameter setting and fit results for JETSET 7.4 ME with
default decays}
\end{table}

\normalsize
\clearpage

\section{Tables of Correlation Coefficients}
\scriptsize
\label{correl_tabs}

\begin{table} [h]
\begin{center}
%
\end{center}
\caption[Table of Correlation Coefficients for JETSET 7.4 PS Fit ]
 {\label{covjt74} Table of Correlation Coefficients for JETSET 7.4 PS Fit}
\end{table}
\begin{table} [h]
\begin{center}
%
\end{center}
\caption[Table of Correlation Coefficients for ARIADNE 4.06 PS Fit]
 {\label{covar46d} Table of Correlation Coefficients for ARIADNE 4.06 PS Fit}
\end{table}
\normalsize
\newpage

\section{Figures of Data and Model Comparisons}
\label{plots}

The figures in this section compare DELPHI data and identified particle spectra
to the following fragmentation models:
\begin{itemize}
\item
JETSET 7.3 DELPHI decays labeled JT 7.3 PS
\item
JETSET 7.4 default decays labeled JT 7.4 PS
\item
ARIADNE 4.06 DELPHI decays labeled AR 4.06
\item
HERWIG 5.8 c default decays labeled H 5.8C
\item
JETSET 7.4 ME default decays labeled JT 7.4 ME
\end{itemize}
The references for the identified particle data can be found in table
\ref{evtfit3}.
For most shape distributions we show on the left side the data as measured
from charged particles only, on the right as measured from charged plus neutral
particles.  The distributions are corrected  to the corresponding final states.
The total correction factor for each distribution is shown in the upper inset.
The lower insets of the plots decipt the relative deviation of the models
to the data. Also shown as shaded area in these insets is the total
experimental error obtained by adding quadratically the systematic and
statistical error in each bin. Except for the point-to-point scatter the error
should be interpreted like a statistical ''$1 \sigma$'' uncertainty.

\newpage

\subsection{Inclusive Distributions}
\begin{figure}[h]
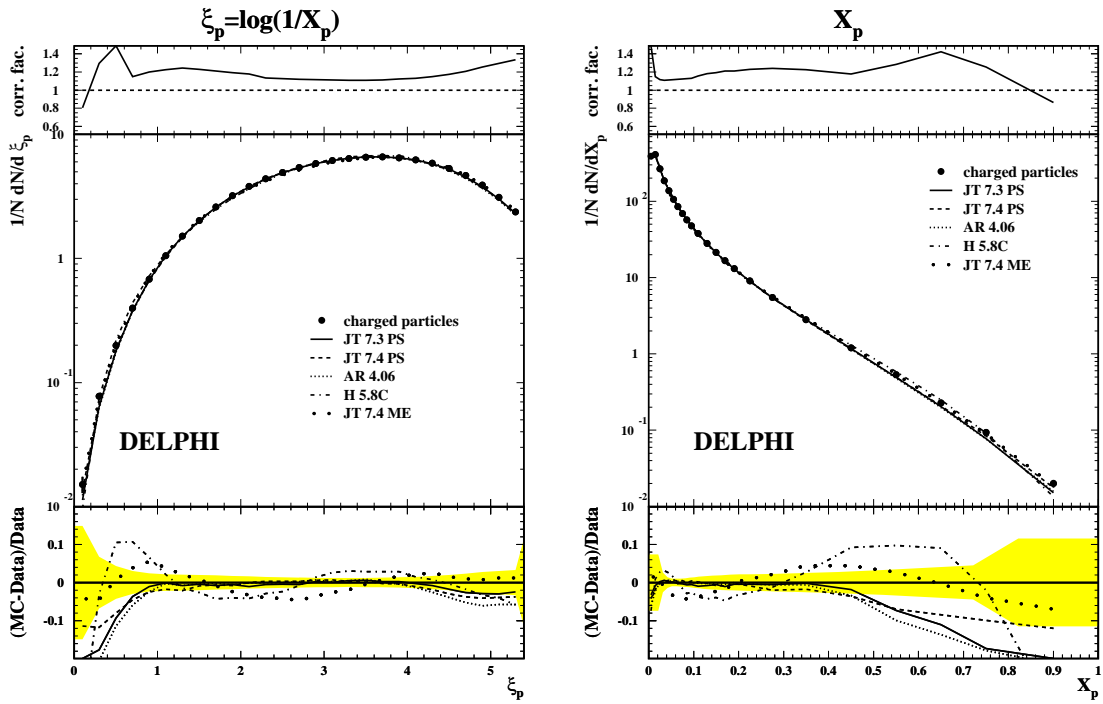

\begin{center}
\vspace{-5.5cm}
\unitlength1cm
 \begin{minipage}[t]{7.5cm}
   \mbox{\epsfig{file=b5_paper_025_e.epsc,width=10.5cm}}
 \end{minipage}
 \begin{minipage}[t]{7.5cm}
   \mbox{\epsfig{file=b5_paper_01023_e.epsc,width=10.5cm}}
 \end{minipage}
\caption[$\xi_p$ and scaled momentum $x_p$]{\label{bildxpp} Distribution of
$\xi_p$ and the scaled momentum $x_p$}
\end{center}
\end{figure}

\begin{figure}
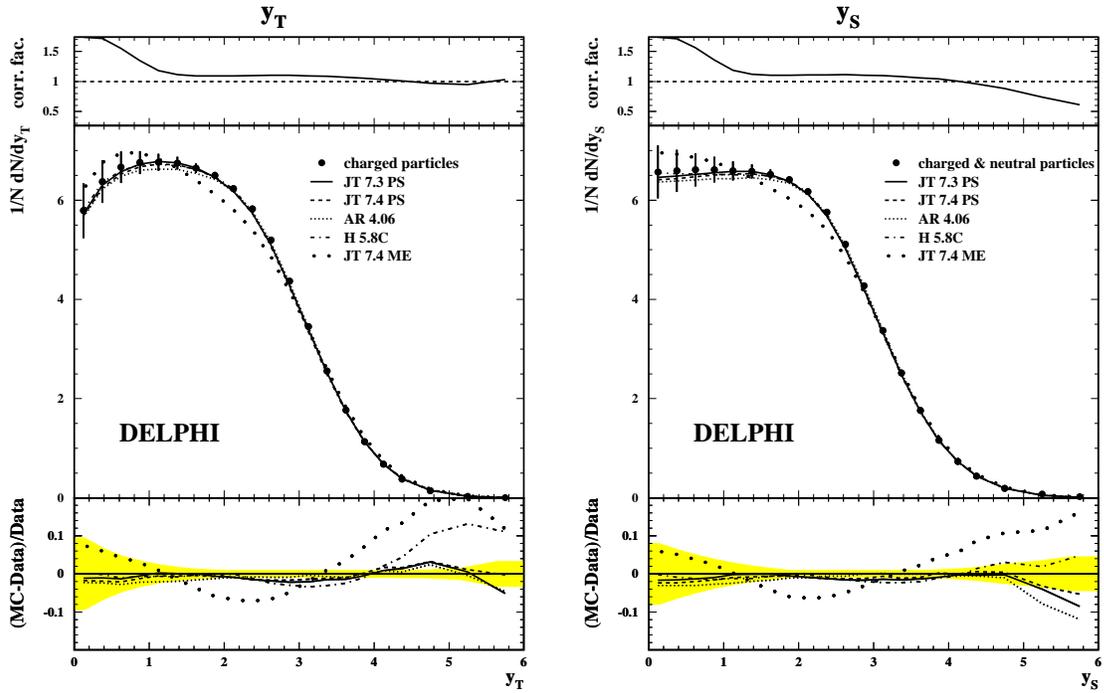

\begin{center}
\vspace{-10.5cm}
\unitlength1cm
  \begin{minipage}[t]{7.5cm}
     \mbox{\epsfig{file=b5_paper_022_e.epsc,width=10.5cm}}
  \end{minipage}
  \begin{minipage}[t]{7.5cm}
     \mbox{\epsfig{file=b5_paper_124_e.epsc,width=10.5cm}}
  \end{minipage}
\caption[Rapidities]
{\label{bildyts} Distribution of the Rapidity $y_T$ and the Rapidity $y_S$}
\end{center}
\end{figure}

\clearpage

\begin{figure}
\begin{center}
\vspace{-5.cm}
\unitlength1cm
 \begin{minipage}[t]{7.5cm}
   \mbox{\epsfig{file=b5_paper_020_e.epsc,width=10.5cm}}
 \end{minipage}
 \begin{minipage}[t]{7.5cm}
   \mbox{\epsfig{file=b5_paper_120_e.epsc,width=10.5cm}}
 \end{minipage}
\caption[$p^{in}_{t}$ with respect to the Thrust axis]{\label{bildptin}
Distribution of
$p^{in}_{t}$ with respect to the Thrust axis}
\end{center}
\end{figure}

\begin{figure}
\begin{center}
\vspace{-5.cm}
\unitlength1cm
  \begin{minipage}[t]{7.5cm}
     \mbox{\epsfig{file=b5_paper_021_e.epsc,width=10.5cm}}
  \end{minipage}
  \begin{minipage}[t]{7.5cm}
     \mbox{\epsfig{file=b5_paper_121_e.epsc,width=10.5cm}}
  \end{minipage}
\caption[$p^{out}_{t}$ with respect to the Thrust axis]{\label{bildptout}
Distribution of
$p^{out}_{t}$ with respect to the Thrust axis}
\end{center}
\end{figure}

\begin{figure}
\begin{center}
\vspace{-5.cm}
\unitlength1cm
 \begin{minipage}[t]{7.5cm}
   \mbox{\epsfig{file=b5_paper_027_e.epsc,width=10.5cm}}
 \end{minipage}
 \begin{minipage}[t]{7.5cm}
   \mbox{\epsfig{file=b5_paper_028_e.epsc,width=10.5cm}}
 \end{minipage}
\caption[$<p^{out}_{t}>$ vs. $x_p$ and $<p_{t}>$ vs. $x_p$]
{\label{bildmean} $<p^{out}_{t}>$ vs. $x_p$ and
$<p_{t}>$ vs. $x_p$}
\end{center}
\end{figure}

\begin{figure}
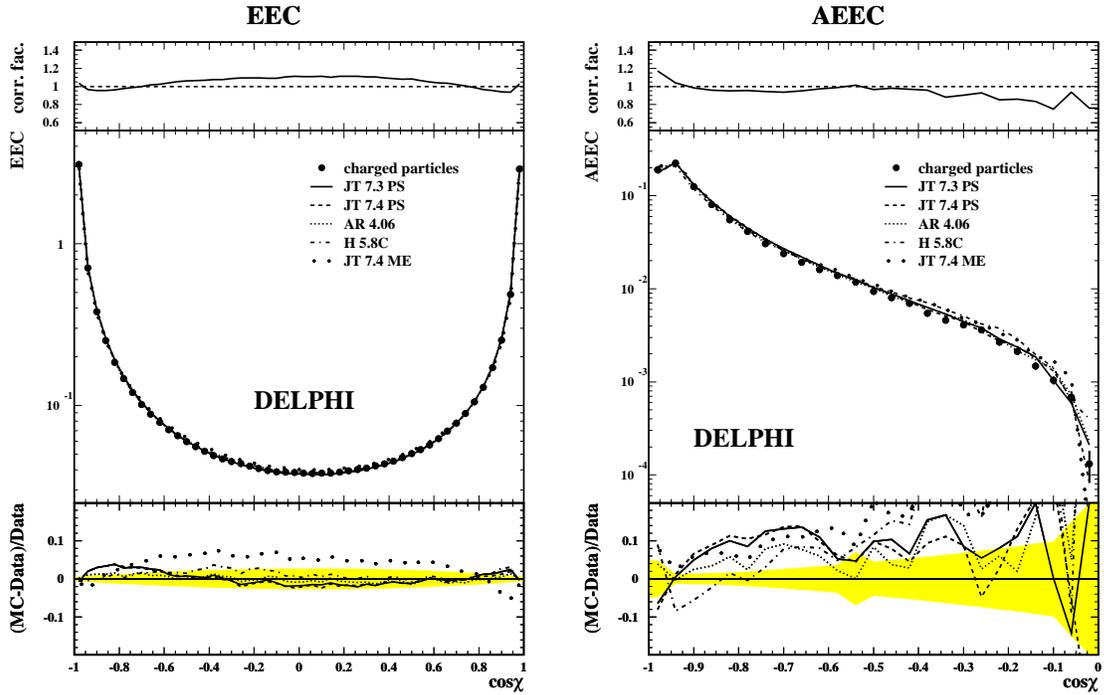

\begin{center}
\vspace{-5.cm}
\unitlength1cm
  \begin{minipage}[t]{7.5cm}
     \mbox{\epsfig{file=b5_paper_040_e.epsc,width=10.5cm}}
  \end{minipage}
  \begin{minipage}[t]{7.5cm}
     \mbox{\epsfig{file=b5_paper_041_e.epsc,width=10.5cm}}
  \end{minipage}
\caption[EEC and AEEC ]{\label{bildeec} Distribution of the Energy Energy
Correlation EEC and the Asymmetry AEEC}
\end{center}
\end{figure}

\clearpage
\clearpage

\subsection{Shape Distributions}

\begin{figure} [h]
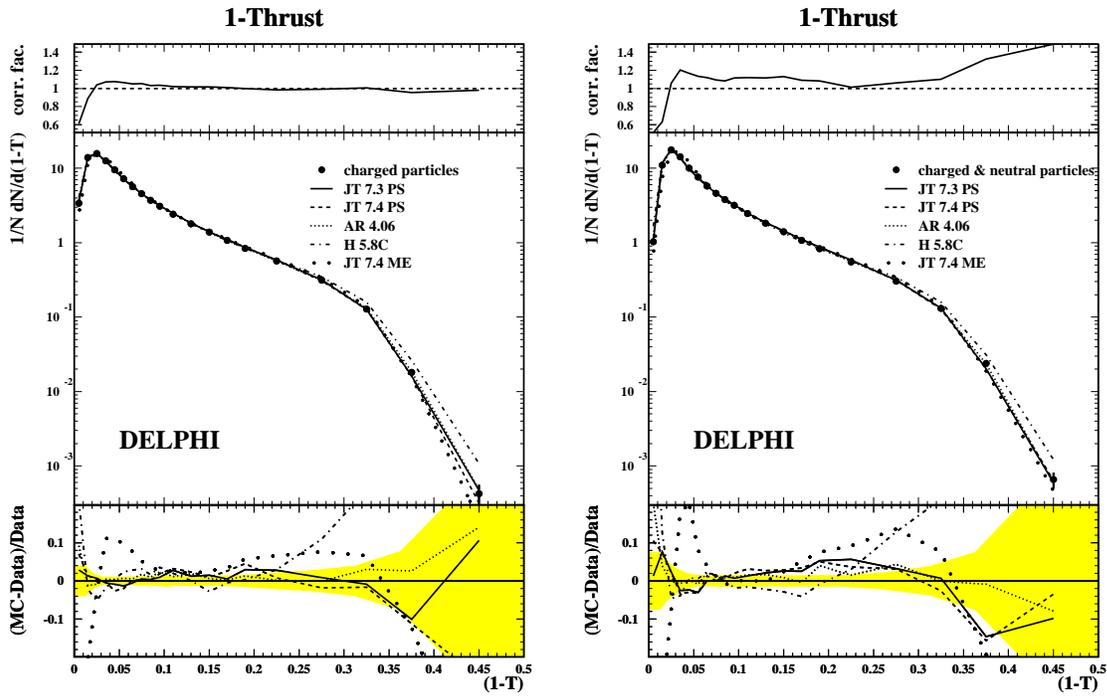

\begin{center}
\vspace{-5.5cm}
\unitlength1cm
 \begin{minipage}[t]{7.5cm}
   \mbox{\epsfig{file=b5_paper_004_e.epsc,width=10.5cm}}
 \end{minipage}
 \begin{minipage}[t]{7.5cm}
   \mbox{\epsfig{file=b5_paper_104_e.epsc,width=10.5cm}}
 \end{minipage}
\caption[1-Thrust]{\label{bildthr} Distribution of 1-Thrust}
\end{center}
\end{figure}

\begin{figure}
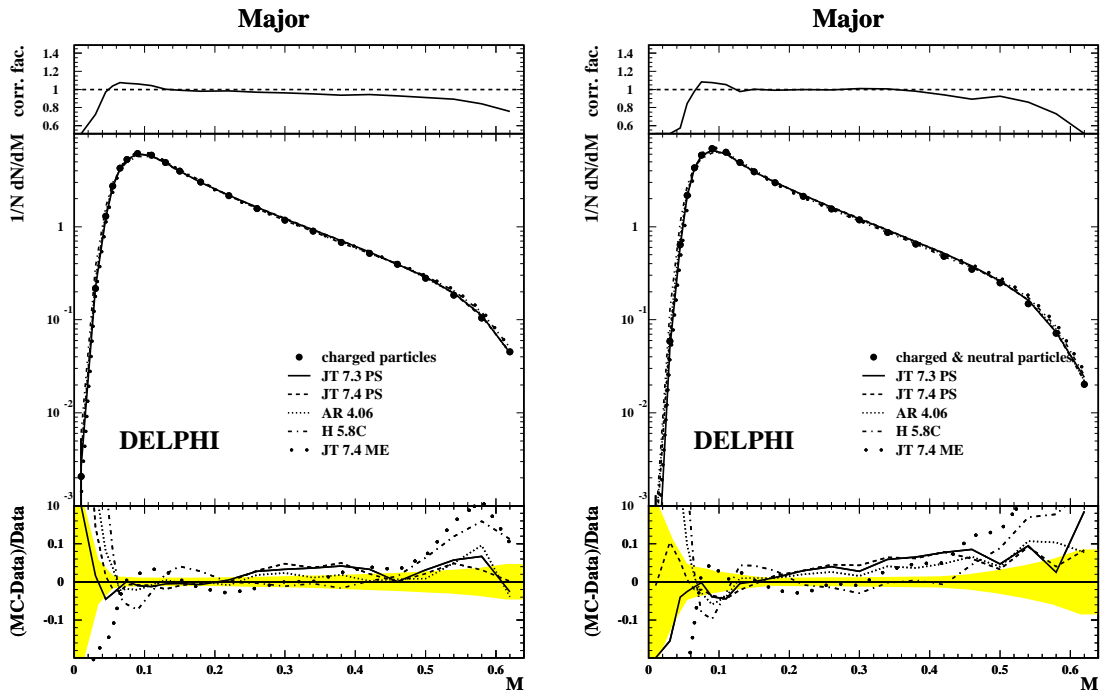

\begin{center}
\vspace{-10.5cm}
\unitlength1cm
  \begin{minipage}[t]{7.5cm}
     \mbox{\epsfig{file=b5_paper_006_e.epsc,width=10.5cm}}
  \end{minipage}
  \begin{minipage}[t]{7.5cm}
     \mbox{\epsfig{file=b5_paper_106_e.epsc,width=10.5cm}}
  \end{minipage}
\caption[Major]{\label{bildmaj} Distribution of Major}
\end{center}
\end{figure}

\clearpage

\begin{figure}
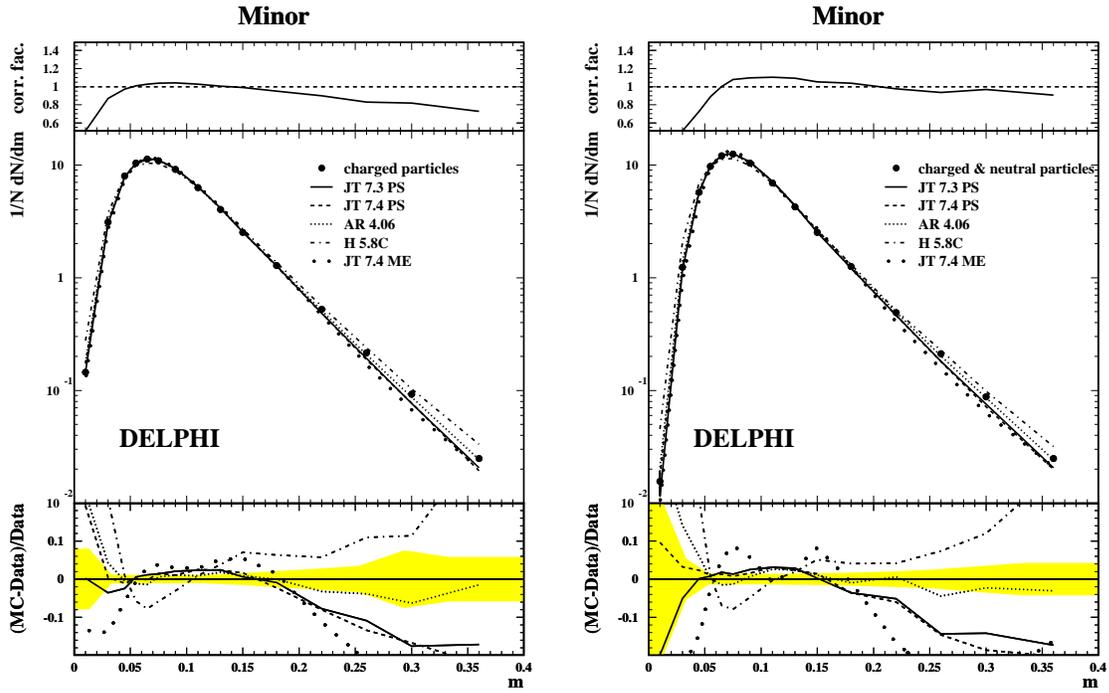

\begin{center}
\vspace{-5.cm}
\unitlength1cm
 \begin{minipage}[t]{7.5cm}
   \mbox{\epsfig{file=b5_paper_007_e.epsc,width=10.5cm}}
 \end{minipage}
 \begin{minipage}[t]{7.5cm}
   \mbox{\epsfig{file=b5_paper_107_e.epsc,width=10.5cm}}
 \end{minipage}
\caption[Minor]{\label{bildmin} Distribution of Minor}
\end{center}
\end{figure}

\begin{figure}
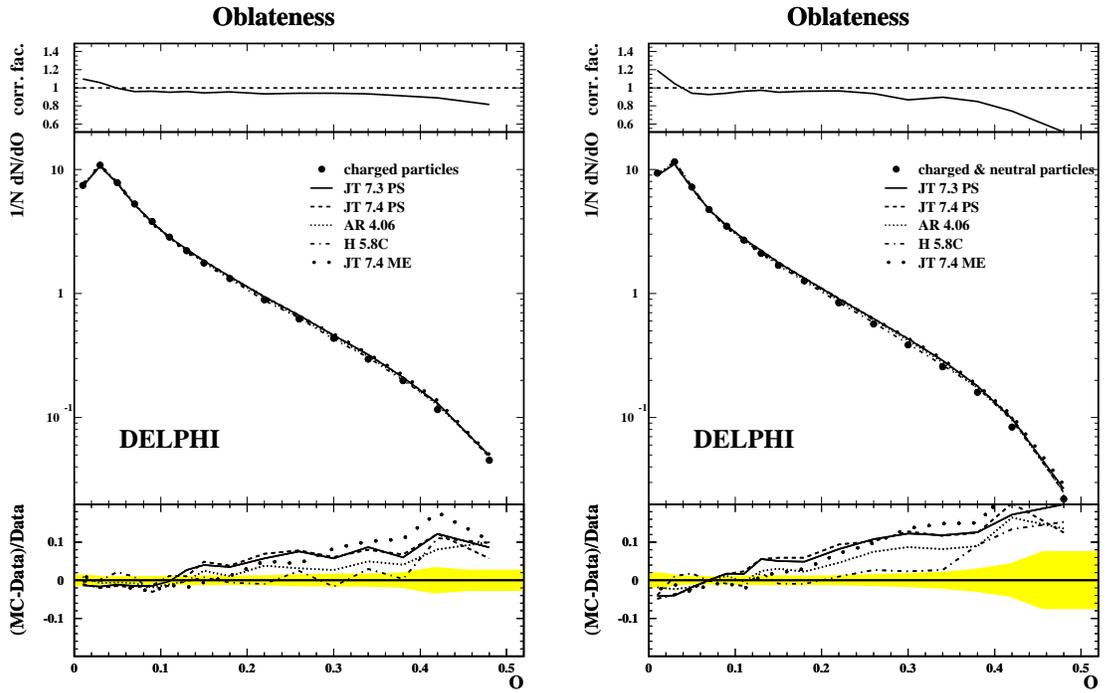

\begin{center}
\vspace{-5.cm}
\unitlength1cm
  \begin{minipage}[t]{7.5cm}
     \mbox{\epsfig{file=b5_paper_005_e.epsc,width=10.5cm}}
  \end{minipage}
  \begin{minipage}[t]{7.5cm}
     \mbox{\epsfig{file=b5_paper_105_e.epsc,width=10.5cm}}
  \end{minipage}
\caption[Oblateness]{\label{bildob} Distribution of the Oblateness}
\end{center}
\end{figure}

\clearpage

\begin{figure}
\begin{center}
\vspace{-5.cm}
\unitlength1cm
 \begin{minipage}[t]{7.5cm}
   \mbox{\epsfig{file=b5_paper_011_e.epsc,width=10.5cm}}
 \end{minipage}
 \begin{minipage}[t]{7.5cm}
   \mbox{\epsfig{file=b5_paper_111_e.epsc,width=10.5cm}}
 \end{minipage}
\caption[Differential 2-Jet Rate Durham]{\label{bild2dd} Distribution of the
Differential 2-Jet Rate
Durham Algorithm}
\end{center}
\end{figure}

\begin{figure}
\begin{center}
\vspace{-5.cm}
\unitlength1cm
  \begin{minipage}[t]{7.5cm}
     \mbox{\epsfig{file=b5_paper_039_e.epsc,width=10.5cm}}
  \end{minipage}
  \begin{minipage}[t]{7.5cm}
     \mbox{\epsfig{file=b5_paper_139_e.epsc,width=10.5cm}}
  \end{minipage}
\caption[Differential 2-Jet Rate Jade]{\label{bild2dj} Distribution of the
Differential 2-Jet Rate
Jade Algorithm}
\end{center}
\end{figure}

\clearpage

\begin{figure}
\begin{center}
\vspace{-5.cm}
\unitlength1cm
 \begin{minipage}[t]{7.5cm}
   \mbox{\epsfig{file=b5_paper_082_e.epsc,width=10.5cm}}
 \end{minipage}
 \begin{minipage}[t]{7.5cm}
   \mbox{\epsfig{file=b5_paper_182_e.epsc,width=10.5cm}}
 \end{minipage}
\caption[Differential 3-Jet Rate Durham]{\label{bild3dd} Distribution of the
Differential 3-Jet Rate
Durham Algorithm}
\end{center}
\end{figure}

\begin{figure}
\begin{center}
\vspace{-5.cm}
\unitlength1cm
  \begin{minipage}[t]{7.5cm}
     \mbox{\epsfig{file=b5_paper_092_e.epsc,width=10.5cm}}
  \end{minipage}
  \begin{minipage}[t]{7.5cm}
     \mbox{\epsfig{file=b5_paper_192_e.epsc,width=10.5cm}}
  \end{minipage}
\caption[Differential 3-Jet Rate Jade]{\label{bild3dj} Distribution of the
Differential 3-Jet Rate
Jade Algorithm}
\end{center}
\end{figure}

\clearpage
\begin{figure}
\begin{center}
\vspace{-5.cm}
\unitlength1cm
 \begin{minipage}[t]{7.5cm}
   \mbox{\epsfig{file=b5_paper_083_e.epsc,width=10.5cm}}
 \end{minipage}
 \begin{minipage}[t]{7.5cm}
   \mbox{\epsfig{file=b5_paper_183_e.epsc,width=10.5cm}}
 \end{minipage}
\caption[Differential 4-Jet Rate Durham]{\label{bild4dd} Distribution of the
Differential 4-Jet Rate
Durham Algorithm}
\end{center}
\end{figure}

\begin{figure}
\begin{center}
\vspace{-5.cm}
\unitlength1cm
  \begin{minipage}[t]{7.5cm}
     \mbox{\epsfig{file=b5_paper_093_e.epsc,width=10.5cm}}
  \end{minipage}
  \begin{minipage}[t]{7.5cm}
     \mbox{\epsfig{file=b5_paper_193_e.epsc,width=10.5cm}}
  \end{minipage}
\caption[Differential 4-Jet Rate Jade]{\label{bild4dj} Distribution of the
Differential 4-Jet Rate
Jade Algorithm}
\end{center}
\end{figure}

\clearpage

\begin{figure}
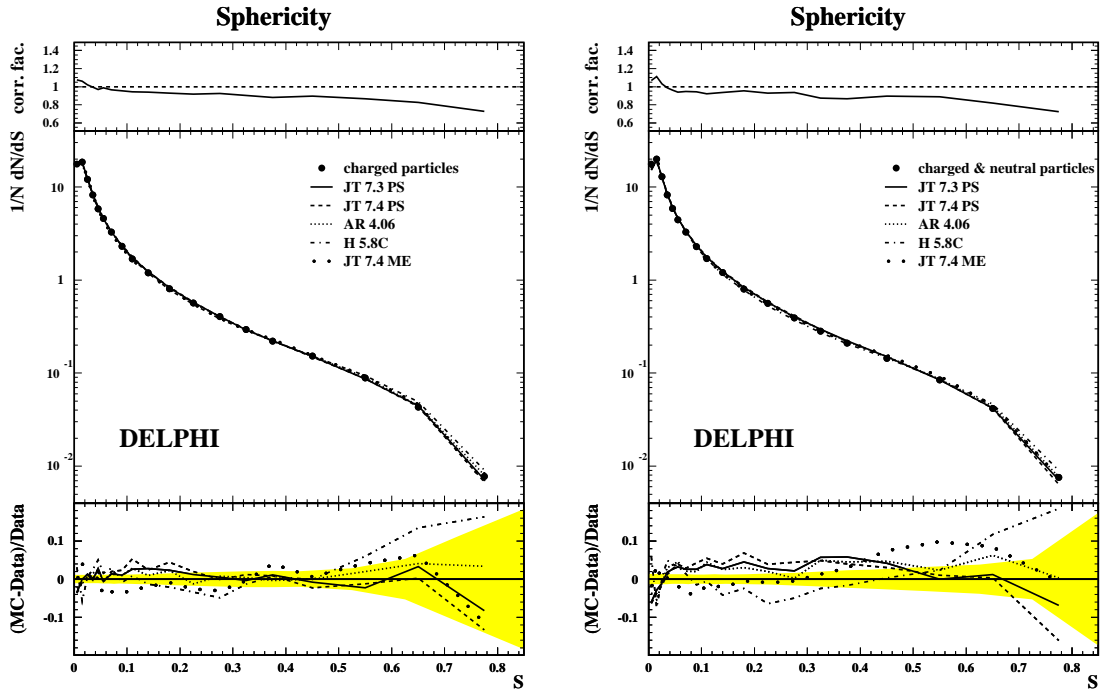

\begin{center}
\vspace{-5.cm}
\unitlength1cm
 \begin{minipage}[t]{7.5cm}
   \mbox{\epsfig{file=b5_paper_002_e.epsc,width=10.5cm}}
 \end{minipage}
 \begin{minipage}[t]{7.5cm}
   \mbox{\epsfig{file=b5_paper_102_e.epsc,width=10.5cm}}
 \end{minipage}
\caption[Sphericity]{\label{bildsph} Distribution of the Sphericity}
\end{center}
\end{figure}

\begin{figure}
\begin{center}
\vspace{-5.cm}
\unitlength1cm
  \begin{minipage}[t]{7.5cm}
     \mbox{\epsfig{file=b5_paper_003_e.epsc,width=10.5cm}}
  \end{minipage}
  \begin{minipage}[t]{7.5cm}
     \mbox{\epsfig{file=b5_paper_103_e.epsc,width=10.5cm}}
  \end{minipage}
\caption[Aplanarity]{\label{bildapl} Distribution of the Aplanarity}
\end{center}
\end{figure}

\clearpage

\begin{figure}
\begin{center}
\vspace{-5.cm}
\unitlength1cm
 \begin{minipage}[t]{7.5cm}
   \mbox{\epsfig{file=b5_paper_042_e.epsc,width=10.5cm}}
 \end{minipage}
 \begin{minipage}[t]{7.5cm}
   \mbox{\epsfig{file=b5_paper_142_e.epsc,width=10.5cm}}
 \end{minipage}
\caption[Planarity]{\label{bildpla} Distribution of the Planarity}
\end{center}
\end{figure}

\begin{figure}
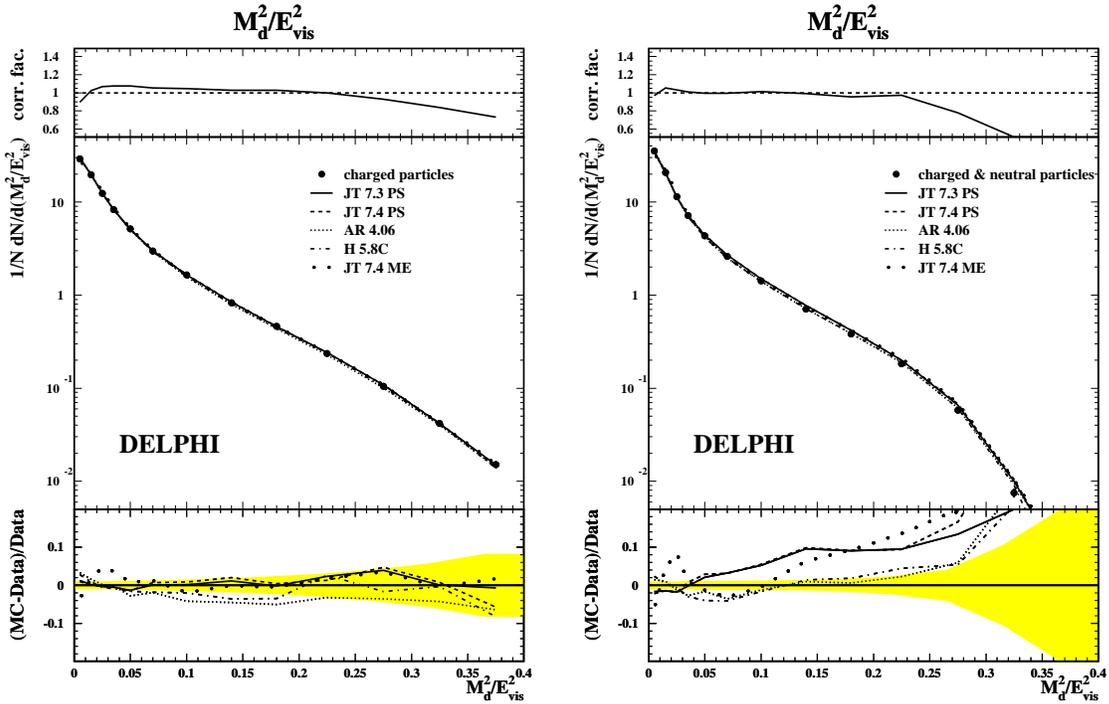

\begin{center}
\vspace{-5.cm}
\unitlength1cm
  \begin{minipage}[t]{7.5cm}
     \mbox{\epsfig{file=b5_paper_010_e.epsc,width=10.5cm}}
  \end{minipage}
  \begin{minipage}[t]{7.5cm}
     \mbox{\epsfig{file=b5_paper_110_e.epsc,width=10.5cm}}
  \end{minipage}
\caption[Difference of Jet masses]{\label{bilddjm} Distribution of the
Difference of Jet masses}
\end{center}
\end{figure}
\clearpage

\begin{figure}
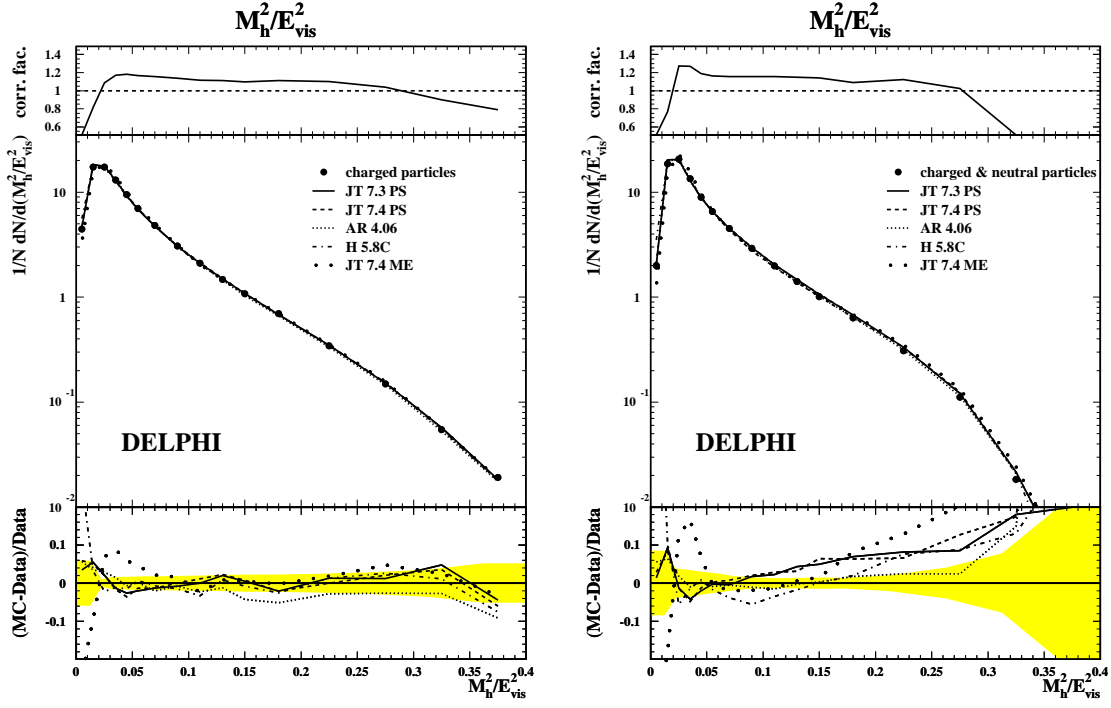

\begin{center}
\vspace{-5.cm}
\unitlength1cm
 \begin{minipage}[t]{7.5cm}
   \mbox{\epsfig{file=b5_paper_008_e.epsc,width=10.5cm}}
 \end{minipage}
 \begin{minipage}[t]{7.5cm}
   \mbox{\epsfig{file=b5_paper_108_e.epsc,width=10.5cm}}
 \end{minipage}
\caption[Heavy Jet mass]{\label{bildhjm} Distribution of the Heavy Jet mass}
\end{center}
\end{figure}

\begin{figure}
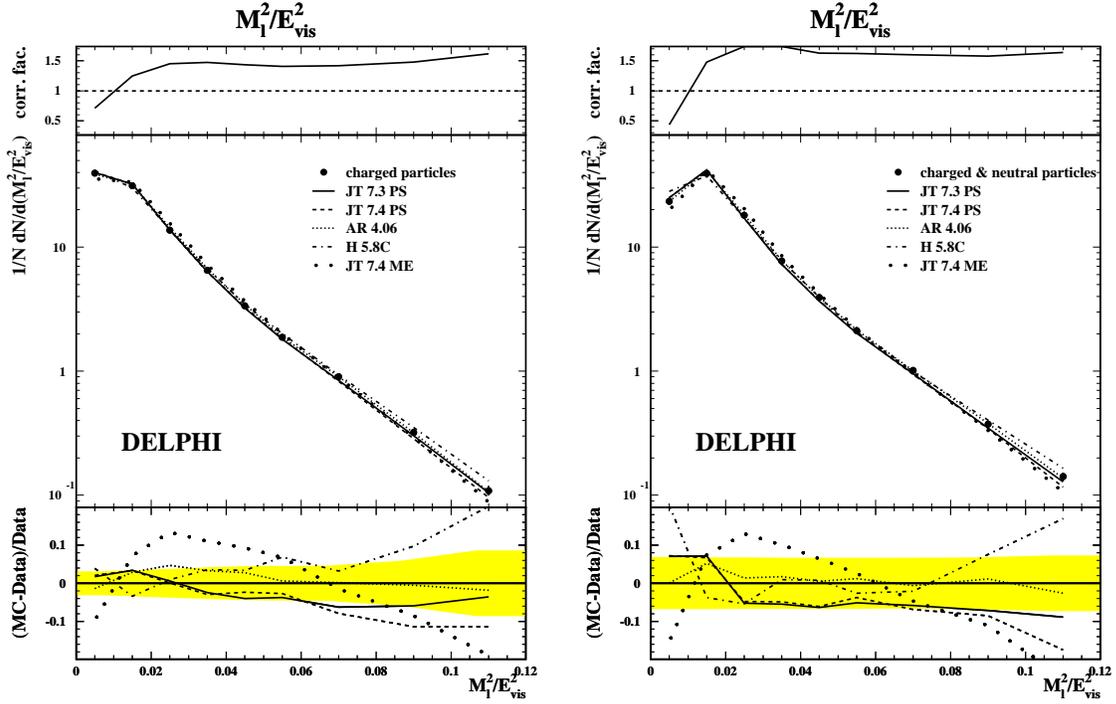

\begin{center}
\vspace{-5.cm}
\unitlength1cm
  \begin{minipage}[t]{7.5cm}
     \mbox{\epsfig{file=b5_paper_009_e.epsc,width=10.5cm}}
  \end{minipage}
  \begin{minipage}[t]{7.5cm}
     \mbox{\epsfig{file=b5_paper_109_e.epsc,width=10.5cm}}
  \end{minipage}
\caption[Light Jet mass]{\label{bildljm} Distribution of the Light Jet mass}
\end{center}
\end{figure}

\clearpage

\begin{figure}
\begin{center}
\vspace{-5.cm}
\unitlength1cm
 \begin{minipage}[t]{7.5cm}
   \mbox{\epsfig{file=b5_paper_043_e.epsc,width=10.5cm}}
 \end{minipage}
 \begin{minipage}[t]{7.5cm}
   \mbox{\epsfig{file=b5_paper_143_e.epsc,width=10.5cm}}
 \end{minipage}
\caption[C-Parameter]{\label{bildcp} Distribution of the C-Parameter}
\end{center}
\end{figure}

\begin{figure}
\begin{center}
\vspace{-5.cm}
\unitlength1cm
  \begin{minipage}[t]{7.5cm}
     \mbox{\epsfig{file=b5_paper_044_e.epsc,width=10.5cm}}
  \end{minipage}
  \begin{minipage}[t]{7.5cm}
     \mbox{\epsfig{file=b5_paper_144_e.epsc,width=10.5cm}}
  \end{minipage}
\caption[D-Parameter]{\label{bilddp} Distribution of the D-Parameter}
\end{center}
\end{figure}

\clearpage

\begin{figure}
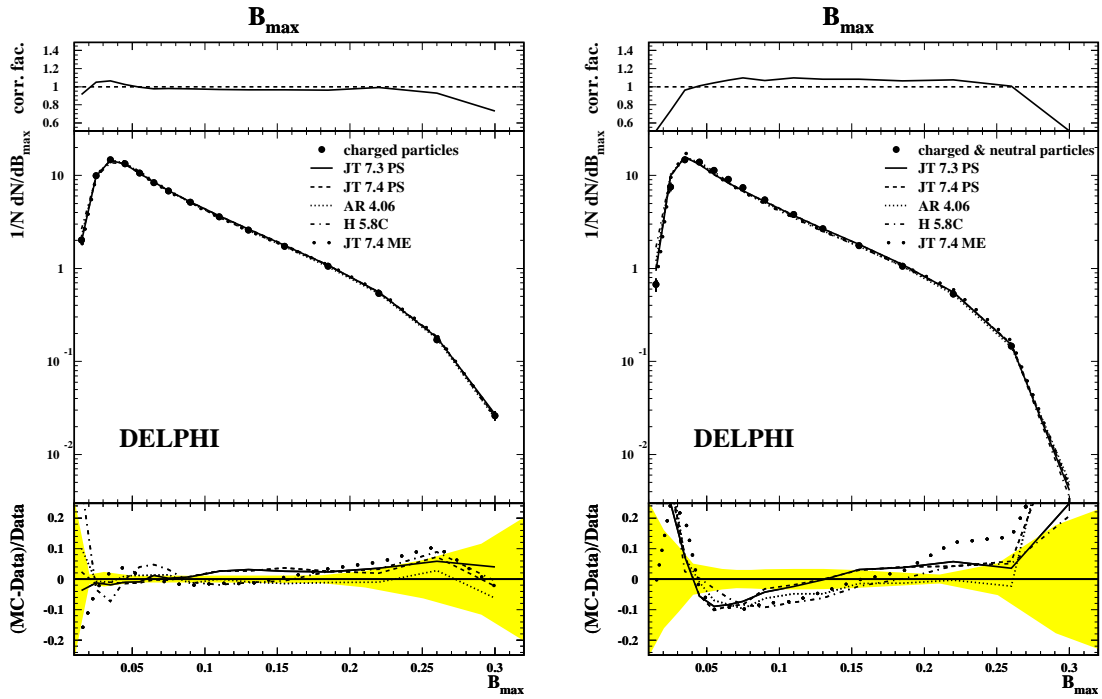

\begin{center}
\vspace{-5.cm}
\unitlength1cm
 \begin{minipage}[t]{7.5cm}
   \mbox{\epsfig{file=b5_paper_035_e.epsc,width=10.5cm}}
 \end{minipage}
 \begin{minipage}[t]{7.5cm}
   \mbox{\epsfig{file=b5_paper_135_e.epsc,width=10.5cm}}
 \end{minipage}
\caption[Wide Jet Broadening] {\label{bildwjb} Distribution of the Wide Jet
Broadening}
\end{center}
\end{figure}

\begin{figure}
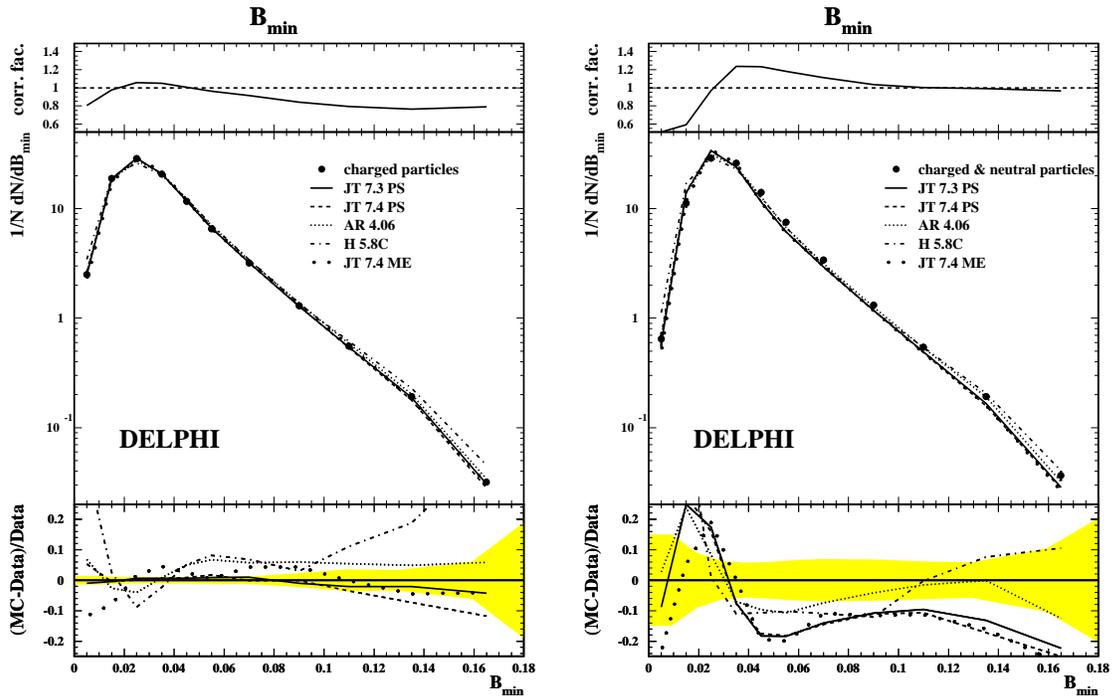

\begin{center}
\vspace{-5.cm}
\unitlength1cm
  \begin{minipage}[t]{7.5cm}
     \mbox{\epsfig{file=b5_paper_036_e.epsc,width=10.5cm}}
  \end{minipage}
  \begin{minipage}[t]{7.5cm}
     \mbox{\epsfig{file=b5_paper_136_e.epsc,width=10.5cm}}
  \end{minipage}
\caption[Narrow Jet Broadening] {\label{bildnjb} Distribution of the
Narrow Jet Broadening}
\end{center}
\end{figure}

\clearpage

\clearpage

\begin{figure}
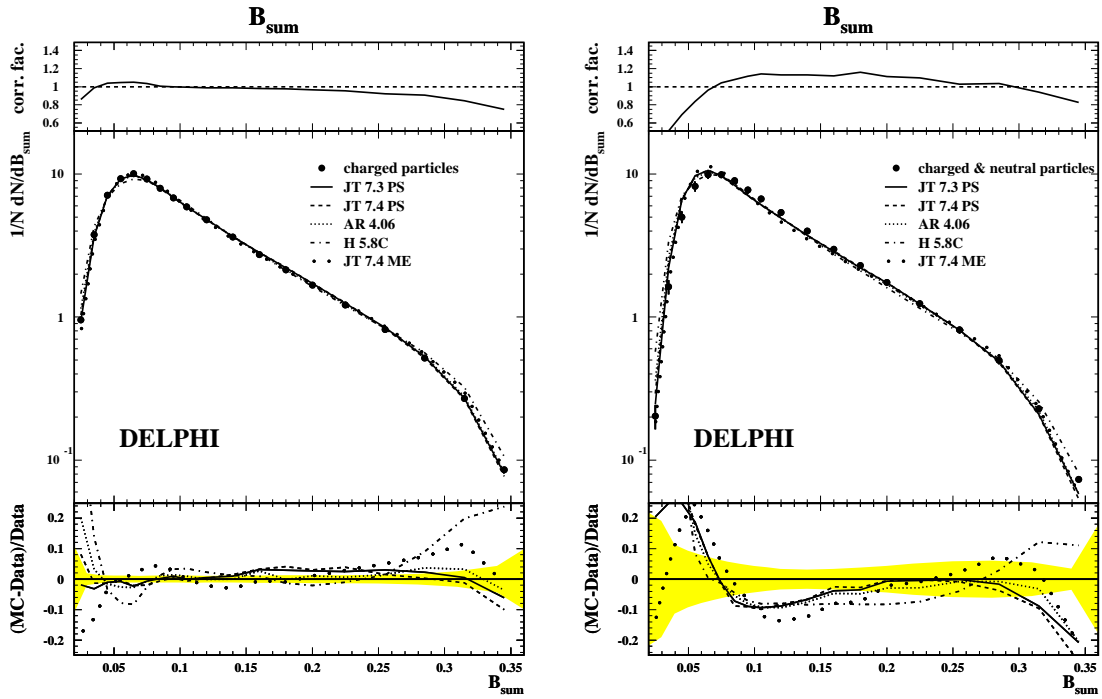

\begin{center}
\vspace{-5.cm}
\unitlength1cm
 \begin{minipage}[t]{7.5cm}
   \mbox{\epsfig{file=b5_paper_037_e.epsc,width=10.5cm}}
 \end{minipage}
 \begin{minipage}[t]{7.5cm}
   \mbox{\epsfig{file=b5_paper_137_e.epsc,width=10.5cm}}
 \end{minipage}
\caption[Total Jet Broadening]
{\label{bildtjb} Distribution of the Total Jet Broadening }
\end{center}
\end{figure}

\begin{figure}
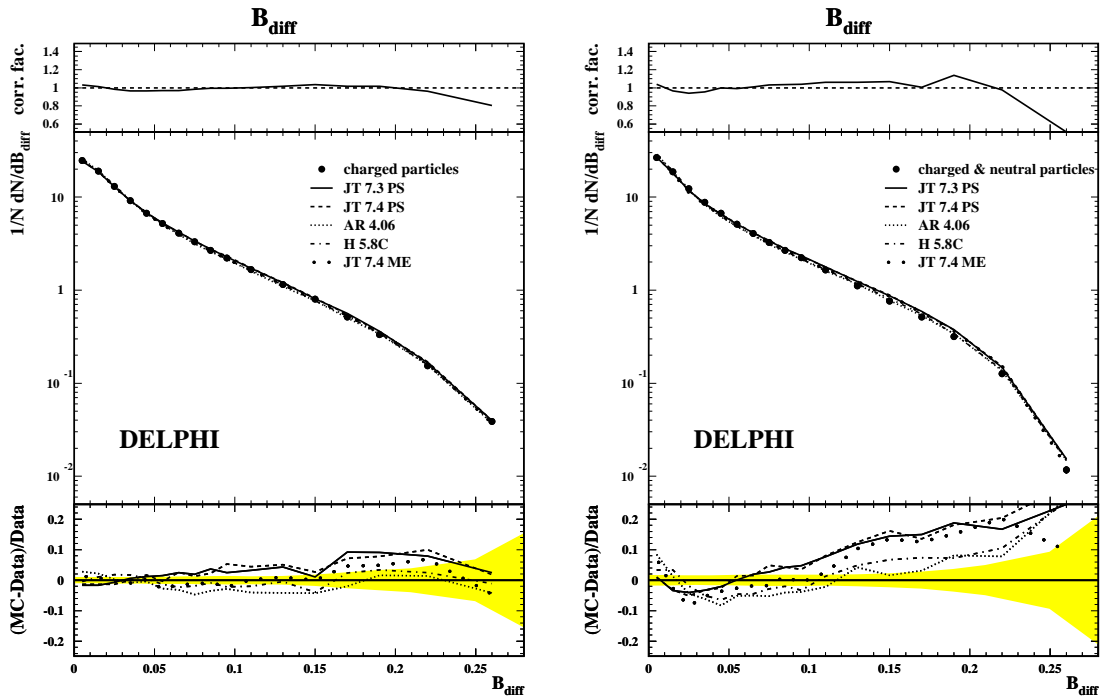

\begin{center}
\vspace{-5.cm}
\unitlength1cm
  \begin{minipage}[t]{7.5cm}
     \mbox{\epsfig{file=b5_paper_038_e.epsc,width=10.5cm}}
  \end{minipage}
  \begin{minipage}[t]{7.5cm}
     \mbox{\epsfig{file=b5_paper_138_e.epsc,width=10.5cm}}
  \end{minipage}
\caption[Difference of the Jet Broadenings]
{\label{bilddjb} Distribution of the Difference of the Jet Broadenings}
\end{center}
\end{figure}

\clearpage

\clearpage

\subsection{Identified Particle Distributions}

\begin{figure}[h]
\begin{center}
%\vspace{-1.0cm}
\unitlength1cm
  \begin{minipage}[t]{6.7cm}
    \mbox{\epsfig{file=k0_cern.epsc,width=7.0cm}}
  \end{minipage}
  \begin{minipage}[t]{6.7cm}
    \mbox{\epsfig{file=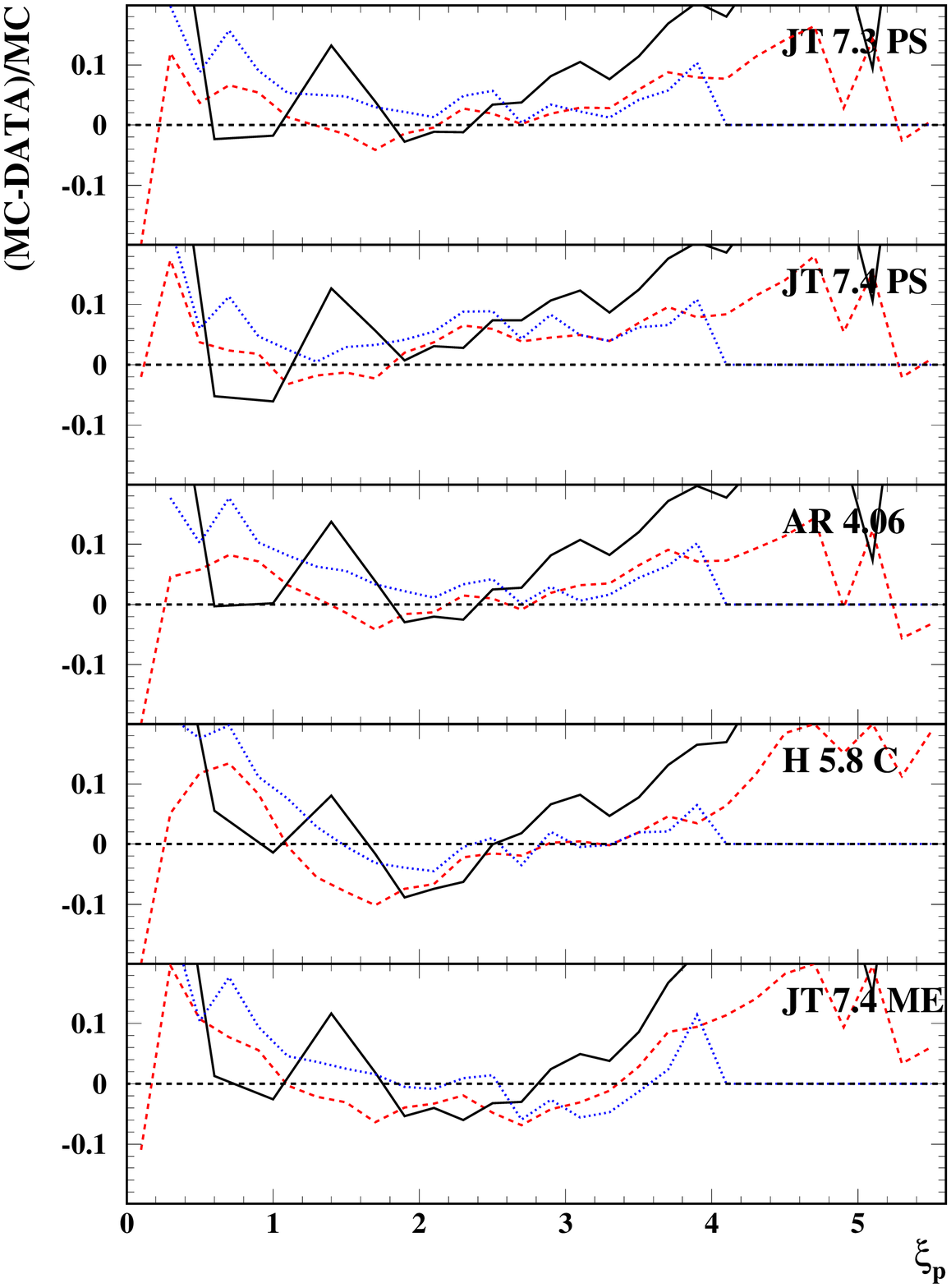,width=7.0cm}}
  \end{minipage}
\caption[$K^0$ spectra]
{\label{bildk0} $\xi_p$ Distribution of $K^0$ }
\end{center}
\end{figure}

\begin{figure}
\begin{center}
\vspace{-5.0cm}
\unitlength1cm
  \begin{minipage}[t]{6.7cm}
    \mbox{\epsfig{file=kpm_cern.epsc,width=7.0cm}}
  \end{minipage}
  \begin{minipage}[t]{6.7cm}
    \mbox{\epsfig{file=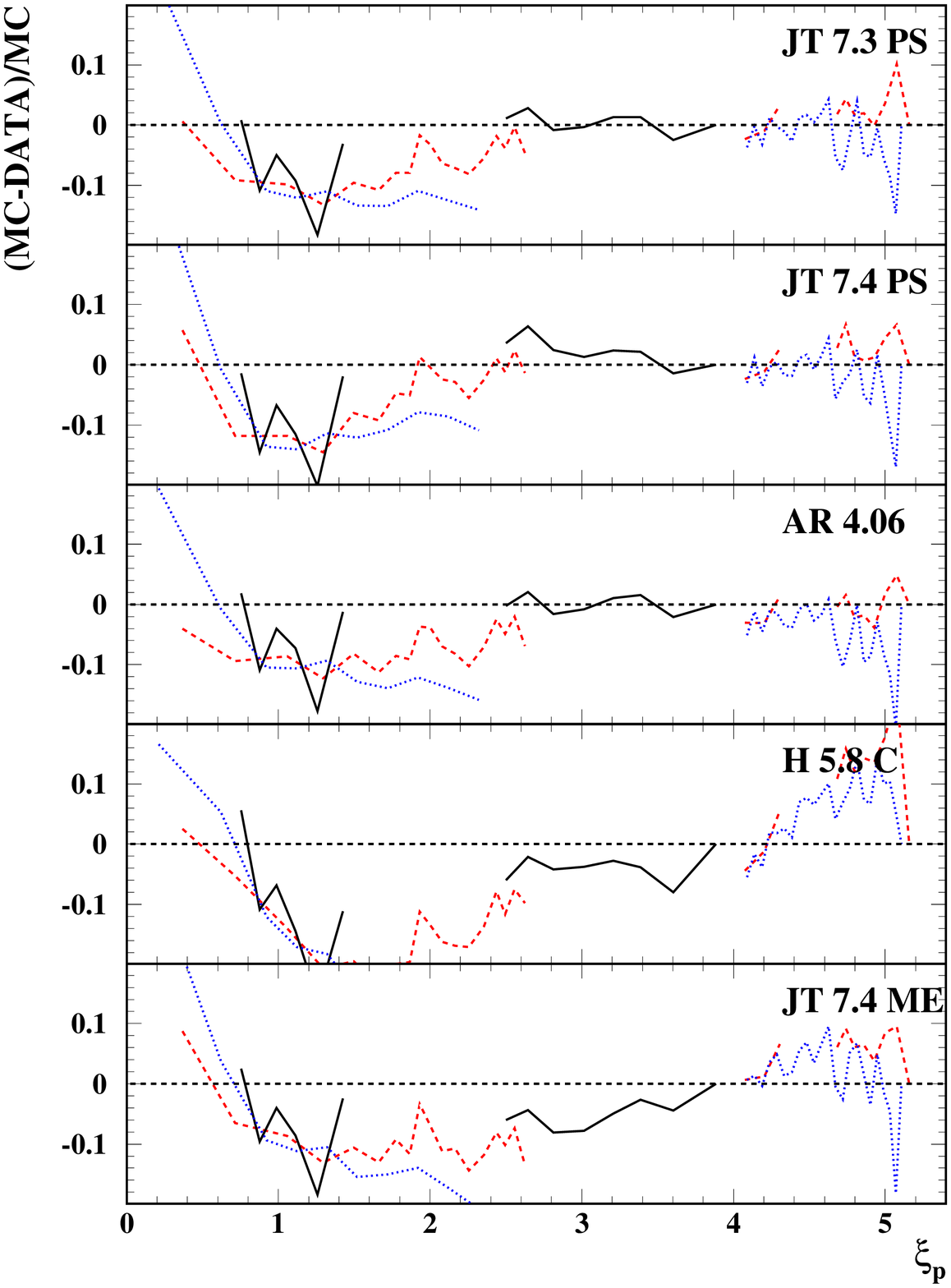,width=7.0cm}}
  \end{minipage}
\caption[$K^{\pm}$ spectra]
{\label{bildkpm} $\xi_p$ Distribution of $K^{\pm}$ }
\end{center}
\end{figure}

\begin{figure}
\begin{center}
\unitlength1cm
  \begin{minipage}[t]{6.7cm}
    \mbox{\epsfig{file=kstar0_cern.epsc,width=7.0cm}}
  \end{minipage}
  \begin{minipage}[t]{6.7cm}
    \mbox{\epsfig{file=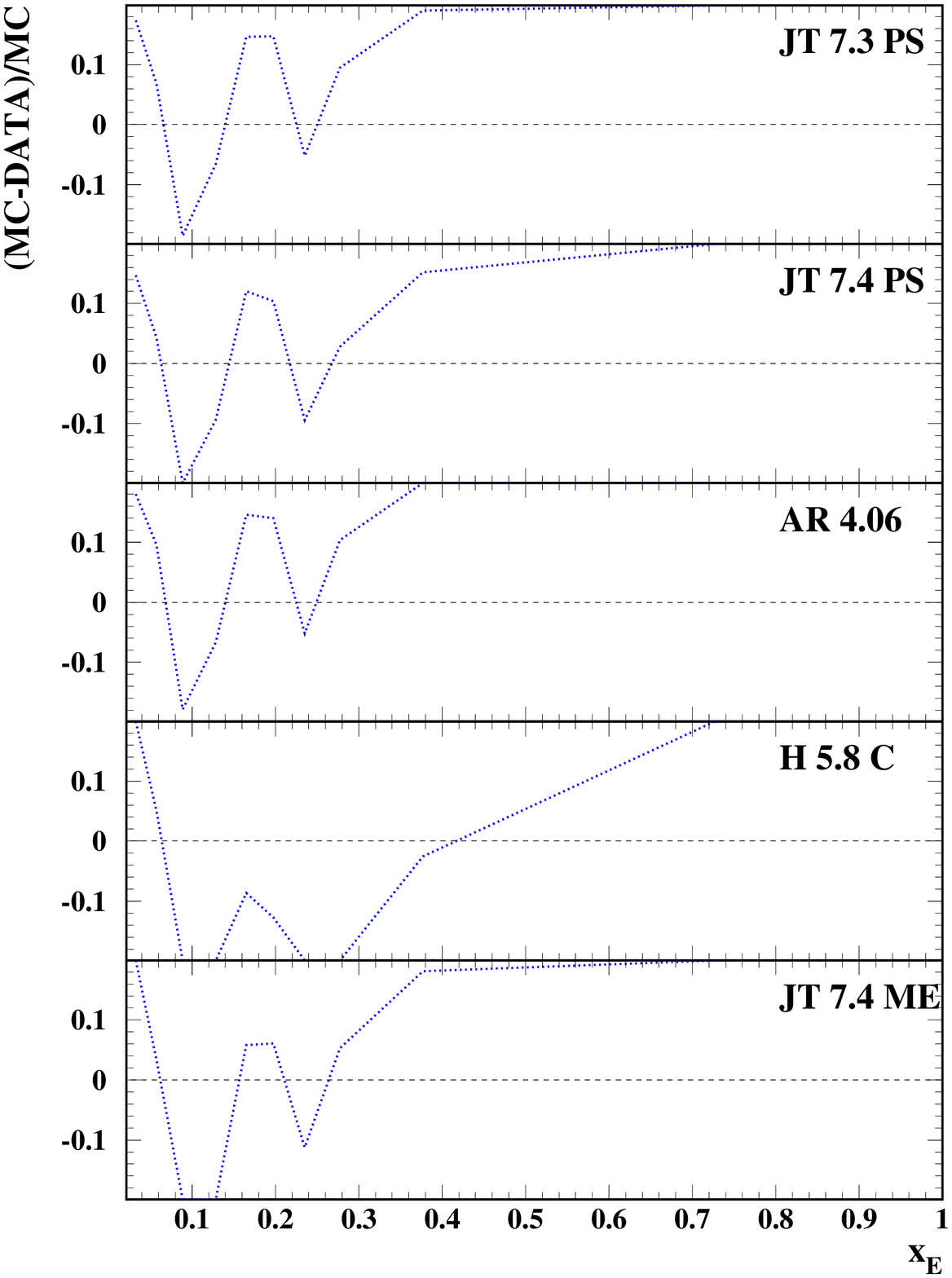,width=7.0cm}}
  \end{minipage}
\caption[$K^{\ast0}$ spectra]
{\label{bildkstar0} $x_E$ Distribution of $K^{\ast0}$}
\end{center}
\end{figure}

\begin{figure}
\begin{center}
\unitlength1cm
  \begin{minipage}[t]{6.7cm}
    \mbox{\epsfig{file=kstarpm_cern.epsc,width=7.0cm}}
  \end{minipage}
  \begin{minipage}[t]{6.7cm}
    \mbox{\epsfig{file=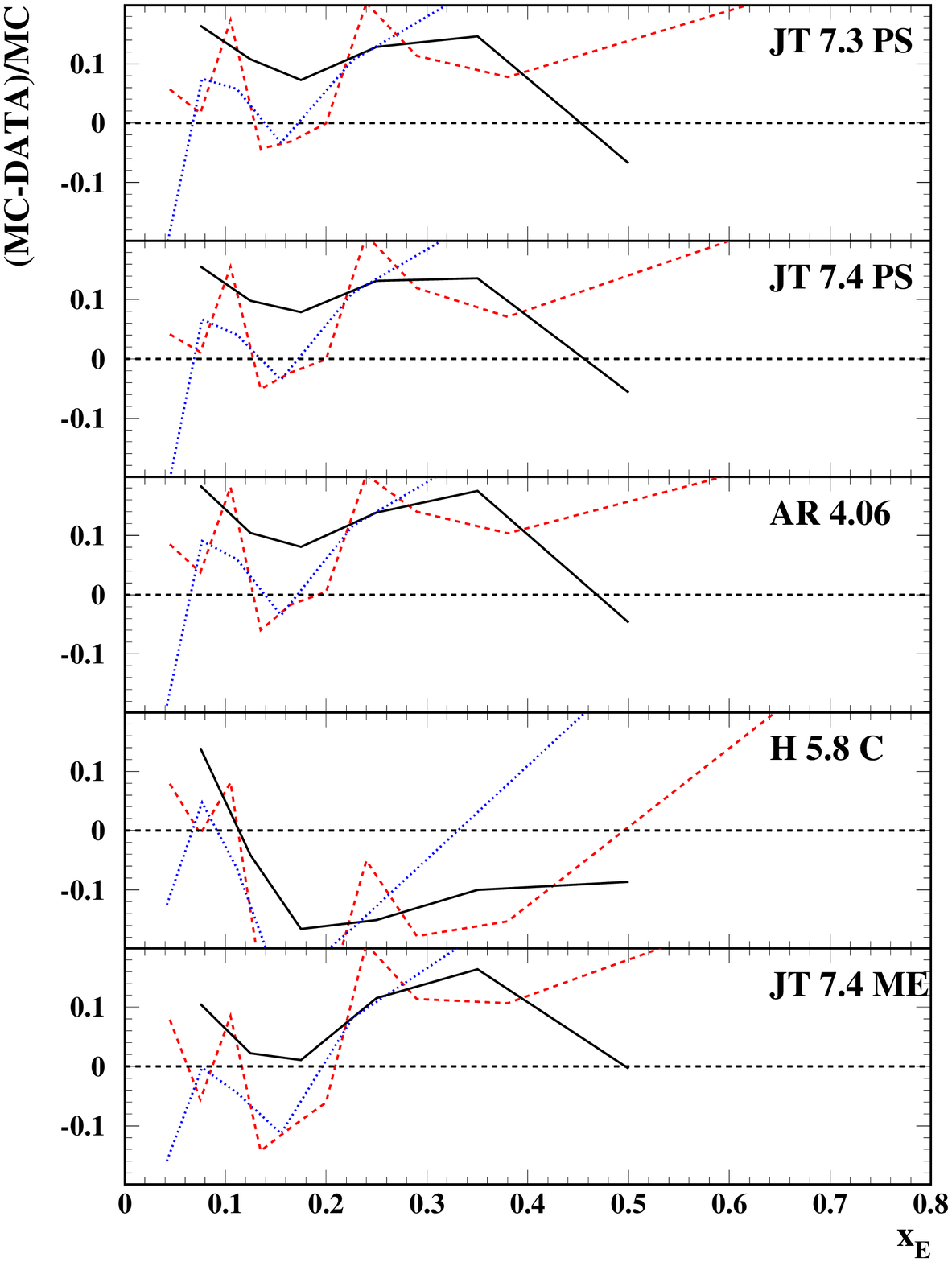,width=7.0cm}}
  \end{minipage}
\caption[$K^{\ast\pm}$ spectra]
{\label{bildkstarpm} $x_E$ Distribution of $K^{\ast\pm}$}
\end{center}
\end{figure}

\begin{figure}
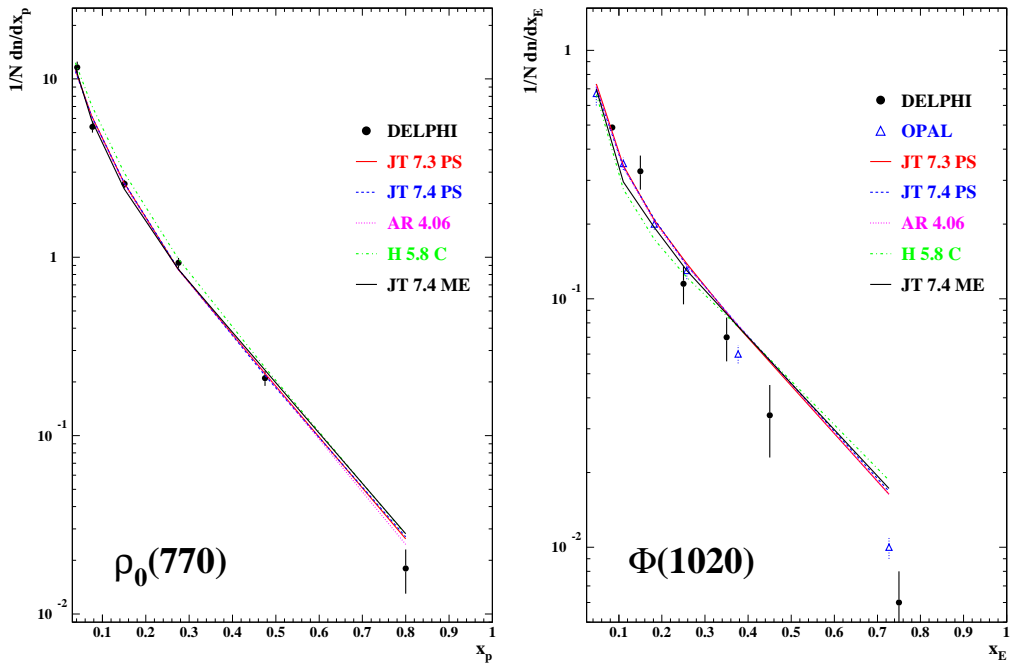

\begin{center}
\unitlength1cm
  \begin{minipage}[t]{6.7cm}
    \mbox{\epsfig{file=rho_cern.epsc,width=7.0cm}}
  \end{minipage}
  \begin{minipage}[t]{6.7cm}
    \mbox{\epsfig{file=phi_cern.epsc,width=7.0cm}}
  \end{minipage}
\caption[$\rho$ and $\phi$ spectra]{\label{bildphi} $x_p$ Distribution of
$\rho$ and $x_E$ Distribution of $\phi$}
\end{center}
\end{figure}

\begin{figure}
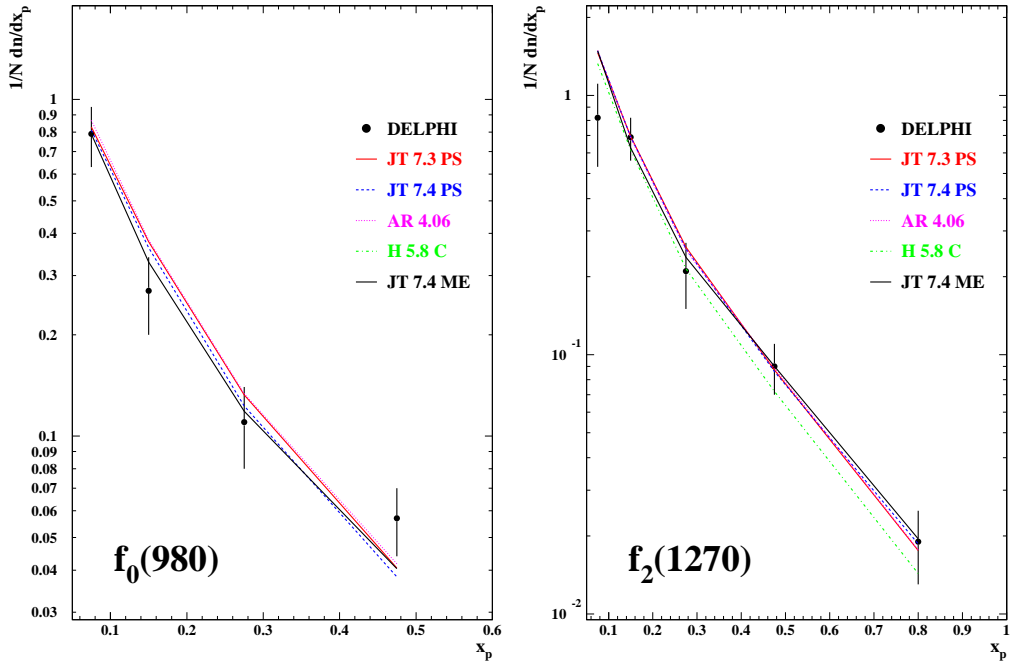

\begin{center}
\unitlength1cm
  \begin{minipage}[t]{6.7cm}
    \mbox{\epsfig{file=f0_cern.epsc,width=7.0cm}}
  \end{minipage}
  \begin{minipage}[t]{6.7cm}
    \mbox{\epsfig{file=f2_cern.epsc,width=7.0cm}}
  \end{minipage}
\caption[$f_0$ and $f_2$ spectra]
{\label{bildf0} $x_p$ Distribution of $f_0$ and $f_2$ }
\end{center}
\end{figure}

\begin{figure} [t]
\begin{center}
\unitlength1cm
  \begin{minipage}[t]{6.7cm}
    \mbox{\epsfig{file=proton_cern.epsc,width=7.0cm}}
  \end{minipage}
  \begin{minipage}[t]{6.7cm}
    \mbox{\epsfig{file=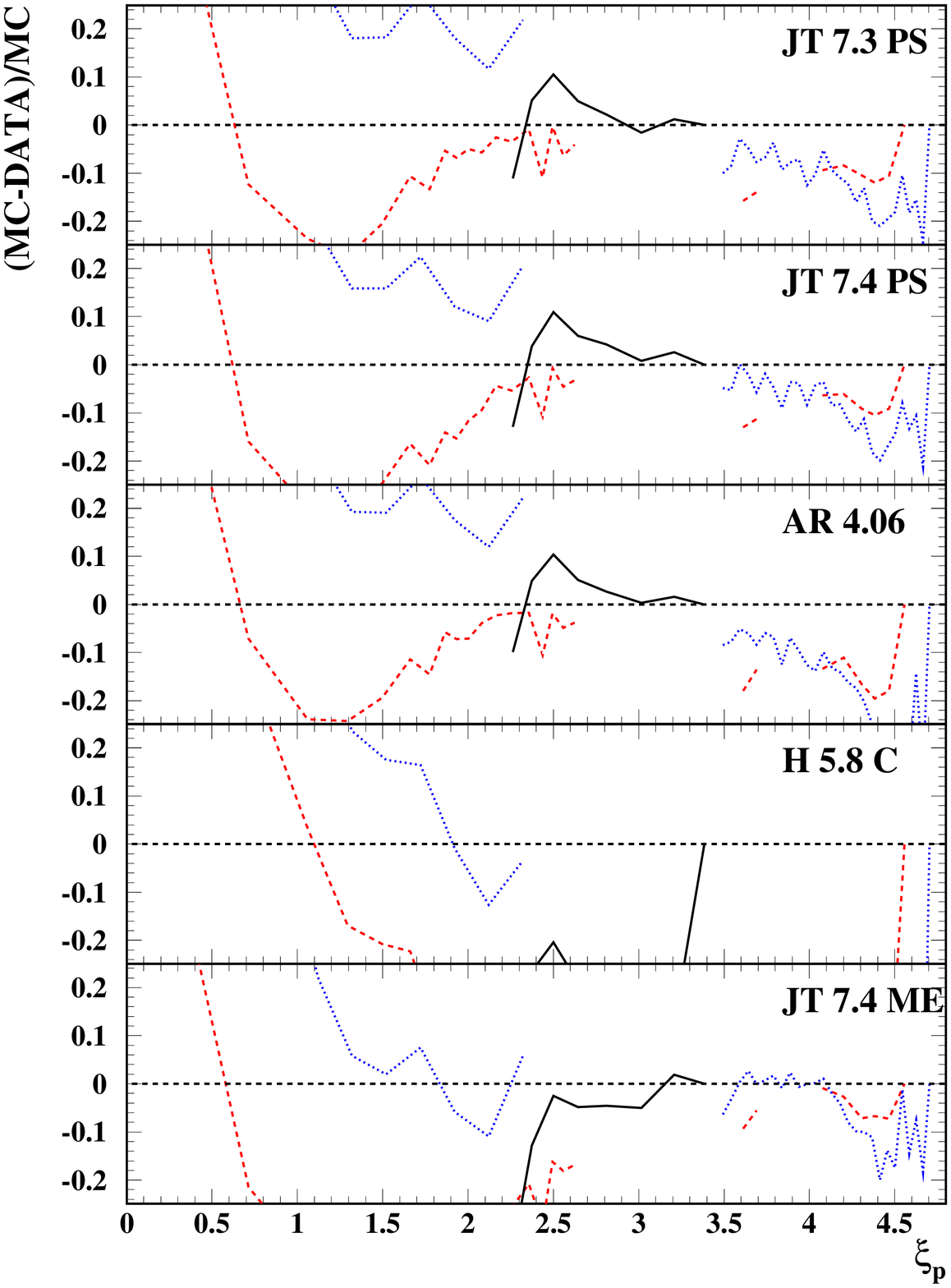,width=7.0cm}}
  \end{minipage}
\caption[p spectra]
{\label{bildproton} $\xi_p$ Distribution proton}
\end{center}
\end{figure}

\begin{figure}
\begin{center}
\unitlength1cm
  \begin{minipage}[t]{6.7cm}
    \mbox{\epsfig{file=lambda_cern.epsc,width=7.0cm}}
  \end{minipage}
  \begin{minipage}[t]{6.7cm}
    \mbox{\epsfig{file=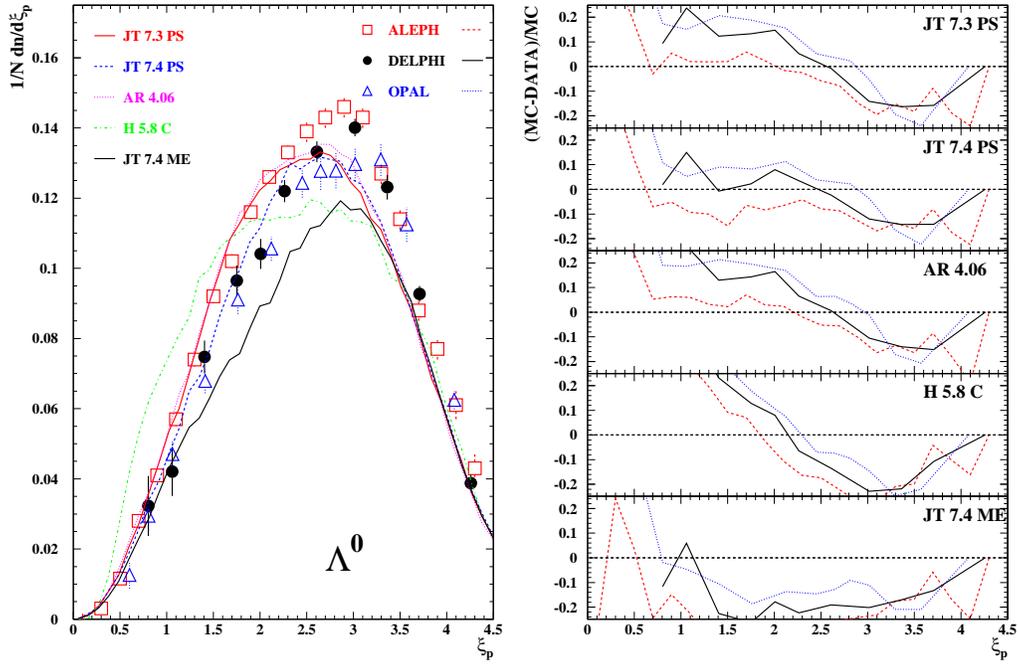,width=7.0cm}}
  \end{minipage}
\caption[$\Lambda$ spectra]
{\label{bildlambda} $xi_p$ Distribution $\Lambda^0$}
\end{center}
\end{figure}

\begin{figure} [t]
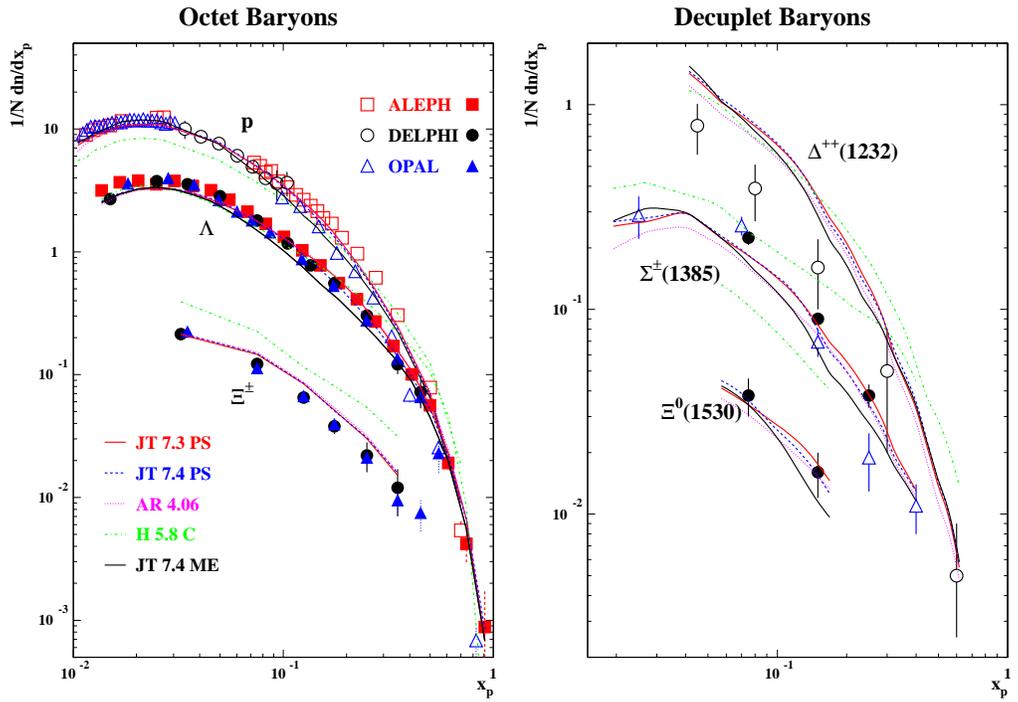

\begin{center}
\vspace{-2cm}
\unitlength1cm
  \begin{minipage}[t]{6.7cm}
    \mbox{\epsfig{file=oktett_cern.epsc,width=7.0cm}}
  \end{minipage}
  \begin{minipage}[t]{6.7cm}
    \mbox{\epsfig{file=dekuplett_cern.epsc,width=7.0cm}}
  \end{minipage}
\caption[Octet and Decuplet Baryons]
{\label{bildbaryons} $x_p$ Distribution for Octet and Decuplet Baryons}
\end{center}
\end{figure}
\clearpage

% ************************************************** Das Quellenverzeichnis


\begin{thebibliography}{99}\small      %\itemsep-.5ex

\bibitem{detector_paper}
DELPHI Coll., P. Aarnio et al., Nucl. Inst. Methods A303 (1991) 187
\bibitem{koralz}
F. Jadach, B.F.L. Ward, Z. Was, Comput. Phys. Commun. 40 (1986) 285;
Nucl. Phys. B253 (1985) 441
\bibitem{delsim}
DELPHI Coll., DELPHI 89-67(1989); DELPHI 89-68(1989).
\bibitem{jetset}
T. Sj\"ostrand, Comput. Phys. Commun. 39 (1986) 347;
T. Sj\"ostrand and M. Bengtsson,  Comput. Phys. Commun. 46 (1987) 367
\bibitem{dymu3}
J.E. Campagne and R. Zitoun, Z. Phys. C 43 (1989) 469
\bibitem{unfolding}
V. Blobel, Unfolding methods in high energy physics experiments, Lecture
given at the 1984 School of Computing, DESY 84-118;
V.B. Anikeev, V.P. Zhigunov, Phys. of Part. and Nucl. 24 Nr. 4 (1993) 424
\bibitem{fuerstenau}
H. F\"urstenau, Ph. D. Thesis, IEKP-KA 92-16
\bibitem{ariadne}
L.\ L\"{o}nnblad, Comp.\ Phys.\ Comm. 71 (1992) 15.
\bibitem{herwig}
G.\ Marchesini et al., Comp.\ Phys.\ Comm. 67 (1992) 465.
\bibitem{pet83}
C. Peterson, D. Schlatter, I. Schmitt and P. Zerwas, Phys. Rev.
D27 (1983) 105
\bibitem{lund_rev}
B.\ Andersson, G.\ Gustafson, G.\ Ingelman, T.\ Sj\"{o}strand,
Phys.\ Rep.  97 (1983) 31.
\bibitem{leningrad}
G. Gustavson, U. Petterson, Nucl. Phys. B308 (1988) 746
\bibitem{GKS}
F. Gutbrod, G. Kramer, G. Schierholz, Z. Phys. C21(1984) 235\\
K. Fabricius et al., Z. Phys C11(1981) 315
\bibitem{ERT}
R.K. Ellis, D.A. Ross, E.A. Terrano, Phys. Rev. Lett. 45 (1980) 1225,\\
Nucl. Phys. B178(1981) 421
\bibitem{sj_yb}
compare discussion and references in:
QCD Generators, B. Bambah et al. CERN 89-08 Vol. 3 p. 143 ff.
\bibitem{stevenson}
P.M. Stevenson, Phys. Rev. D23 (1981) 2916
\bibitem{herwig2}
compare discussion and references in:
QCD Generators, B. Bambah et al. CERN 89-08 Vol. 3 p. 235 ff.
\bibitem{ak0}
ALEPH Coll., D. Busculic et al., Z. Phys. C 64, 361 (1994)
\bibitem{a_fit}
ALEPH Coll., D. Decamp et al., Z. Phys. C 55, 209 (1992)
\bibitem{t_fit}
TASSO Coll., M. Althoff et al., Z. Phys. C 26 (1984) 157;
TASSO Coll., W. Braunschweig et al. Z. Phys. C 41 (1988) 359
\bibitem{l_fit}
L3 Coll., B. Adeva et al., Z. Phys. C 55 (1992) 39
\bibitem{svd}
W.H. Press, B.P. Flannery, S.A. Teukolsky, W.T. Vetterling, Numerical Recipes,
{\sl The Art of Scientific Computing}, Cambridge University Press
\bibitem{nag}
NAG library MK16, NAG ltd., Oxford, GBR.
\bibitem{minuit}
F. James, M. Goossens, {\sl MIUNIT, Function Minimization and
Error Analysis}, Reference Manual, CERN Program Library Long Writeup D506
(1992).
\bibitem{o_fit}
M.Z. Akrawy et al., Z. Phys. C 47 (1991) 505
\bibitem{mean_x_heavy}
V. Gibson, Charm and Beauty Hadron Production at $\sqrt{s} \approx M_Z$,
Proc. XXVII ICHEP, Glasgow 1994, P.J. Bussey and I. G. Knowles (ed.)
\bibitem{ak0}
ALEPH Collaboration, Production of $K^{\circ}$ and $\Lambda$ in Hadronic Z
Decays, CERN-PPE/94-74, submitted to Z. Phys. C.
\bibitem{oB**}
OPAL Coll., R. Akers et al. CERN-PPE/94-206, submitted to Z. Phys. C.
\bibitem{dB**}
DELPHI Coll., P. Abreu et al., Phys. Lett 345B (1995) 598
\bibitem{ap}
ALEPH Coll., D. Busculic et al., CERN-PPE/94-201, submitted to Z. Phys. C.
\bibitem{ok0}
OPAL Coll., R. Akers et al., CERN-PPE/95-24 (1995), submitted to Z. Phys. C.
\bibitem{op}
OPAL Coll., R. Akers et al., Z. Phys. C.63 (1994) 181
\bibitem{dp}
DELPHI Coll., P. Abreu et al., Nucl. Phys. B444 (1995) 3
\bibitem{okstar0}
OPAL Coll., Inclusive strange vector and tensor meson production in
hadronic $Z^{\circ}$ decays, preliminary results presented at ICHEP94 Glasgow.
\bibitem{akstar}
ALEPH Coll., preliminary results presented at ICHEP94 Glasgow, GLS0548.
Holger Hepp (ALEPH Coll.),\\
Inklusive Produktion von geladenen $K^*$-Mesonen
in hadronischen Z-Zerf\"allen, Diplomarbeit, HD-IHEP 93-04.
\bibitem{ostr_bary}
OPAL Collaboration, Strange Baryon Production and Correlations in Hadronic
$Z^{\circ}$ Decays, preliminary results presented at ICHEP94 Glasgow.
\bibitem{dlambda}
DELPHI Coll., P. Abreu et al., Phys. Lett. 318B (1993) 249
\bibitem{dstrange}
DELPHI Collaboration, Strange Baryon Production in $Z^{\circ}$ Hadronic Decays.
preliminary results presented at ICHEP94 Glasgow
\bibitem{drho}
DELPHI Coll., P. Abreu et al., Z. Phys. C65 (1995) 587
\bibitem{lomega}
L3 Coll., Y. Pei, talk given at the CERN PPE Seminar
\bibitem{leta}
L3 Coll., O. Adriani et al., Phys. Lett. 286 B (1992) 403.
L3 Coll., M. Acciari et al., Phys. Lett.  328 B (1994) 223
\bibitem{aeta}
ALEPH Coll., D. Buskulic et al., Phys. Lett. 292 B (1992) 210.
ALEPH Coll., Production Rates of $\eta$ and $\eta'$ in Hadronic Z
Decays, preliminary results presented at ICHEP94 Glasgow.
\bibitem{dlambda}
DELPHI Coll., P. Abreu et al., CERN-PPE/95-39, submitted to Z. Phys. C.
\bibitem{pdg}
Review of Particle Properties 1994, Phys. Rev. D50 (1994) 3
\bibitem{bepaper}
DELPHI Coll., P. Abreu et al., Z. Phys. C63 (1994)17, Phys. Lett. 323B (1994)
242, CERN-PPE/95-77, submitted to Phys. Lett.
\bibitem{ada}
A. De Angelis, J. Phys. G19, 1233 (1993)
\bibitem{multipl}
OPAL Collaboration, M.Z. Akrawy Z. Phys. C 47 (1990) 505,\\
ALEPH Collaboration, D. Decamp et al., Phys. Lett. 273B(1991)181,\\
DELPHI Collaboration, A. Abreu et al., Z. Phys. C 50(1991)185,\\
OPAL Collaboration, P.D. Acton et al.,  Z. Phys. C 53 (1992) 539.
\bibitem{b-inclusive}
DELPHI Coll. P. Abreu et al., Phys. Lett B347 (1995) 447
\bibitem{deltapp}
Delphi Coll., P. Abreu et al. DELPHI 95-50 PHYS 488, contribution
to the eps conference on High Energy Physics, Brussels 1995
\bibitem{thrust}
S. Brandt et al., Phys. Lett. 12 (64) 57;
E. Fahri, Phys. Rev. Lett. 39 (1977) 1587
\bibitem{sphericity}
J.D. Bjorken, S. Brodsky: Phys. Rev. D1 (1970) 1416
\bibitem{c_d}
R.K. Ellis, D.A. Ross, A.E. Terrano: Nucl. Phys. B178 (1981) 421
G. Parisi: Phys. Lett 74B (1978) 65.
J.F. Donohue, F.E. Low, S.Y. Pi: Phys. Rev. D20 (1979) 2759
\bibitem{clavelli}
L. Clavelli: Phys. Lett. B85 (1979) 111.
%\bibitem{fox}
%G.C. Fox, S. Wolfram: Nucl. Phys. B149 (1979) 413
\bibitem{jade}
JADE Coll., W. Bartel et al.: Z. Phys. C33 (1986) 23
JADE Coll., S. Bethke et al.: Phys. Lett. B213 (1988) 235
\bibitem{smolik}
L. Smolik, DESY 90-089
\bibitem{eec}
C.L. Basham, L.S. Brown, S.D. Ellis, S.T. Love:
Phys. Rev. D17 (1978) 2298,
Phys. Rev. Lett. 41 (1978) 1585,
Phys. Rev. D19 (1979) 2018
\bibitem{bbw}
S. Catani, G. Turnock, B.R. Webber, CERN-TH.6640/92 (1992)
\end{thebibliography}
\end{document}